\documentclass[aps, 11pt,prr,citeautoscript,superscriptaddress,nofootinbib]{revtex4}
\usepackage[utf8]{inputenc}
\usepackage{graphicx, float, pdfpages, physics, dsfont}
\usepackage{appendix}
\usepackage{xcolor, color}
\usepackage{epsfig}
\usepackage{verbatim}
\usepackage{bm}
\usepackage{mathrsfs}
\usepackage{yfonts}

\usepackage{times}
\usepackage{amsmath, amsthm, amssymb, amsfonts}
\usepackage{pst-all}
\usepackage{leftidx}
\usepackage{lmodern}

\newcommand{\Sgh}{\mathcal{S}^{({\bf g,h};z)}_{(a_{\bf g}, \mu)(b_{\bf h}, \nu)}}
\newcommand{\Tgh}{\mathcal{T}^{({\bf g,h};z)}_{(a_{\bf g}, \mu)(b_{\bf g}, \nu)}}
\newcommand{\Cgh}{\mathcal{C}^{({\bf g,h};z)}_{(a_{\bf g}, \mu)(b_{\bf \bar{h}\bar{g}h}, \nu)}}
\newcommand{\Qgh}{\mathcal{Q}^{({\bf g,h};z)}_{(a_{\bf g}, \mu)(b_{\bf g\bar{h}\bar{g}h g \bar{h}gh\bar{g}}, \nu)}}

\newcommand{\mb}[1]{\mathbf{#1}}
\newcommand{\gineq}[2]{\vcenter{\hbox{\includegraphics[width=#2cm]{#1}}}}

\begin{document}
\title{$G$-Crossed Modularity of Symmetry Enriched Topological Phases}
\author{Arman Babakhani}
\affiliation{Department of Chemistry, University of California, Santa Barbara, California 93106 USA}
\affiliation{Information Sciences Institute, Marina Del Rey, CA, 90292, USA}
\author{Parsa Bonderson}
\affiliation{Microsoft Station Q, Santa Barbara, California 93106-6105 USA}
\date{\today}
\begin{abstract}
The universal properties of $(2+1)$D topological phases of matter enriched by a symmetry group $G$ are described by $G$-crossed extensions of unitary modular tensor categories (UMTCs).
While the fusion and braiding properties of quasiparticles associated with the topological order are described by a UMTC, the $G$-crossed extensions further capture the properties of the symmetry action, fractionalization, and defects arising from the interplay of the symmetry with the topological order.
We describe the relation between the $G$-crossed UMTC and the topological state spaces on general surfaces that may include symmetry defect branch lines and boundaries that carry topological charge.
We define operators in terms of the $G$-crossed UMTC data that represent the mapping class transformations for such states on a torus with one boundary, and show that these operators provide projective representations of the mapping class groups.
This allows us to represent the mapping class group on general surfaces and ensures a consistent description of the corresponding symmetry enriched topological phases on general surfaces.
Our analysis also enables us to prove that a faithful $G$-crossed extension of a UMTC is necessarily $G$-crossed modular.
\end{abstract}
\maketitle


\section{Introduction}

Gapped quantum systems at zero temperature may manifest distinct topologically ordered phases of matter~\cite{Wen04,Nayak08}.
Topological phases are characterized by emergent phenomena that cannot be distinguished by local observables and do not require nor depend on symmetries of the system.
As such, these are universal properties of the phase that are robust to local perturbations, which can only change when the system undergoes a phase transition.
These topological phenomena include ground state degeneracies that depend only on the topology of the system and quasiparticle excitations that exhibit exotic exchange statistics.
For $(2+1)$D topological phases, the universal properties of the emergent quasiparticles, such as their fusion and exchange statistics, are captured by algebraic structures known as unitary modular tensor categories (UMTCs)~\cite{Moore89b,Turaev94,Bakalov01}.
Moreover, the UMTC describing a topological phase can be used to define a $(2+1)$D topological quantum field theory (TQFT), which describes the universal properties of the system on manifolds of arbitrary topology.
In this way, the UMTC encodes all the universal properties of a $(2+1)$D topological phase, except for the value of the chiral central charge, which it only encodes mod 8.

When topological order exists in a system with symmetry, the interplay between the two allows for a richer array of topological properties, giving rise to possibly distinct symmetry enriched topological (SET) phases.
This enriched structure includes that action of the symmetry on topological charges and states, fractionalization of symmetry quantum numbers carried by topologically nontrivial objects, and symmetry defects.
For $(2+1)$D SET phases with on-site unitary symmetries, the universal properties of these features are captured by algebraic structures known as $G$-crossed UMTCs~\cite{Turaev2000,turaev2010,Kirillov2004,Barkeshli2019}.
The symmetry defects in such theories have a similar notion of fusion, with a $G$-grading imposed, and a generalized notion of braiding, which incorporates the symmetry action and fractionalization, and extends them to defects.

A defining property of a UMTC is the notion of modularity, which is a non-degeneracy condition of braiding, i.e. the vacuum charge is the unique topological charge (quasiparticle type) which has trivial braiding with all topological charges.
Physically, this implies that braiding distinguishes between all topological charge values.
Equivalently, modularity of a unitary braided tensor category (UBTC) can be succinctly expressed as unitarity of a certain topological invariant known as the $S$-matrix, together with the condition that the number of topological charge types (simple objects) is finite.
Modularity implies that the $S$ and $T$ matrices of the theory provide a projective representation of the modular group, i.e. the mapping class group of the torus, yielding well-defined basis transformations of the state space.
More generally, it allows the UMTC to be used to define a $(2+1)$D TQFT~\cite{Turaev94,Bakalov01} which constitutes the low-energy effective theory of the topological phase.
In particular, this can be used to describe the topological state spaces and operations on surfaces of arbitrary topology and topological charge content, e.g. ascribed to boundaries or quasiparticles.

Modularity for $G$-crossed categories was discussed in Refs.~\onlinecite{Turaev2000,turaev2010,Kirillov2004,Barkeshli2019}.~\footnote{Refs.~\onlinecite{Turaev2000,turaev2010,Kirillov2004,Barkeshli2019} use three slightly different definitions of $G$-crossed modularity.
In this paper, we introduce yet another slightly different definition: a $G$-crossed UMTC is defined to be a $G$-crossed UBTC $\mathcal{B}_{G}^{\times}$ for which $\lvert\mathcal{B}_{g}\lvert$, the number of topological charges (simple objects) in each ${\bf g}$-sector, is finite for all ${\bf g}\in G$, and the operator $\boldsymbol{S}$ defined by Eq.~\eqref{eq:modular_S} is unitary.
We will show that these different definitions of $G$-crossed modularity are equivalent under the condition that $\mathcal{B}_{G}^{\times}$ is faithful, that is $\lvert\mathcal{B}_{g}\lvert\neq 0$ for all ${\bf g}\in G$.}
In Ref.~\onlinecite{Barkeshli2019}, $G$-crossed UMTCs were used to describe the topological state spaces of bosonic SET phases on the torus in the presence of symmetry defect branch lines, and also provide the representation of modular transformations on these states.
In this paper, we extend the analysis of $G$-crossed modularity to describe the topological state spaces and representations of the mapping class transformations of SET phases on the torus with a boundary, which can carry nontrivial topological charge as well as have symmetry defect branch lines terminating on it.
This allows one to describe the topological state space and operations of SET phases on surfaces of arbitrary topology and topological charge content in the presence of symmetry defects and branch lines.
Additionally, our analysis allows us to provide a direct proof that a $G$-crossed UBTC $\mathcal{B}_{G}^{\times}$ is $G$-crossed modular if and only if it is a faithful $G$-crossed extension of a UMTC $\mathcal{B}_{\bf 0}$.

\section{States and Modular Transformations on a Torus}
\label{sec:torus}

In this section, we review the description of topological ground states and modular transformations of SET phases on a 2D torus with no boundary, using a corresponding $G$-crossed UMTC $\mathcal{B}_{G}^{\times}$, following Ref.~\onlinecite{Barkeshli2019}.
We provide a brief review of the $G$-crossed UBTC formalism in Appendix~\ref{sec:review}.

\begin{figure}[t!]
    \centering
    \includegraphics[width=5.5cm]{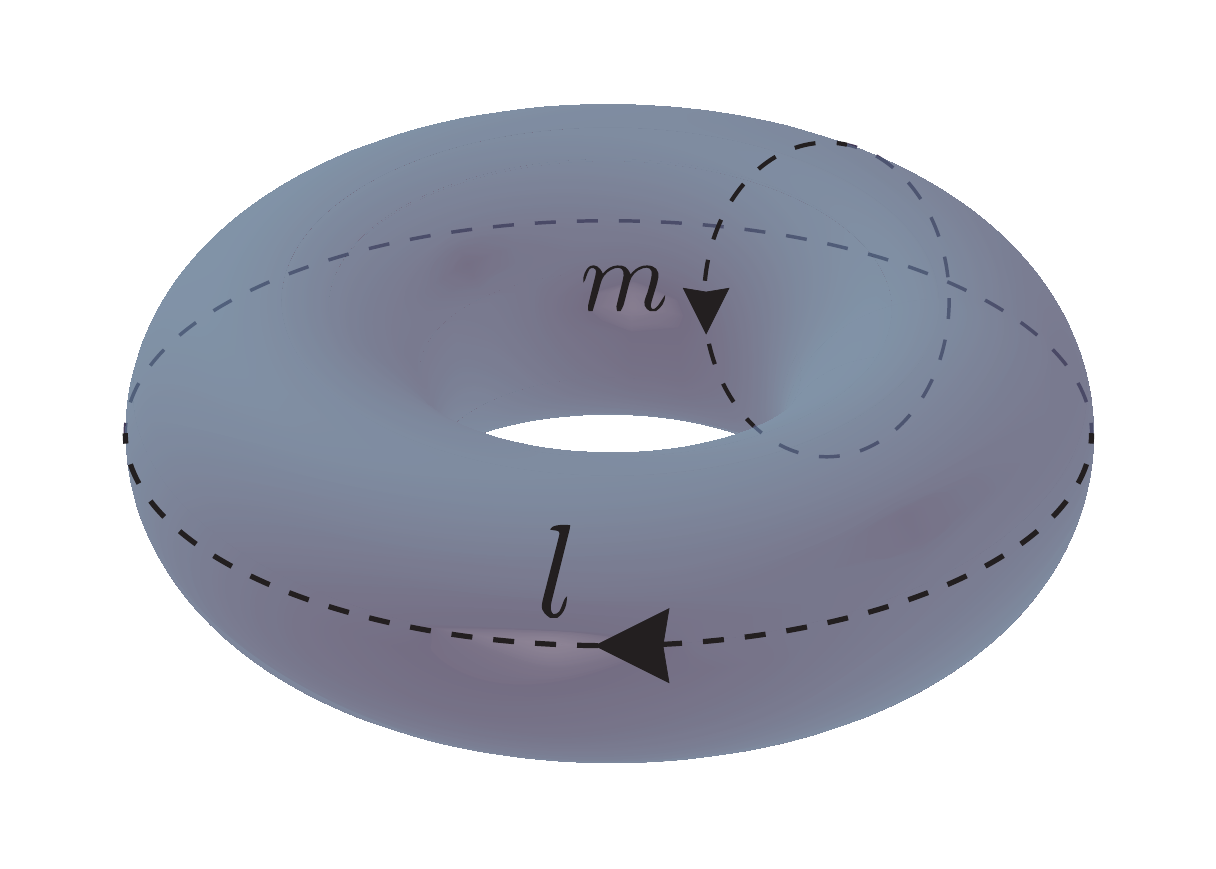}
    \caption{Longitudinal ($l$) and meridional ($m$) oriented generating cycles of a 2D torus for a particular embedding in 3D space.}
    \label{fig:torus_lm}
\end{figure}

A torus can be specified by an ordered pair of oriented generating cycles with intersection number $+1$.
We write a particular choice of such an ordered pair as $(l,m)$, which can be thought of as the longitudinal and meridional cycles for a particular embedding of the torus in 3D space, as shown in Fig.~\ref{fig:torus_lm}. 
Every choice of ordered pair of generating cycles with intersection number $+1$ can be related to each other through linear transformations
\begin{eqnarray}
\label{eq:modular_trans}
\begin{pmatrix}
   l' \\
   m'
\end{pmatrix}
=
\begin{bmatrix}
    \alpha & \beta \\
    \gamma & \delta
\end{bmatrix}
\begin{pmatrix}
   l \\
   m
\end{pmatrix} 
=
\begin{pmatrix}
   \alpha l + \beta m \\
   \gamma l + \delta m
\end{pmatrix}
,
\end{eqnarray}
where $\alpha, \beta, \gamma, \delta \in \mathbb{Z}$ and $\alpha \delta - \beta \gamma =1$.
These conditions on the linear transformation ensure generating pairs are mapped to generating pairs while preserving the intersection number.
These transformations form the modular group $\text{SL}(2,\mathbb{Z})$, which is isomorphic to the mapping class group of the torus, i.e. the automorphisms of the surface modulo the continuous deformations.

Denoting a compact, orientable surface with genus $g$ and $n$ boundary components as $\Sigma_{g,n}$, we can write the mapping class group of the torus $\Sigma_{1,0}$ in the presentation
\begin{equation}
\text{MCG}(\Sigma_{1,0}) \cong \langle \mathfrak{s}, \mathfrak{t} \, \lvert (\mathfrak{st})^3 = \mathfrak{s}^2, \mathfrak{s}^4=\mathds{1} \rangle
\label{eq:modulargroup}
\end{equation}
The generators used for this presentation can be chosen to correspond to the following linear transformations
\begin{align}
\mathfrak{s} \cong
\begin{bmatrix}
0 & -1 \\
1 & 0
\end{bmatrix}
,\qquad 
\mathfrak{t} \cong 
\begin{bmatrix}
1 & 1 \\
0 & 1
\end{bmatrix}
.
\end{align}
The generator $\mathfrak{s}$ interchanges the two generating cycles, with a relative sign that preserves the $+1$ intersection number.
The generator $\mathfrak{t}$, which is a ``Dehn twist'' around the second cycle, shears the torus.

As with topological phases without symmetry, a basis for the ground state space on a torus in the presence of symmetry defect branch lines is defined with respect to a choice of ordered pair of generating cycles.
Since there are no boundaries, the symmetry branch lines have no endpoints. (For these purposes, we treat bulk quasiparticles and defects as boundaries.)
As such, any configuration of branch lines can be deformed to a configuration with one branch loop around each of the two generating cycles.
We say that the system is in the $({\bf g,h})$-sector with respect to the pair $(l,m)$, when it can be deformed to a configuration with a ${\bf g}$-branch loop around the $l$ cycle and a ${\bf h}$-branch loop around the $m$ cycle.
The group elements ${\bf g}$ and ${\bf h}$ must commute, i.e. ${\bf gh} = {\bf hg}$, for such a configuration, as otherwise it would result in a nontrivial residual defect (branch line endpoint) that would require a boundary.
One can envision processes that transform the system between different $({\bf g,h})$-sectors by pair-creating defects, transporting them around nontrivial cycles, and then pair-annihilating the defects.
In particular, the $({\bf g,h})$-sector is obtained from the $({\bf 0,0})$-sector by creating a ${\bf h}$-${\bf \bar{h}}$ pair of defects, transporting the ${\bf h}$-defect around the $m$ cycle (once in the positive direction), and annihilating the pair of defects, and then creating a ${\bf g}$-${\bf \bar{g}}$ pair of defects, transporting the ${\bf g}$-defect around the $l$ cycle and then annihilating the pair of defects.

\begin{figure}[t!]
    \includegraphics[scale=.4]{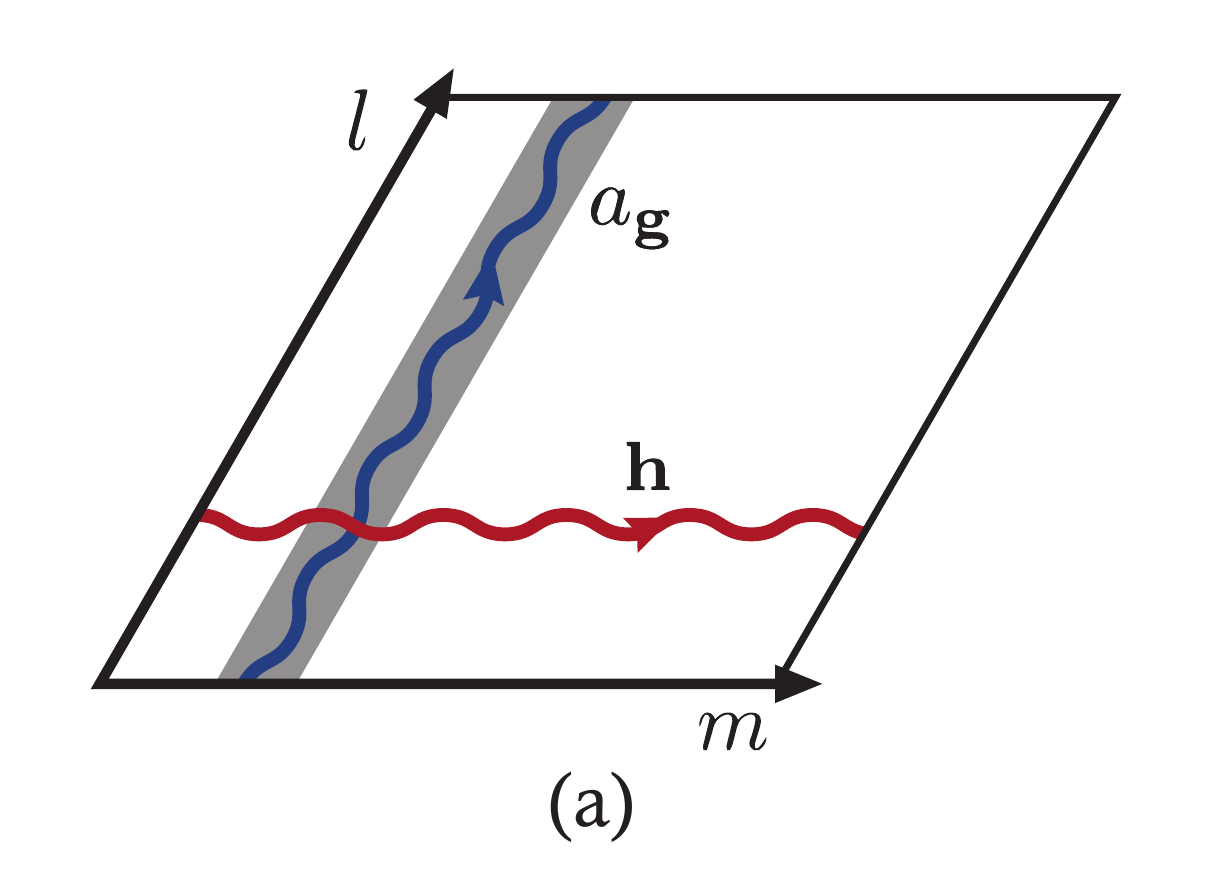}
    \includegraphics[scale=.4]{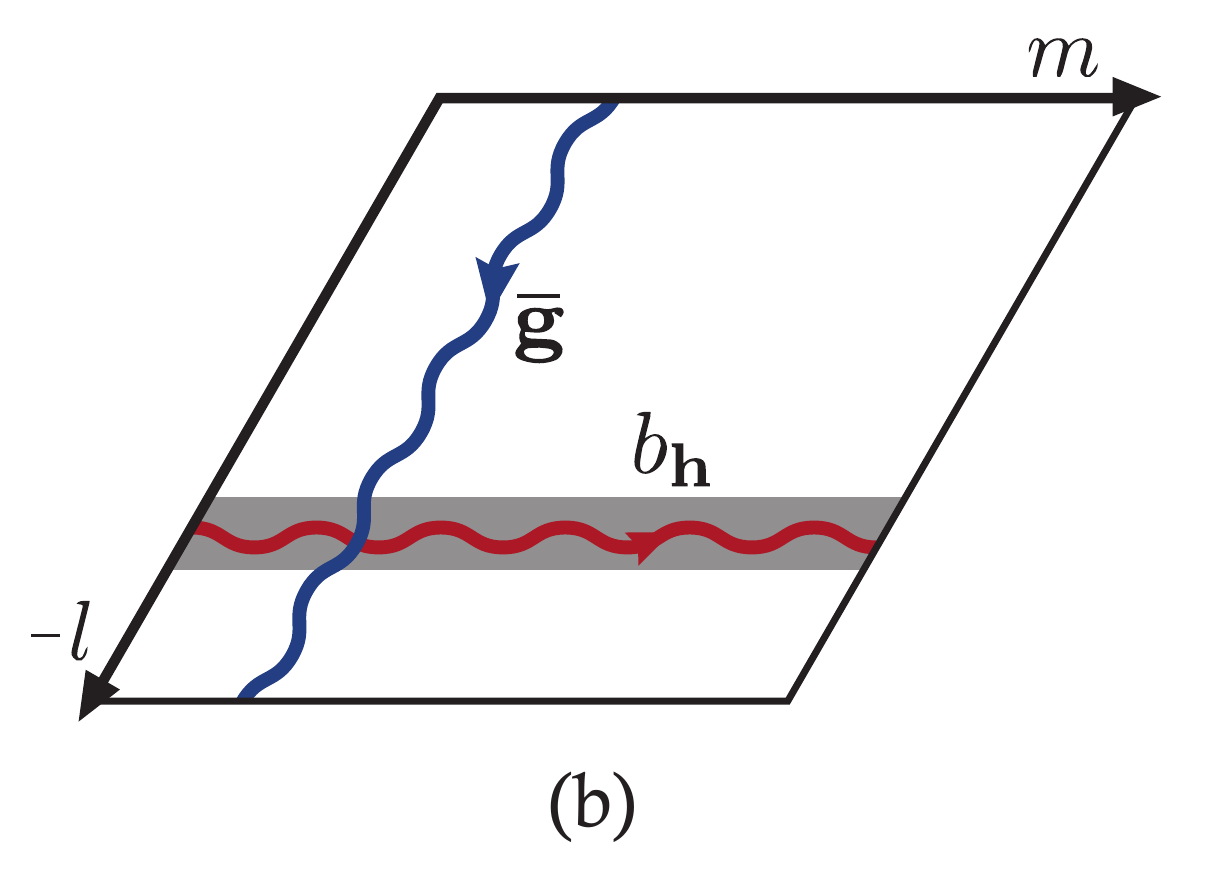}
    \qquad
    \includegraphics[scale=.4]{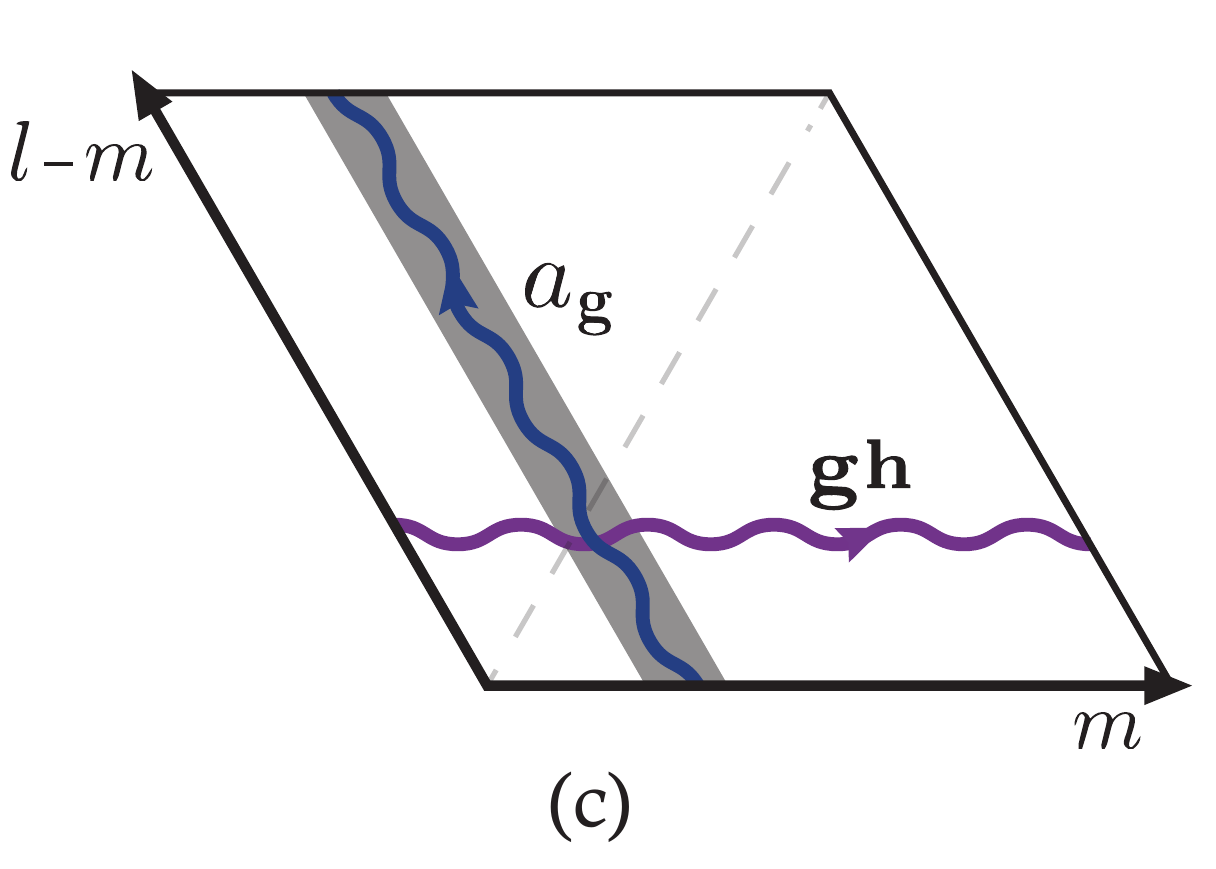}
    \caption{Planar representations of the topological ground states of the torus in the presence of symmetry defect branch lines for difference choices of pairs of generating cycles.
    The wavy lines in color represent defect branch lines, while the grey ribbons correspond to quasiparticle/defect ribbon operators of a specified topological charge.
    (a) Ground state $\ket{\Phi_{a_{\bf g}}}_{(l,m)}^{(\bf g,h)}$ in the $({\bf g,h})$ sector with respect to $(l,m)$.
    (b) Ground state $\ket{\Phi_{b_{\bf h}}}_{(m,-l)}^{(\bf h,\bar{g})}$ in the $({\bf h,\bar{g}})$ sector with respect to $(m,-l)$.
    (c) Ground state $\ket{\Phi_{a_{\bf g}}}_{(l-m,m)}^{(\bf gh,h)}$ in the $({\bf gh,h})$ sector with respect to $(l-m,m)$.
    The displayed configurations correspond to the same sector of defect branch line configurations, represented with respect to different pairs of generating cycles.
    The basis states with respect to these different choices of pairs of generating cycles are related by $G$-crossed modular transformations $\mathcal{S}$ and $\mathcal{T}$, as specified in Eqs.~(\ref{eq:modular_trans_S}) and (\ref{eq:modular_trans_T}).
    }
    \label{fig:2d_noz}
\end{figure}

We denote the topological ground state space on the torus in a given $({\bf g,h})$-sector as $\mathcal{H}_{\Sigma_{1,0}}^{({\bf g,h})}$.
We can write basis states for a given $({\bf g,h})$-sector with respect to $(l,m)$ in a similar manner as for a topological phase without symmetry.
In particular, we can write orthonormal basis states for the $({\bf g,h})$-sector as $\ket{\Phi_{a_{\bf g}}}_{(l,m)}^{(\bf g,h)}$, where $a_{\bf g} \in \mathcal{B}_{\bf g}^{\bf h}$. 
Here, $\mathcal{B}_{\bf g}^{\bf h} = \{ a_{\bf g} \in \mathcal{B}_{\bf g} \lvert  \,^{\bf h} a_{\bf g} = a_{\bf g}  \}$  is the subset of ${\bf g}$-defect topological charges that are ${\bf h}$-invariant.
It follows that
\begin{align}
\text{dim}\left( \mathcal{H}_{\Sigma_{1,0}}^{({\bf g,h})} \right) = \left\lvert  \mathcal{B}_{\bf g}^{\bf h} \right\lvert
.
\end{align}
The state $\ket{\Phi_{a_{\bf g}}}_{(l,m)}^{(\bf g,h)}$ corresponds to the configuration shown in Fig.~\ref{fig:2d_noz}(a), where there is a defect ribbon operator loop of definite charge $a_{\bf g}$ around the $l$ cycle.
For this state, if a topological charge measurement is performed around the $m$ cycle (which measures the topological charge passing through the loop $m$), the measurement outcome will have a definite outcome of $a_{\bf g}$.
On the other hand, the topological charge value passing through the loop $l$ would be in a superposition for this state, reflecting the fact that quasiparticle/defect ribbon operators that loop around the two different cycles are non-commuting operators, i.e. cannot be simultaneously diagonalized.
One can envision obtaining the state $\ket{\Phi_{a_{\bf g}}}_{(l,m)}^{(\bf g,h)}$ from $\ket{\Phi_{0_{\bf 0}}}_{(l,m)}^{(\bf 0,0)}$ by creating a ${\bf h}$-${\bf \bar{h}}$ pair of defects from vacuum (the topological charges of these are unimportant), transporting the ${\bf h}$-defect around the $m$ cycle, and annihilating the pair of defects, and then creating a $a_{\bf g}$-$\overline{a_{\bf g}}$ pair of defects from vacuum, transporting the $a_{\bf g}$-defect around the $l$ cycle and then annihilating the pair of defects.
Annihilation is not a unitary or deterministic process; rather, one should think of it as a fusion operation, involving a topological charge measurement of the pair, for which the fusion and measurement outcome is $0$, the vacuum fusion channel.
If we obtain an undesired fusion outcome from this process, we can simply repeat the preparation process until we obtain the desired fusion outcome $0$.
Since the $a_{\bf g}$-defect will cross the ${\bf h}$-branch line in this process, it must be ${\bf h}$-invariant to be able to pair-annihilate, which is why we required $a_{\bf g} \in \mathcal{B}_{\bf g}^{\bf h}$.
Otherwise, the resulting charge $^{\bf \bar{h}}a_{\bf g}$ could not fuse with $\overline{a_{\bf g}}$ into vacuum, i.e. there would necessarily be a nontrivial quasiparticle charge remaining.

We can represent these basis states for the topological ground state space on the torus diagrammatically as
\begin{align}
 \ket{ \Phi_{a_{\bf g}}}^{(\bf g,h)}_{(l,m)} = \, \gineq{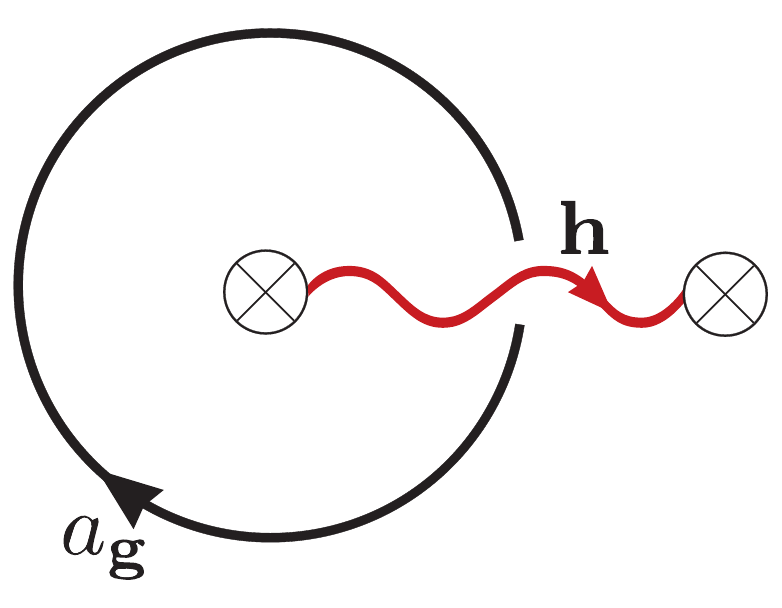}{2.5}
.
\end{align}
The $\otimes$ in the diagram represents the presence of a nontrivial cycle that one cannot na\"ively isotopy lines past.
We include a ${\bf h}$-branch line around that nontrivial cycle to track the fact that as the $a_{\bf g}$-ribbon operator traverses the cycle $l$, it crosses the cycle $m$ with the ${\bf h}$-branch loop around it.
We draw a wavy line to represent the ${\bf h}$-branch loop, indicating that it acts on the line passing under it, but otherwise does not actually contribute to the state space.
In this case, explicitly drawing this branch line in the diagrammatic representation may seem excessive, since we restrict to charges $a_{\bf g} \in \mathcal{B}_{\bf g}^{\bf h}$, but it becomes necessary later, when we include boundaries.

In the same manner, we can write a basis with respect to a different choice of generating cycles $(l',m')$.
When we make such changes of basis, related by Eq.~(\ref{eq:modular_trans}), the $({\bf g,h})$ sector with respect to $(l,m)$ corresponds to the $({\bf g',h'})$ sector with respect to $(l',m')$, where ${\bf g}' = {\bf g}^{\alpha} {\bf h}^{-\gamma}$ and ${\bf h}' = {\bf g}^{-\beta} {\bf h}^{\delta}$.
The basis states with respect to these choices are related by a linear transformation representing the corresponding mapping class group transformation.
For the generators $\mathfrak{s}$ and $\mathfrak{t}$, we write these changes of basis as
\begin{align}
\ket{\Phi_{a_{\bf g}}}^{(\bf g,h)}_{(l,m)} &= \sum\limits_{b \in \mathcal{B}_{\bf h}^{\bf \bar{g}}} \mathcal{S}^{(\bf g,h)}_{a_{\bf g} b_{\bf h}} \ket{\Phi_{b_{\bf h}}}^{(\bf h, \bar{g})}_{(m,-l)} 
\label{eq:modular_trans_S}
\\ 
&= \sum\limits_{b \in \mathcal{B}_{\bf g}^{\bf gh}} \mathcal{T}^{(\bf g,h)}_{a_{\bf g} b_{\bf g}} \ket{\Phi_{b_{\bf g}}}^{(\bf g, gh)}_{(l-m,m)} 
.
\label{eq:modular_trans_T}
\end{align}
The basis states written here correspond to the configurations shown in Fig.~\ref{fig:2d_noz}.

In Ref.~\onlinecite{Barkeshli2019}, it was shown that defining these transformations to be
\begin{align}
\mathcal{S}^{(\bf g,h)}_{a_{\bf g} b_{\bf h}} &= \frac{S_{a_{\bf g} b_{\bf h}}}{ U_{\bf h}(a_{\bf g} , \overline{a_{\bf g}}; 0) }
\label{eq:modular_S_0}
,  
\\ 
\mathcal{T}^{(\bf g,h)}_{a_{\bf g} b_{\bf g}} &= \eta_{a_{\bf g}}({\bf g,h}) \theta_{a_{\bf g}} \delta_{a_{\bf g} b_{\bf g}}
,
\label{eq:modular_T_0}
\end{align}
yields a projective representation of the modular group when the topological $S$-matrix is $G$-graded unitary.
(See Appendix~\ref{sec:review} for definitions of the quantities on the right hand sides of these equations.)
In particular, these definitions satisfy $(\mathcal{S} \mathcal{T})^{3} = \Theta_{\bf 0} \mathcal{S}^{2}$ and $\mathcal{S}^{4} = \openone$, where
\begin{align}
\Theta_{\bf 0} = \frac{1}{\mathcal{D}_{\bf 0}} \sum_{a \in \mathcal{B}_{\bf 0}} d_{a}^{2} \theta_{a}
,
\end{align}
is a phase related to the chiral central charge, as long as $\mathcal{B}_{\bf 0}$ is modular. 
We note that we must have $\lvert\mathcal{B}_{\bf g}^{\bf h}\lvert= \lvert\mathcal{B}_{\bf h}^{\bf \bar{g}}\lvert=\lvert\mathcal{B}_{\bf g}^{\bf gh}\lvert$ for the dimension of the state space to be the same for configurations related by modular transformations.
These conditions follow from the $G$-graded unitarity of the $S$-matrix and the fact that $\mathcal{B}_{\bf g}^{\bf gh}= \mathcal{B}_{\bf g}^{\bf h}$ when ${\bf gh}={\bf hg}$, since ${\bf g}$-defects are invariant under ${\bf g}$-action.

\section{States and Mapping Class Transformations on a Torus with a Boundary}
\label{sec:torus_with_boundary}

In the case of surfaces with boundaries, the situation becomes more complicated, as the boundaries can carry nontrivial topological charge and even have defect branch lines ending on them.
In order to deal with this, we first examine the case of a torus with a single boundary component.
The mapping class groups for surfaces with boundaries are defined using automorphisms that restrict to the identity along the boundary, i.e. each point on the boundary maps back to itself.
The mapping class group of the torus with one boundary component $\Sigma_{1,1}$ is given by the presentation
\begin{equation}
\text{MCG}(\Sigma_{1,1}) \cong \langle \mathfrak{s}, \mathfrak{t} \, \lvert (\mathfrak{st})^3 = \mathfrak{s}^2 \rangle
.
\label{eq:toruswboundaryMCG}
\end{equation}
This is almost the same as the mapping class group for the torus with no boundary, except it does not include the condition that $\mathfrak{s}^4$ is equal to identity.
In this case, one can show that the center of $\text{MCG}(\Sigma_{1,1})$ is isomorphic to $\mathbb{Z}$ and generated by $\mathfrak{s}^2$.
We can also see that $\mathfrak{s}^4= \mathfrak{t}_{q}^{-1}$, where $\mathfrak{t}_{q}$ is a Dehn twist around a cycle $q$ that is homotopic to the boundary.

We can describe the torus with a boundary in a similar manner as the torus without a boundary, using a pair of oriented generating cycles $(l,m)$ with intersection number $+1$.
However, accounting for the boundary requires additional information that tracks the twisting of the surface with respect to the boundary, generated by $\mathfrak{t}_{q}$.
This can done by choosing a point on the boundary and a oriented path between this point and a point along the first cycle of the generating pair, and associating a label $w$ to this.
We then let the mapping class generators transform $(l,m)$ as in Eq.~(\ref{eq:modular_trans}), while transforming $w$ to $w'=w - \frac{1}{4}$ under $\mathfrak{s}$ and $w'=w$ under $\mathfrak{t}$.
Clearly $\frac{1}{4}$ differences of $w$ do not provide well-defined comparisons by themselves, since they are always accompanied by different values of $(l,m)$.
However, they are chosen so that application of $\mathfrak{s}^{-4}$ yields $w'=w + 1$ while returning to $(l',m')=(l,m)$, i.e. this quantity tracks the twisting around the cycle $q$ corresponding to $\mathfrak{t}_{q}$, as intended.

\begin{figure}[t!]
    \includegraphics[scale=1]{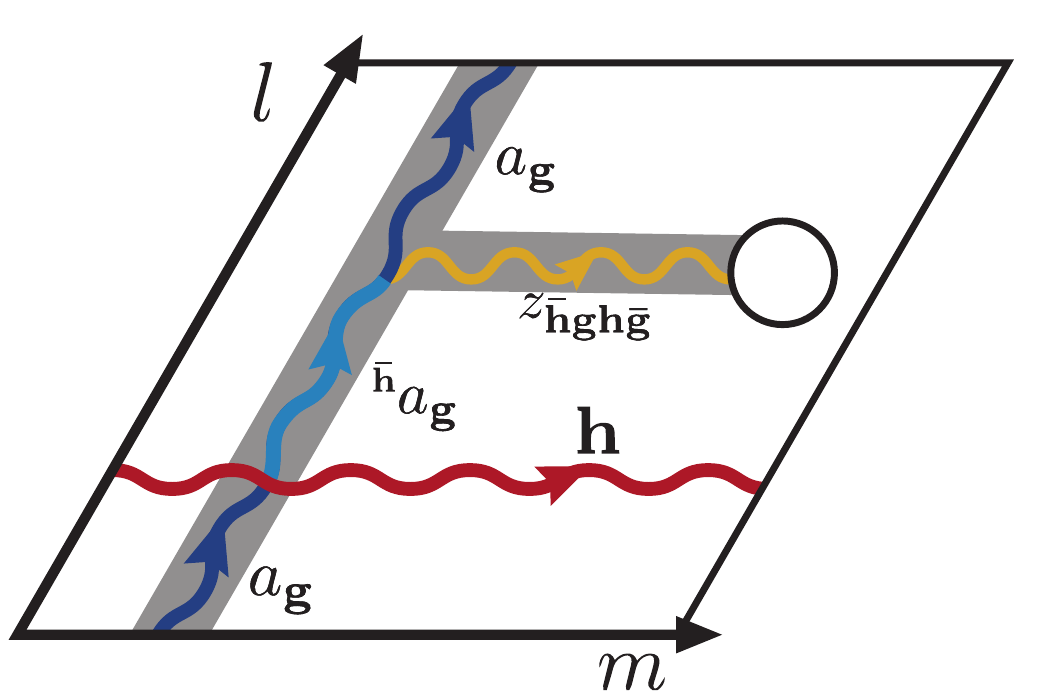}
    \caption{ 
    Planar representation of the topological ground states of the torus with a boundary in the presence of symmetry defect branch lines.
    The boundary can carry a potentially nontrivial topological charge $z$, which may correspond to a nontrivial defect.
    The wavy lines in color represent defect branch lines, while the grey ribbons correspond to quasiparticle/defect ribbon operators of a specified topological charge.
    The configuration shown corresponds to the ground states $\ket{\Phi_{a_{\bf g} ; z_{\bf \bar{h}gh\bar{g}}, \mu}}_{(l,m;w)}^{({\bf g,h})}$ in the $({\bf g,h} ; z_{\bf \bar{h}gh\bar{g}})$ sector with respect to $(l,m;w)$, where $w$ tracks the twisting of the bulk configuration with respect to the boundary.
    In this case, ${\bf g}$ and ${\bf h}$ are not required to commute, though the boundary necessarily carries a ${\bf \bar{h}gh\bar{g}}$-defect charge.
    } 
    \label{fig:torus_2d}
\end{figure}

In order to describe states on the torus with a boundary, we must specify the topological charge on the boundary, as well as the configuration of defect branch lines.
We can deform the branch lines into a configuration with a branch line looping around each of the generating cycles $l$ and $m$, but now there can also be a nontrivial branch line connecting these loops to the boundary.
We make a consistent choice of the path of this connecting line to be from the loop around the first cycle $l$ of the generating pair to boundary.
Since the boundary can carry a defect charge, the symmetry group elements assigned to the defect branch lines around the generating cycles need not commute.
We say that the system is in the $({\bf g,h} ; z_{\bf \bar{h}gh\bar{g}})$-sector with respect to $(l,m;w)$ when the boundary carries topological charge $z_{\bf \bar{h}gh\bar{g}}$, and the defect branch lines can be deformed into the following configuration: the generating cycle $m$ has a ${\bf h}$-defect branch line around it, the boundary defect line connecting the cycle $l$ and the boundary has value ${\bf \bar{h}gh\bar{g}}$, and the cycle $l$ has a branch line around it with value ${\bf g}$ between the junction with the boundary defect line and the ${\bf h}$-defect branch line, and value ${\bf \bar{h}gh}$ between the ${\bf h}$-defect branch line and the junction with the boundary defect line.
This configuration is shown in Fig.~\ref{fig:torus_2d}.

We denote the topological ground state space on the torus with a boundary in a given $({\bf g,h} ; z_{\bf \bar{h}gh\bar{g}})$-sector as $\mathcal{H}_{\Sigma_{1,1}}^{({\bf g,h} ; z_{\bf \bar{h}gh\bar{g}})}$.
Using the conventions and normalizations of Ref.~\onlinecite{Bonderson2017} for representing states on general surfaces, with modifications to account for the inclusion of symmetry defects, we can write orthonormal basis states for the $({\bf g,h} ; z_{\bf \bar{h}gh\bar{g}})$-sector with respect to $(l,m;w)$ as
\begin{equation}
\ket{\Phi_{a_{\bf g} ; \,  z_{\bf \bar{h}gh\bar{g}}, \mu}}_{(l,m;w)}^{({\bf g,h})}
=\ket{(a_{\bf g})^{\bf h} ; \overline{z_{\bf \bar{h}gh\bar{g}}}, \mu}
\ket{\overline{z_{\bf \bar{h}gh\bar{g}}}, z_{\bf \bar{h}gh\bar{g}};0}
=\frac{1}{d_z^{1/4}}\gineq{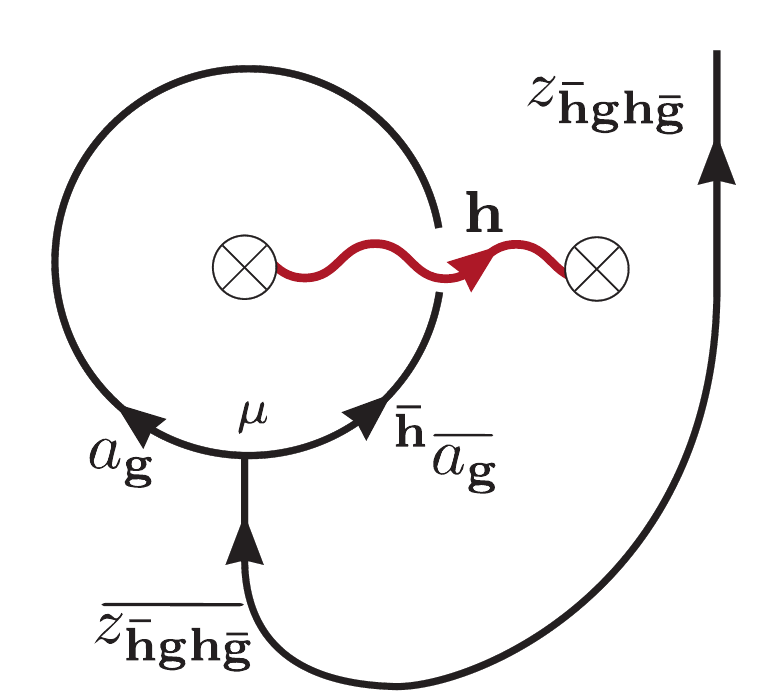}{3.5}
,
\label{eq:state_ag}
\end{equation}
where the index $\mu \in \{1,\ldots,N_{a_{\bf g}, \,^{\bf \bar{h}}\overline{a_{\bf g}}}^{\overline{z_{\bf \bar{h}gh\bar{g}}}} \}$ is associated with the fusion multiplicity of fusing $a_{\bf g}$ with $^{\bf \bar{h}}\overline{a_{\bf g}}$ to obtain $\overline{z_{\bf \bar{h}gh\bar{g}}}$.
This corresponds to the configuration shown in Fig.~\ref{fig:torus_2d}, where the defect ribbon operator around the $l$ cycle has definite charge value $a_{\bf g}$ on the segment between the junction with the boundary defect line and the ${\bf h}$-defect branch line, and definite charge value $^{\bf \bar{h}}a_{\bf g}$ between the ${\bf h}$-defect branch line and the junction with the boundary defect line, while the boundary line defect ribbon operator has definite charge value $z_{\bf \bar{h}gh\bar{g}}$.
The fusion multiplicity can be interpreted as the number of distinct (orthogonal) ways one can form a ribbon operator trijunction for the given topological charges.
From this, we see that
\begin{align}
\text{dim}\left( \mathcal{H}_{\Sigma_{1,1}}^{({\bf g,h};z_{\bf \bar{h}gh\bar{g}})} \right) = \sum_{a \in \mathcal{B}_{\bf g}} N_{a_{\bf g}, \,^{\bf \bar{h}}\overline{a_{\bf g}}}^{\overline{z_{\bf \bar{h}gh\bar{g}}}}
= \sum_{a \in \mathcal{B}_{\bf g}} N_{\,^{\bf \bar{h}}a_{\bf g}, \overline{a_{\bf g}}}^{z_{\bf \bar{h}gh\bar{g}}} 
.
\end{align}
We note that this can potentially be zero for a given $({\bf g,h};z_{\bf \bar{h}gh\bar{g}})$-sector, which indicates that such a sector has no topological ground states.
For example, when $z=0$, we see that $\text{dim}\left( \mathcal{H}_{\Sigma_{1,1}}^{({\bf g,h};0)} \right) =0$ when ${\bf gh} \neq {\bf hg}$.
For $z=0$, we also find that $a_{\bf g} \in \mathcal{B}_{\bf g}^{\bf h}$ in order to have $N_{a_{\bf g}, \,^{\bf \bar{h}}\overline{a_{\bf g}}}^{\overline{z_{\bf \bar{h}gh\bar{g}}}} \neq 0$, recovering $\text{dim}\left( \mathcal{H}_{\Sigma_{1,1}}^{({\bf g,h};0)} \right) = \lvert\mathcal{B}_{\bf g}^{\bf h}\lvert$ when ${\bf gh} = {\bf hg}$, as for the torus with no boundary.

For the state $\ket{\Phi_{a_{\bf g} ; \,  z_{\bf \bar{h}gh\bar{g}}, \mu}}_{(l,m;w)}^{({\bf g,h})}$ in Eq.~(\ref{eq:state_ag}), if a topological charge measurement is performed around the $m$ cycle in the region where it crosses a ${\bf g}$-branch line, the measurement outcome will have a definite outcome of $a_{\bf g}$; if a topological charge measurement is performed around the $m$ cycle in the region where it crosses a ${\bf \bar{h}gh}$-branch line, the measurement outcome will have a definite outcome of $^{\bf \bar{h}}a_{\bf g}$.
We emphasize that such measurements will generally require a separate way of determining whether one is performing the measurement in the region corresponding to a ${\bf g}$-branch line or a ${\bf \bar{h}gh}$-branch line in order to properly calibrate the measurement.
A topological charge measurement performed around a cycle homotopic to the boundary will have a definite measurement outcome of $z_{\bf \bar{h}gh\bar{g}}$.
The topological charge value passing through the loop $l$ would generally be in a superposition for this state.
One can envision obtaining the state $\ket{\Phi_{a_{\bf g} ; \,  z_{\bf \bar{h}gh\bar{g}}, \mu}}_{(l,m;w)}^{({\bf g,h})}$ from $\ket{\Phi_{0_{\bf 0} ; \,  0}}_{(l,m;w)}^{({\bf 0,0})}$ by creating a ${\bf h}$-${\bf \bar{h}}$ pair of defects from vacuum, transporting the ${\bf h}$-defect around the $m$ cycle, and annihilating the pair of defects, and then creating a $\overline{z_{\bf \bar{h}gh\bar{g}}}$-$z_{\bf \bar{h}gh\bar{g}}$ pair of defects from vacuum, transporting the $z_{\bf \bar{h}gh\bar{g}}$-defect to the boundary, and then splitting the $\overline{z_{\bf \bar{h}gh\bar{g}}}$-defect into a $a_{\bf g}$-$^{\bf \bar{h}}\overline{a_{\bf g}}$ pair of defects, transporting the $a_{\bf g}$-defect around the $l$ cycle and then annihilating this pair of defects.
Alternatively, one could create a ${\bf h}$-${\bf \bar{h}}$ pair of defects from vacuum, transport the ${\bf h}$-defect around the $m$ cycle, and annihilate the pair of defects, and then create a $a_{\bf g}$-$\overline{a_{\bf g}}$ pair of defects from vacuum, transport the $a_{\bf g}$-defect around the $l$ cycle and then fuse the pair of defects to obtain a $z_{\bf \bar{h}gh\bar{g}}$-defect, which is then transported to the boundary.
As is the case for pair-annihilation, the fusion process will generally have a probabilistic outcome; if a specific outcome $z$ is desired, we repeat the preparation process until it is obtained.

\begin{figure}[t!]
    \includegraphics[scale=.65]{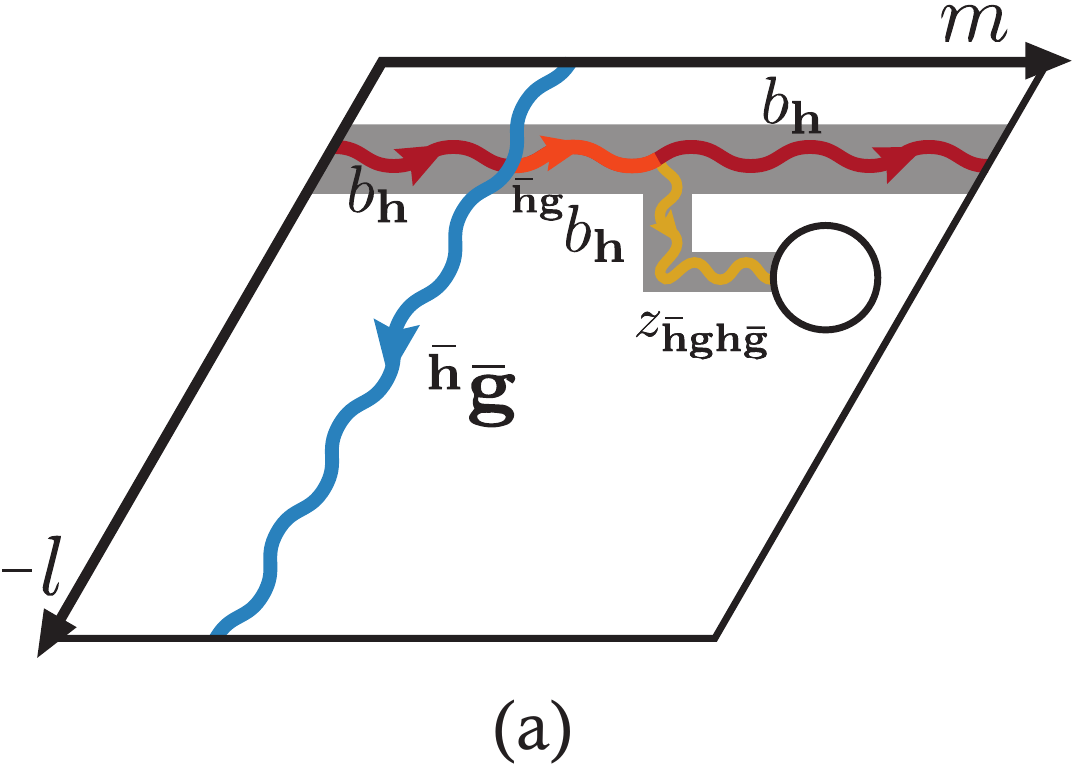}
    \qquad \qquad
    \includegraphics[scale=.65]{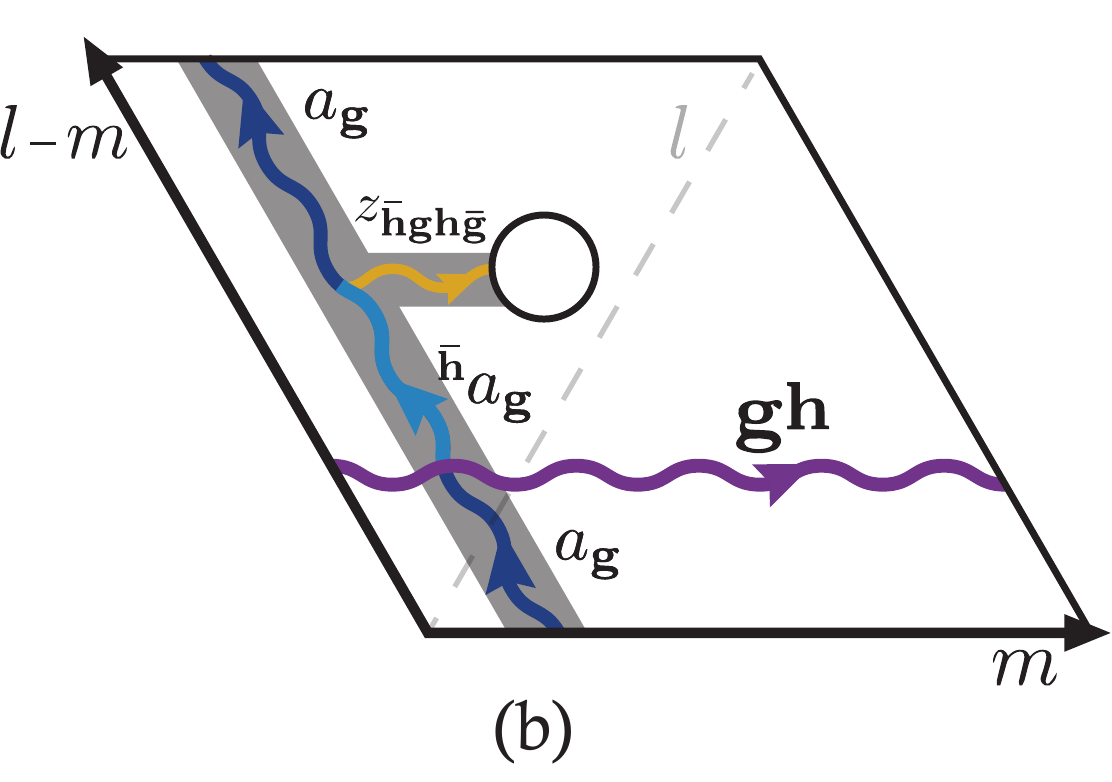}
    \caption{
    Planar representation of the topological ground states of the torus with a boundary in the presence of symmetry defect branch lines.
    (a) The configuration shown corresponds to the ground states $\ket{\Phi_{b_{\bf h} ; z_{\bf \bar{h}gh\bar{g}}, \mu}}_{(m,-l;w+\frac{1}{4})}^{({\bf h,\bar{h}\bar{g}h})}$ in the $({\bf h,\bar{h}\bar{g}h} ; z_{\bf \bar{h}gh\bar{g}})$ sector with respect to $(m,-l;w+\frac{1}{4})$.
    (b) The configuration shown corresponds to the ground states $\ket{\Phi_{a_{\bf g} ; \,  z_{\bf \bar{h}gh\bar{g}}, \mu}}_{(l-m,m;w)}^{({\bf g,gh})}$ in the $({\bf g,gh} ; z_{\bf \bar{h}gh\bar{g}})$ sector with respect to $(l-m,m;w)$.
    These basis states are related to those described in Fig.~\ref{fig:torus_2d} by $G$-crossed mapping class transformations, as specified in Eqs.~(\ref{eq:modular_trans_Sz}) and (\ref{eq:modular_trans_Tz}).
    The defect ribbon operator ends at the same marked point on the boundary for all of these configurations and states, reflecting the fact that the mapping class transformations leave the boundary fixed. 
    } 
    \label{fig:torus_2d_ST}
\end{figure}

We can write basis states with respect to a different choice of $(l',m';w')$.
These basis changes can again be generated by the mapping class transformations, so we focus on the choices related by the generators $\mathfrak{s}$ and $\mathfrak{t}$ of the mapping class group.
We write the basis states corresponding to the configuration shown in Fig.~\ref{fig:torus_2d_ST}(a) with respect to $(m,-l;w+\frac{1}{4})$ as
\begin{equation}
\ket{\Phi_{b_{\bf h} ; \,  z_{\bf \bar{h}gh\bar{g}}, \mu}}_{(m,-l;w+\frac{1}{4})}^{({\bf h,\bar{h}\bar{g}h})}
=\ket{(b_{\bf h})^{\bf \bar{h}\bar{g}h} ; \overline{z_{\bf \bar{h}gh\bar{g}}}, \mu}
\ket{\overline{z_{\bf \bar{h}gh\bar{g}}}, z_{\bf \bar{h}gh\bar{g}};0}
=\frac{1}{d_z^{1/4}}\gineq{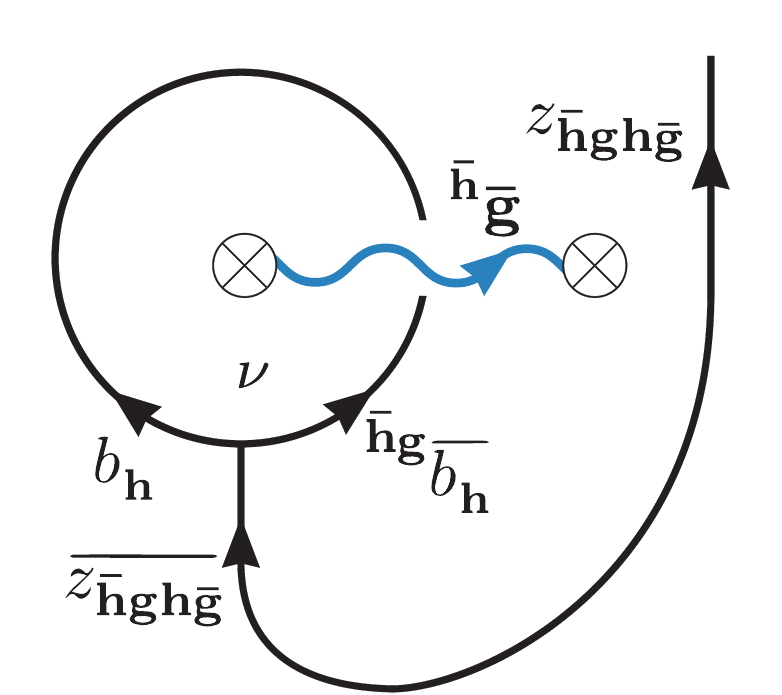}{3.5}
.
\label{eq:state_bh}
\end{equation}
We emphasize that, since ${\bf g}$ and ${\bf h}$ are not required to commute, the group labels of defect lines can change in a more complicated manner under mapping class transformations than in the case of the torus with no boundary.
We similarly write the basis states corresponding to the configuration shown in Fig.~\ref{fig:torus_2d_ST}(b) with respect to $(l-m,m;w)$ as \begin{equation}
\ket{\Phi_{a_{\bf g} ; \,  z_{\bf \bar{h}gh\bar{g}}, \mu}}_{(l-m,m;w)}^{({\bf g,gh})}
=\ket{(a_{\bf g})^{\bf gh} ; \overline{z_{\bf \bar{h}gh\bar{g}}}, \mu}
\ket{\overline{z_{\bf \bar{h}gh\bar{g}}}, z_{\bf \bar{h}gh\bar{g}};0}
=\frac{1}{d_z^{1/4}}\gineq{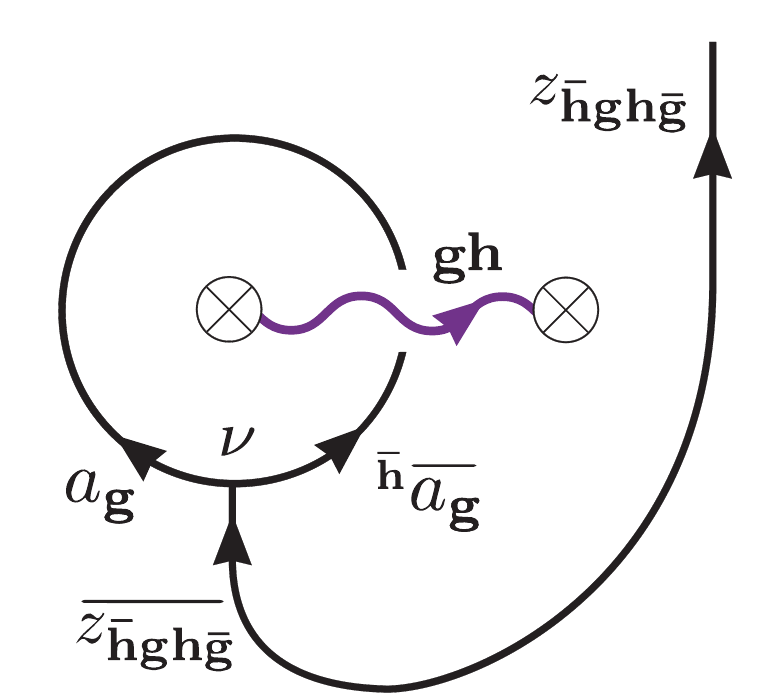}{3.5}
.
\label{eq:state_bg}
\end{equation}
An important point in writing different basis choices is that the boundary is fixed under mapping class transformations.
In particular, this means the $z_{\bf \bar{h}gh\bar{g}}$-defect ribbon operator must attach to the boundary at the same marked point for all the different bases.
We see in Fig.~\ref{fig:torus_2d_ST} that $w+\frac{1}{4}$ corresponds to a ``quarter twist'' of the boundary defect line as compared to $w$.

Similar to the case of a torus without boundary, we write the change of basis transformations corresponding to the $\mathfrak{s}$ and $\mathfrak{t}$ generators of the mapping class group as
\begin{align}
\ket{\Phi_{a_{\bf g} ; \,  z_{\bf \bar{h}gh\bar{g}}, \nu}}_{(l,m;w)}^{({\bf g,h})}
&= \sum_{b_{\bf h}, \nu} \mathcal{S}_{(a_{\bf g},\mu)(b_{\bf h},\nu)}^{({\bf g,h};z_{\bf \bar{h}gh\bar{g}})} 
\ket{\Phi_{b_{\bf h} ; \,  z_{\bf \bar{h}gh\bar{g}}, \nu}}_{(m,-l;w+\frac{1}{4})}^{({\bf h,\bar{h}\bar{g}h})}
\label{eq:modular_trans_Sz}
\\
&=\sum_{b_{\bf g}, \nu} \mathcal{T}_{(a_{\bf g},\mu)(b_{\bf g},\nu)}^{({\bf g,h};z_{\bf \bar{h}gh\bar{g}})}
\ket{\Phi_{b_{\bf g} ; \,  z_{\bf \bar{h}gh\bar{g}}, \nu}}_{(l-m,m;w)}^{({\bf g,gh})}
.
\label{eq:modular_trans_Tz}
\end{align}
From here on, we let the restrictions to the appropriate subsets of topological charge labels in the sums be imposed implicitly by the definitions of the basis sets described above.

We will show that defining the matrix elements of these transformations to be
\begin{align}
\mathcal{S}_{(a_{\bf g},\mu)(b_{\bf h},\nu)}^{({\bf g,h};z_{\bf \bar{h}gh\bar{g}})} &=
\frac{S_{(a_{\bf g},\mu)(b_{\bf h},\nu)}^{(z_{\bf \bar{h}gh\bar{g}})}}{ U_{\bf h}(a_{\bf g} , \overline{a_{\bf g}}; 0 ) }
= \frac{1}{U_{\bf h}(a_{\bf g} , \overline{a_{\bf g}}; 0 ) 
\mathcal{D}_{\bf 0} \sqrt{d_{z_{\bf \bar{h}gh\bar{g}}}} }
\gineq{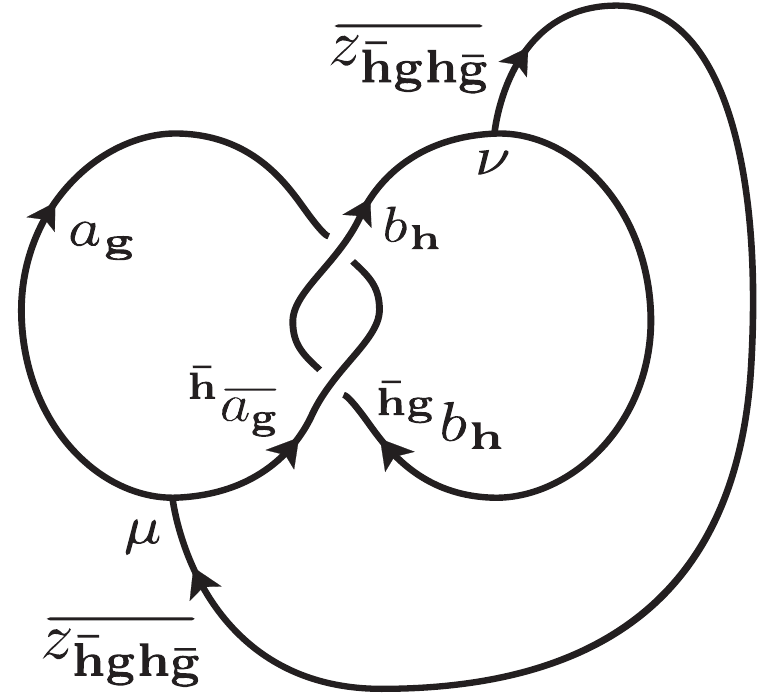}{3.5} 
\label{eq:Sz}
\end{align}
and
\begin{align}
\mathcal{T}_{(a_{\bf g},\mu)(b_{\bf g},\nu)}^{({\bf g,h};z_{\bf \bar{h}gh\bar{g}})} &= \eta_{a_{\bf g}}({\bf g,h}) \theta_{a_{\bf g}} \delta_{a_{\bf g} b_{\bf g}}
=  \frac{\eta_{a_{\bf g}}({\bf g,h}) \delta_{a_{\bf g} b_{\bf g}}}{ d_{a_{\bf g}}  \sqrt{d_{z_{\bf \bar{h}gh\bar{g}}} }}
\gineq{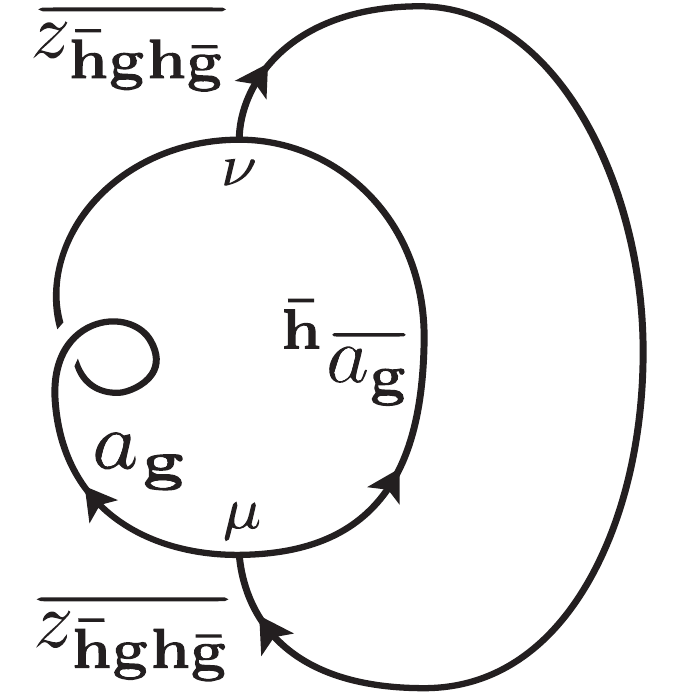}{2.5}
\label{eq:Tz}
\end{align}
yields a projective representation of the mapping class group when the topological $S^{(z)}$-matrix is $G$-graded unitary.
In particular, these satisfy $(\mathcal{S} \mathcal{T})^{3} = \Theta_{\bf 0} \mathcal{S}^{2}$.
Moreover, we will prove that the $S^{(z)}$-matrix is $G$-graded unitary for all $z$ if and only if $\mathcal{C}_{\bf 0}$ is a UMTC.
We note that when $z=0$, Eqs.~(\ref{eq:Sz}) and (\ref{eq:Tz}) equal the modular $\mathcal{S}^{({\bf g,h})}$ and $\mathcal{T}^{({\bf g,h})}$ of the torus with no boundary given in Eqs.~(\ref{eq:modular_S_0}) and (\ref{eq:modular_T_0}).

\begin{figure}[t!]
    \includegraphics[scale=0.65]{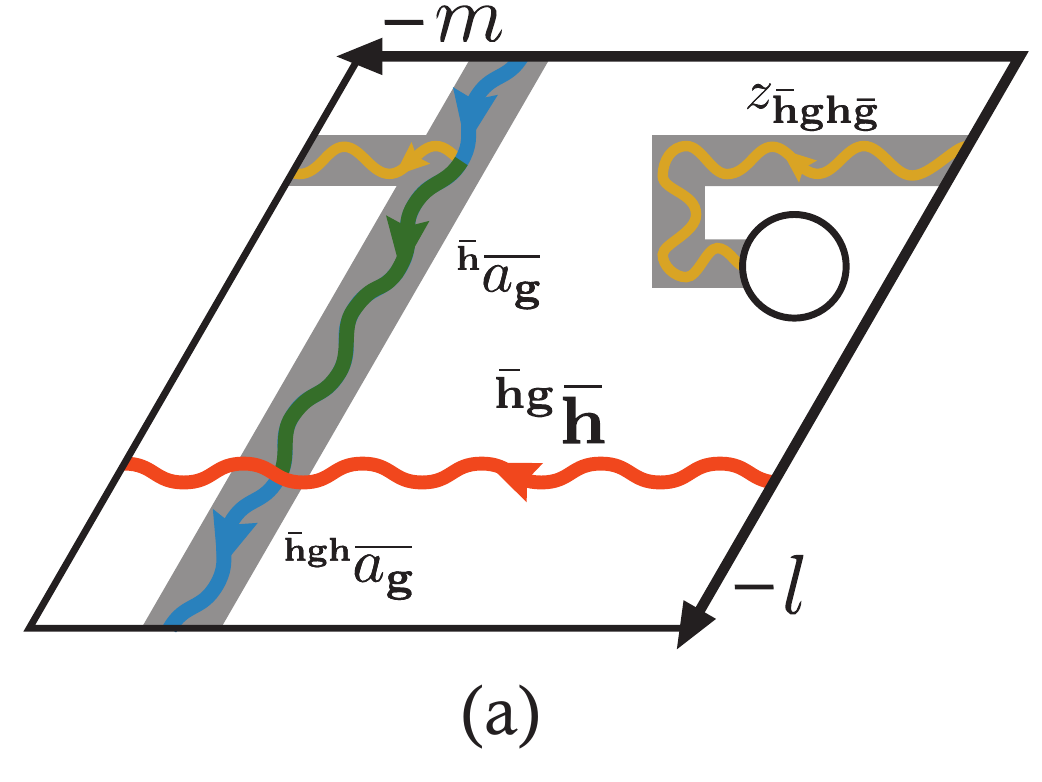}
    \qquad \qquad
    \includegraphics[scale=0.65]{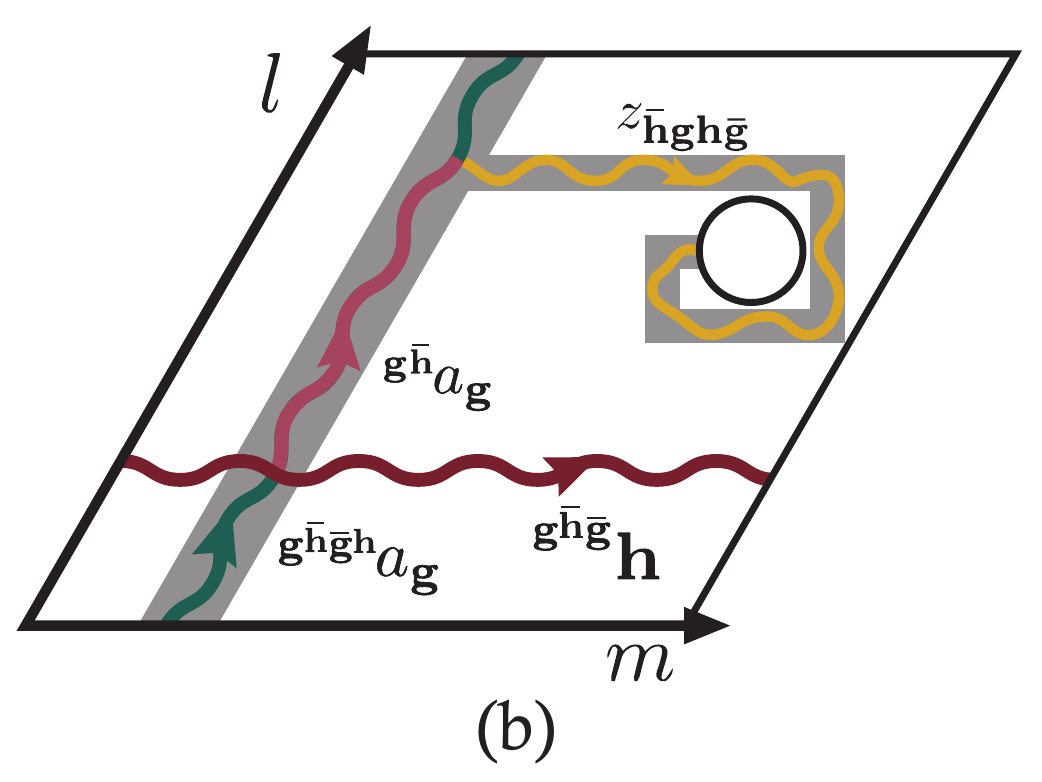}
    \caption{
    Planar representation of the topological ground states of the torus with a boundary in the presence of symmetry defect branch lines.
    (a) The configuration shown corresponds to the ground states $\ket{\Phi_{^{\bf \bar{h}} \overline{a_{\bf g}} ; \,  z_{\bf \bar{h}gh\bar{g}}, \mu}}_{(-l,-m;w+\frac{1}{2})}^{({\bf \bar{h}\bar{g}h,\bar{h}g\bar{h}\bar{g}h})}$ in the $({\bf \bar{h}\bar{g}h,\bar{h}g\bar{h}\bar{g}h} ; z_{\bf \bar{h}gh\bar{g}})$ sector with respect to $(-l,-m;w+\frac{1}{2})$.
    (b) The configuration shown corresponds to the ground states $\ket{\Phi_{^{\bf g\bar{h}\bar{g}h} a_{\bf g} ; z_{\bf \bar{h}gh\bar{g}}, \mu}}_{(l,m;w-1)}^{(^{\bf g\bar{h}\bar{g}h}{\bf g},^{\bf g\bar{h}\bar{g}}{\bf h})}$ in the $(^{\bf g\bar{h}\bar{g}h}{\bf g},^{\bf g\bar{h}\bar{g}}{\bf h} ; z)$ sector with respect to $(l,m;w-1)$.
    These basis states are related to those described in Fig.~\ref{fig:torus_2d} by $G$-crossed mapping class transformations, as specified in Eqs.~(\ref{eq:modular_trans_Cz}) and (\ref{eq:modular_trans_Qz}).
    }
    \label{fig:torus_2d_CQ}
\end{figure}

Two other mapping class transformations that are of particular significance for our analysis are $\mathfrak{c} = \mathfrak{s}^2$ and $\mathfrak{t}_{q} = \mathfrak{s}^{-4}$.
We will write the respective basis transformations representing these as
\begin{align}
\ket{\Phi_{a_{\bf g} ; \,  z_{\bf \bar{h}gh\bar{g}}, \nu}}_{(l,m;w)}^{({\bf g,h})}
&= \sum_{b_{\bf \bar{h}\bar{g}h}, \nu} \mathcal{C}_{(a_{\bf g},\mu)(b_{\bf \bar{h}\bar{g}h},\nu)}^{({\bf g,h};z_{\bf \bar{h}gh\bar{g}})} 
\ket{\Phi_{b_{\bf \bar{h}\bar{g}h} ; \,  z_{\bf \bar{h}gh\bar{g}}, \mu}}_{(-l,-m;w+\frac{1}{2})}^{({\bf \bar{h}\bar{g}h,\bar{h}g\bar{h}\bar{g}h})}
\label{eq:modular_trans_Cz}
\\
&=\sum_{b_{\bf  g\bar{h}\bar{g}h g \bar{h}gh\bar{g}}, \nu} \mathcal{Q}_{(a_{\bf g},\mu)(b_{\bf  g\bar{h}\bar{g}h g \bar{h}gh\bar{g}},\nu)}^{({\bf g,h};z_{\bf \bar{h}gh\bar{g}})}
\ket{\Phi_{b_{\bf  g\bar{h}\bar{g}h g \bar{h}gh\bar{g}} ; z_{\bf \bar{h}gh\bar{g}}, \mu}}_{(l,m;w-1)}^{(^{\bf g\bar{h}\bar{g}h}{\bf g}, \,^{\bf g\bar{h}\bar{g}h}{\bf h})}
.
\label{eq:modular_trans_Qz}
\end{align}
The corresponding planar configurations for these basis states are shown in Fig.~\ref{fig:torus_2d_CQ}.
We see that configurations differing by a full Dehn twist around the boundary, e.g. $w-1$ compared to $w$, can have different group labels and topological charges on the defect lines and ribbon operators around the same cycles.
This results from the action of the boundary defect's ${\bf \bar{h}gh\bar{g}}$-branch line on the system as a consequence of the twisting.
We will compute the matrix elements of these mapping class transformations using the definition of $\mathcal{S}^{({\bf g,h};z_{\bf \bar{h}gh\bar{g}})}$.
For $z=0$, which requires ${\bf gh}={\bf hg}$ and $a_{\bf g}\in\mathcal{B}_{\bf g}^{\bf h}$, these will have the resulting values
\begin{align}
\label{eq:C_z0}
\mathcal{C}_{a_{\bf g} b_{\bf \bar{g}}}^{({\bf g,h};0_{\bf 0})} &= \frac{1}{ U_{\bf h}(\overline{b_{\bf \bar{g}}}, b_{\bf \bar{g}}; 0) \eta_{b_{\bf \bar{g}}}({\bf h,\bar{h}}) } \delta_{\overline{a_{\bf g}}, b_{\bf \bar{g}}}
,\\
\label{eq:Q_z0}
\mathcal{Q}_{a_{\bf g} b_{\bf g}}^{({\bf g,h};0_{\bf 0})} &= \delta_{a_{\bf g}, b_{\bf g}}
,
\end{align}
matching the case of the torus with no boundary described in Ref.~\onlinecite{Barkeshli2019}.

\section{States and Mapping Class Transformations on a General Surface with Boundaries}

We can now extend the discussion of topological states and mapping class transformations to general orientable surfaces.
We consider a surface $\Sigma_{g,n}$ with genus $g$ and $n$ boundary components.
We can construct a surface $\Sigma_{g,n}$ from $g$ copies of a torus with one boundary, $\Sigma_{1,1}$, and a sphere with $g+n$ boundaries, $\Sigma_{0,g+n}$, by gluing each torus' boundary together with a boundary of the sphere.
We can also construct these surfaces from a collection of spheres with three or fewer boundary components by gluing together boundary components in a similar manner, now including the possibility of gluing together two boundary components of the same sphere in order to create genus.
In order to have a consistent orientation, boundary components that are glued together must have opposite orientations.

\begin{figure}[t!]
    \includegraphics[scale=0.75]{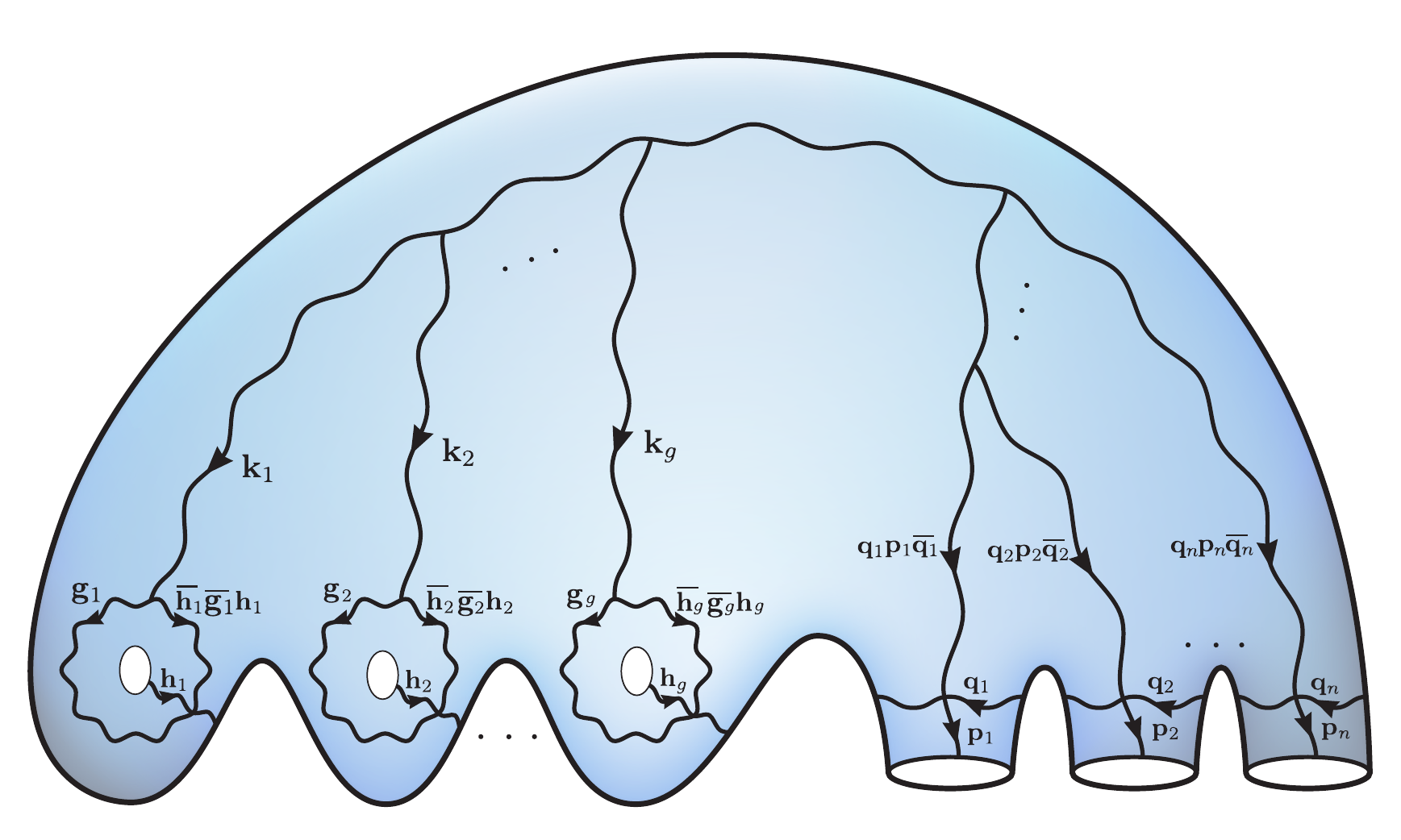}
    \caption{
    An orientable surface $\Sigma_{g,n}$ with genus $g$ and $n$ boundary components with symmetry defect branch lines labeled by elements of the symmetry group $G$.
    The branch line labels must respect fusion rules at trijunctions corresponding to group multiplication.
    (As such, ${\bf k}_j = {\bf g}_j {\bf \bar{h}}_j {\bf \bar{g}}_j {\bf h}_j$ and Eq.~\eqref{eq:general_surface_group_condition} must hold for the configuration shown.
    A general configuration of defect branch lines on the surface can be deformed into some choice of standard configuration, which we will take to be the configuration shown here.
    A corresponding basis for the topological ground state space of $\Sigma_{g,h}$ with this configuration of symmetry defect branch lines can be expressed diagrammatically as in Eq.~\eqref{eq:general_basis}.
    }
    \label{fig:3d_general}
\end{figure}

We will use a similar diagrammatic description of the topological states on general surfaces as was presented in Ref.~\onlinecite{Bonderson2017} using UMTCs for topological phases without symmetry.
In order to generalize this to SET phases, we need to allow for general configurations of symmetry defect branch lines on the surface.
A general configuration of branch lines can be deformed into a conventional standard configuration, which we take as shown in 
Fig.~\ref{fig:3d_general}.
We choose this standard configuration by envisioning the surface as $g$ copies of $\Sigma_{1,1}$ glued onto $\Sigma_{0,g+n}$.
Just as in Sec.~\ref{sec:torus_with_boundary}, for the $j$th $\Sigma_{1,1}$ we pick an ordered pair of generating cycles $(l_j , m_j)$, as well as a path from the $l_j$ cycle to the boundary, then we let there be a ${\bf h}_j$-defect branch line looping around the $m_j$ cycle and a ${\bf g}_j$-defect line around the $l_j$ cycle until it crosses the ${\bf h}_j$-defect branch at which point it becomes an ${\bf \bar{h}}_j {\bf g}_j {\bf h}_j$-branch line.
Where this branch line closes back on itself, it is intersected by the defect branch line connecting the boundary to the $l_j$ cycle, which must carry the group element ${\bf k}_j = {\bf g}_j {\bf \bar{h}}_j {\bf \bar{g}}_j {\bf h}_j$ in order for the tri-junction to respect group multiplication.
The $g$ boundaries of $\Sigma_{0,g+n}$ onto which the tori are glued have ${\bf k}_j$-branch lines ending on them to precisely match up with the branch lines ending on the corresponding tori boundaries.
For the remaining $n$ boundaries of $\Sigma_{0,g+n}$, we assume a ${\bf p}_r$-defect branch line ends on the $r$th boundary.
There can also be defect branch lines encircling each of these $n$ boundaries (i.e. looping a cycle homotopic to the boundary), which we take to have group label ${\bf q}_r$.
In this way, the branch line leading to the $r$th boundary has label ${\bf q}_r {\bf p}_r {\bf \bar{q}}_r$ prior to crossing the ${\bf q}_r$-branch loop.
Finally, all the branch lines running to the boundaries of $\Sigma_{0,g+n}$ are organized into a fusion tree.
Only defect branch line configurations that respect the fusion rules of $G$, i.e. group multiplication, are allowed.
This yields the condition
\begin{align}
\label{eq:general_surface_group_condition}
\prod_{j=1}^{g} {\bf k}_j \times \prod_{r=1}^{n} {\bf q}_r {\bf p}_r {\bf \bar{q}}_r = {\bf 0}.
\end{align}

\begin{figure}[t!]
    \includegraphics[scale=0.6]{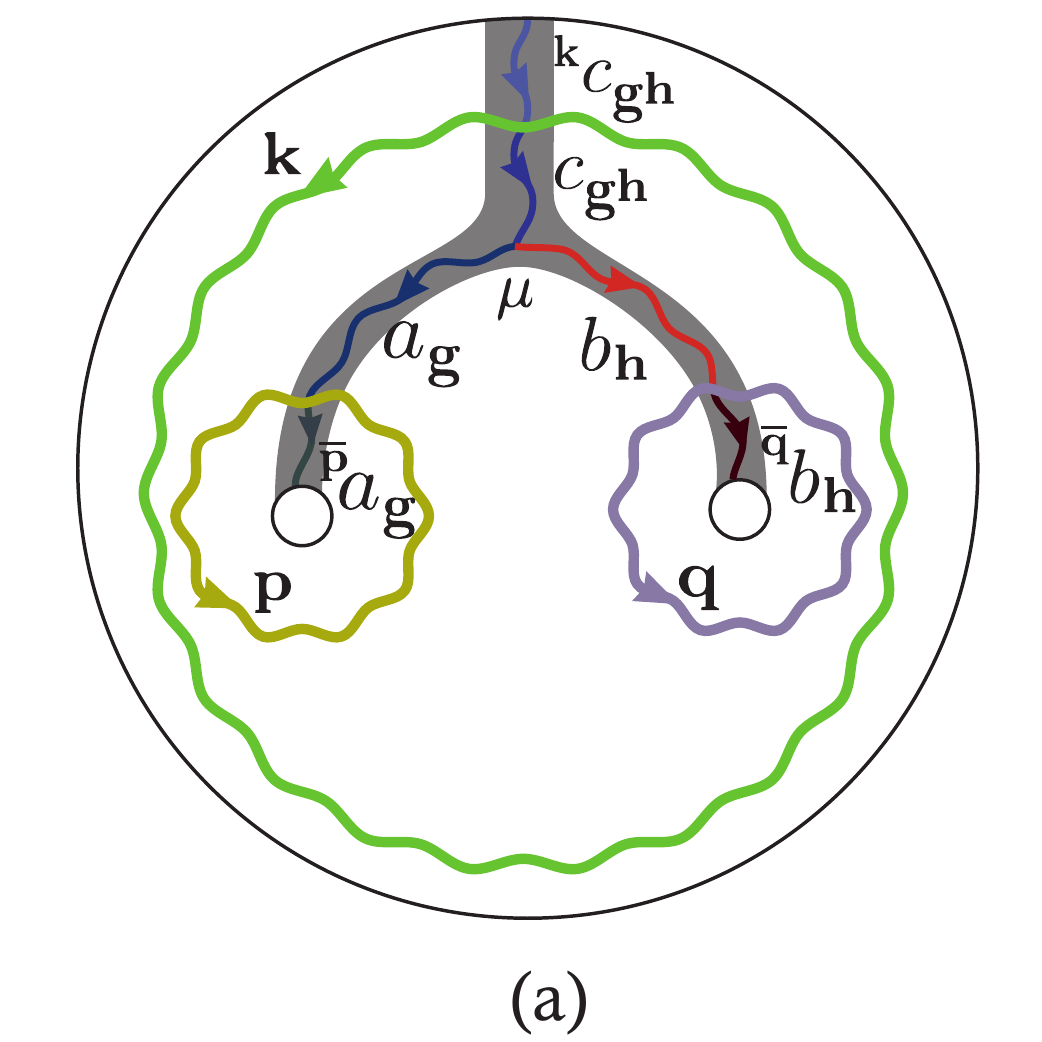}
    \qquad \qquad
    \includegraphics[scale=0.6]{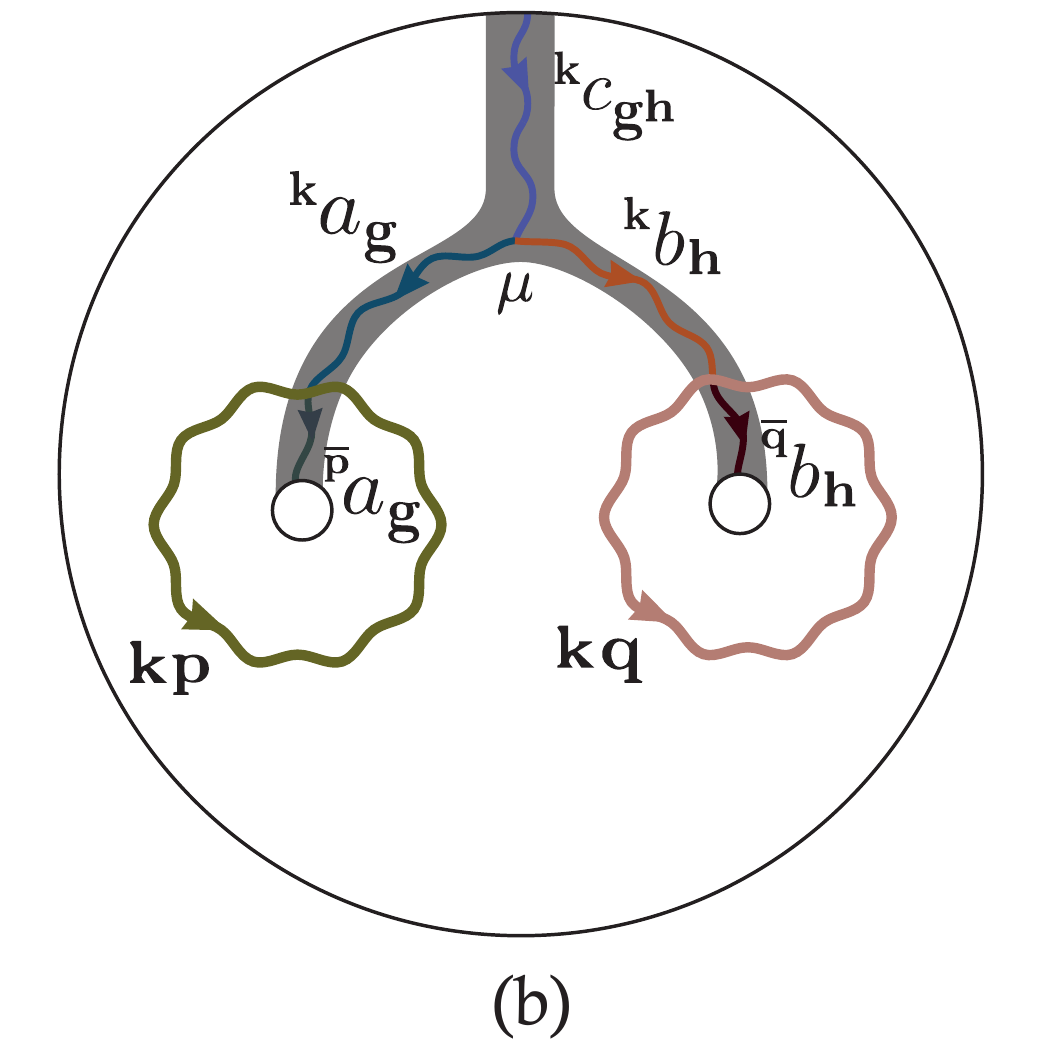}
    \caption{
    An orientable surface $\Sigma_{0,3}$ with symmetry defect branch lines in the standard configuration (a) can be deformed into a configuration (b) branch loops around only two of the three cycles that are homotopic to the boundaries.
    The basis states corresponding to these two configurations are related by unitary transformations given by $\eta_{^{\bf k}a}({\bf k,p}) \eta_{^{\bf k}b}({\bf k,q}) U_{\bf k}(^{\bf k}a,\,^{\bf k}b;\,^{\bf k}c)$.
    }
    \label{fig:3boundary_relation}
\end{figure}

\begin{figure}[t!]
    \includegraphics[scale=0.6]{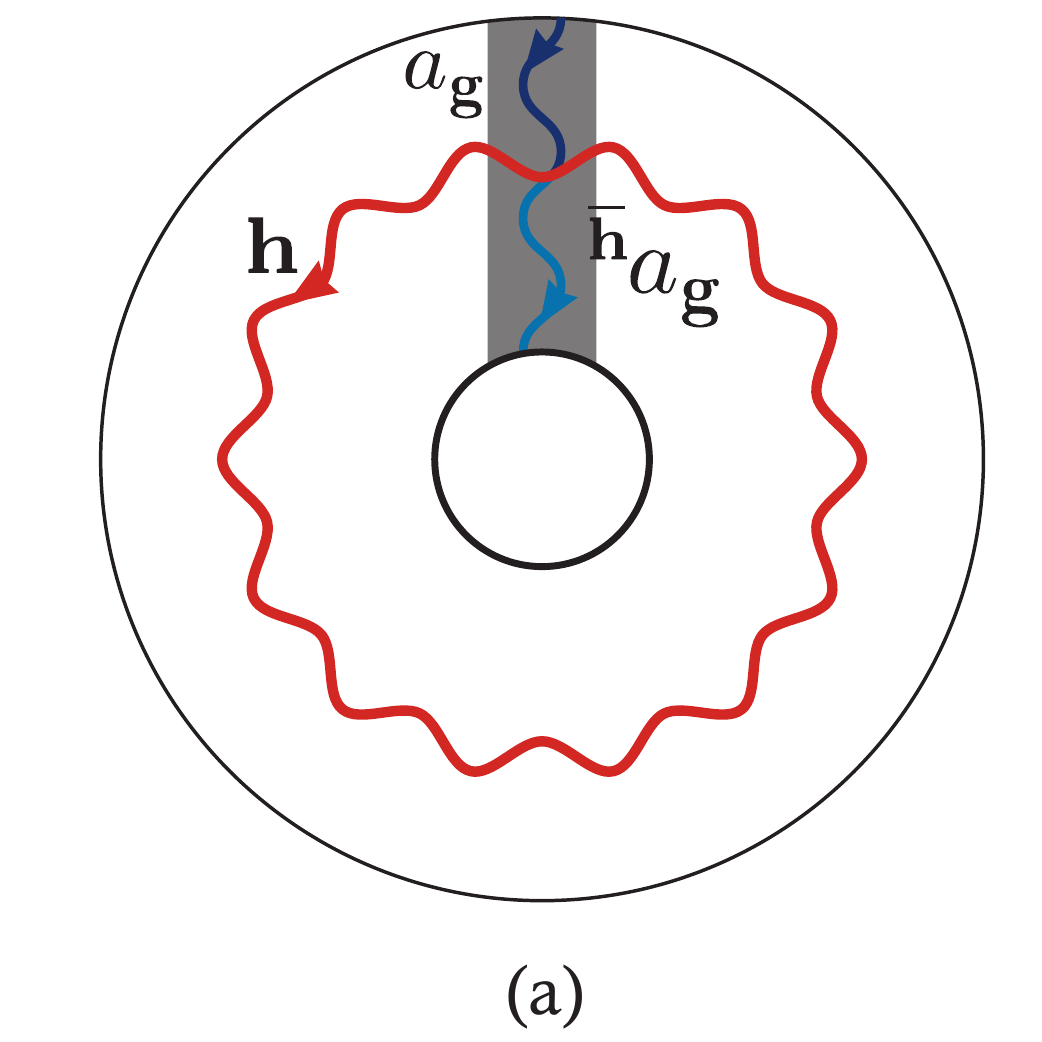}
    \qquad \qquad
    \includegraphics[scale=0.6]{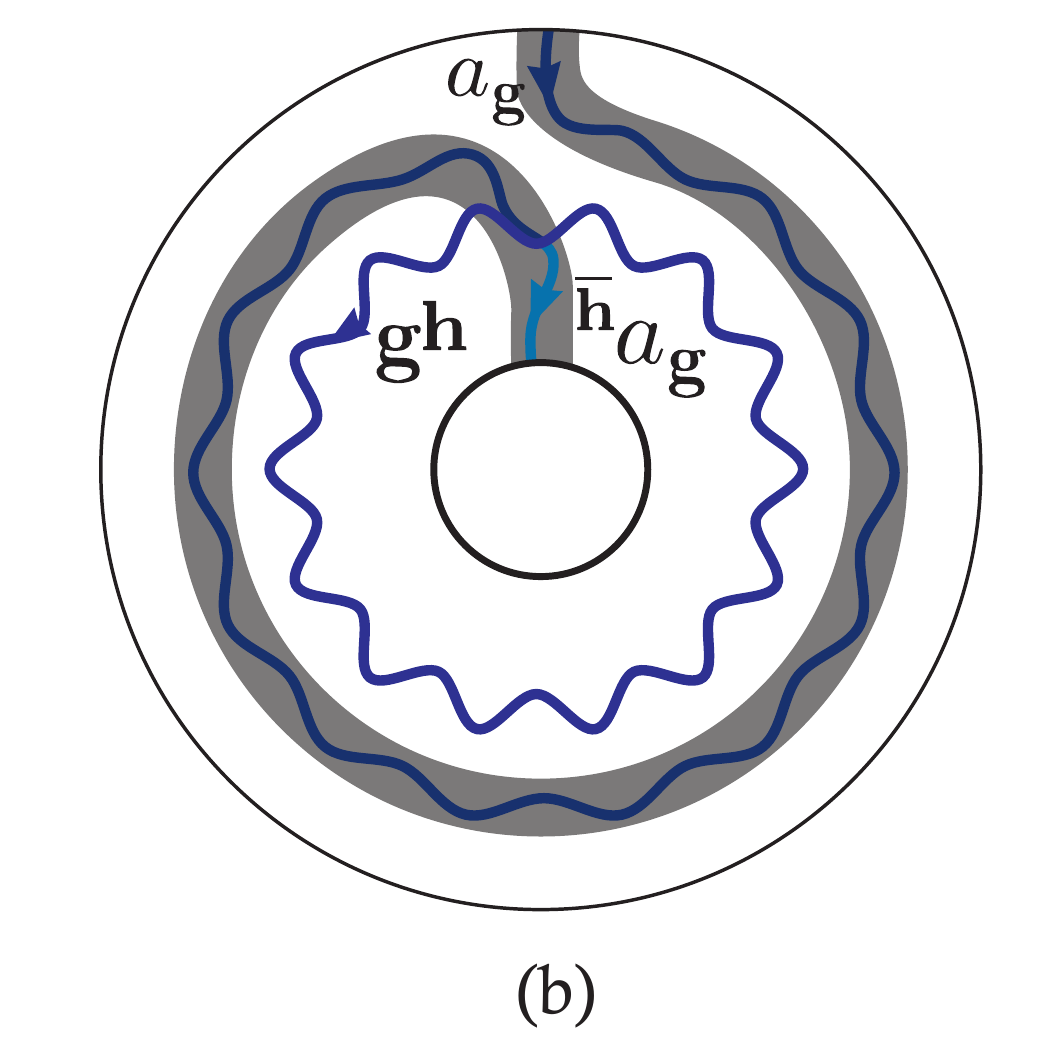}
    \caption{
    An annulus $\Sigma_{0,2}$ with symmetry defect branch lines.
    The basis states for the topological ground state space are specified by a defect ribbon operator connecting the two boundaries.
    For the ribbon operators shown here, the outer boundary carries topological charge $\overline{a_{\bf g}}$ and the inner boundary carries topological charge $^{\bf \bar{h}}a_{\bf g}$.
    The two configurations shown in (a) and (b) are related by a Dehn twist, and their corresponding basis states are related by the unitary transformation given in Eq.~\eqref{eq:Dehn}.
    }
\label{fig:annulus_twist}
\end{figure}

In order to understand why no other branch loops need to be included in our standard configurations, we observe that the branch line configuration of Fig.~\ref{fig:3boundary_relation}(a) can be deformed into that of Fig.~\ref{fig:3boundary_relation}(b).
This allows one to push all branch loops around cycles complementary to the fusion tree out to the leaves of the tree.
In other words, the branch line configuration for $\Sigma_{0,n}$ can simply be written as a fusion tree of $n$ branch lines together with ${\bf q}_r$-branch loops that are homotopic to each boundary, intersecting the fusion tree at its leaves.
In fact, we only really need to keep $n-1$ of the ${\bf q}_r$-branch loops around boundaries, since one of them can be pushed onto the other branch line.
A simple example of this is given by the annulus $\Sigma_{0,2}$ in Fig.~\ref{fig:annulus_twist}.
However, it is useful to write all of the boundaries in a similar manner for the case of general surfaces.
This also applies for pushing a branch loop homotopic to the boundary of $\Sigma_{1,1}$ onto the branch lines around the genus.
In particular, if we had $\Sigma_{1,1}$ with $({\bf g},{\bf h})$ around the $(l,m)$ cycles, as in Fig.~\ref{fig:torus_2d}, but with an extra ${\bf q}$-branch loop around the cycle homotopic to the boundary, we could push the ${\bf q}$-branch loop onto the handle, resulting in $({\bf \bar{q}gq},{\bf \bar{q}hq})$ around the $(l,m)$ cycles.

As in Sec.~\ref{sec:torus_with_boundary}, we describe the topological states on a general surface with defect branch lines by further introducing defect ribbon operators along various branch lines.
The topological charges assigned to the different ribbon segments have group labels that match those of the corresponding defect branch lines.
As with the torus, the ribbon operator is only applied to one of the cycles of the generating pair for each genus handle, and the path with a branch line connecting that cycle to the boundary.
The loop around the complementary cycle of a generating pair does not have a ribbon operator applied to it (whether or not it has a nontrivial branch line around it).
Similarly, ribbon operators are applied to the paths with branch lines connecting boundaries in the fusion tree, while the loops that are homotopic to boundaries do not have ribbon operators applied to them (whether or not they have nontrivial branch lines around them).
Translating this for the standard configuration into the diagrammatic representation, the basis states of the topological state space can be written as
\begin{align}
\frac{1}{(d_{z_1} \cdots d_{z_n})^{1/4}}
\gineq{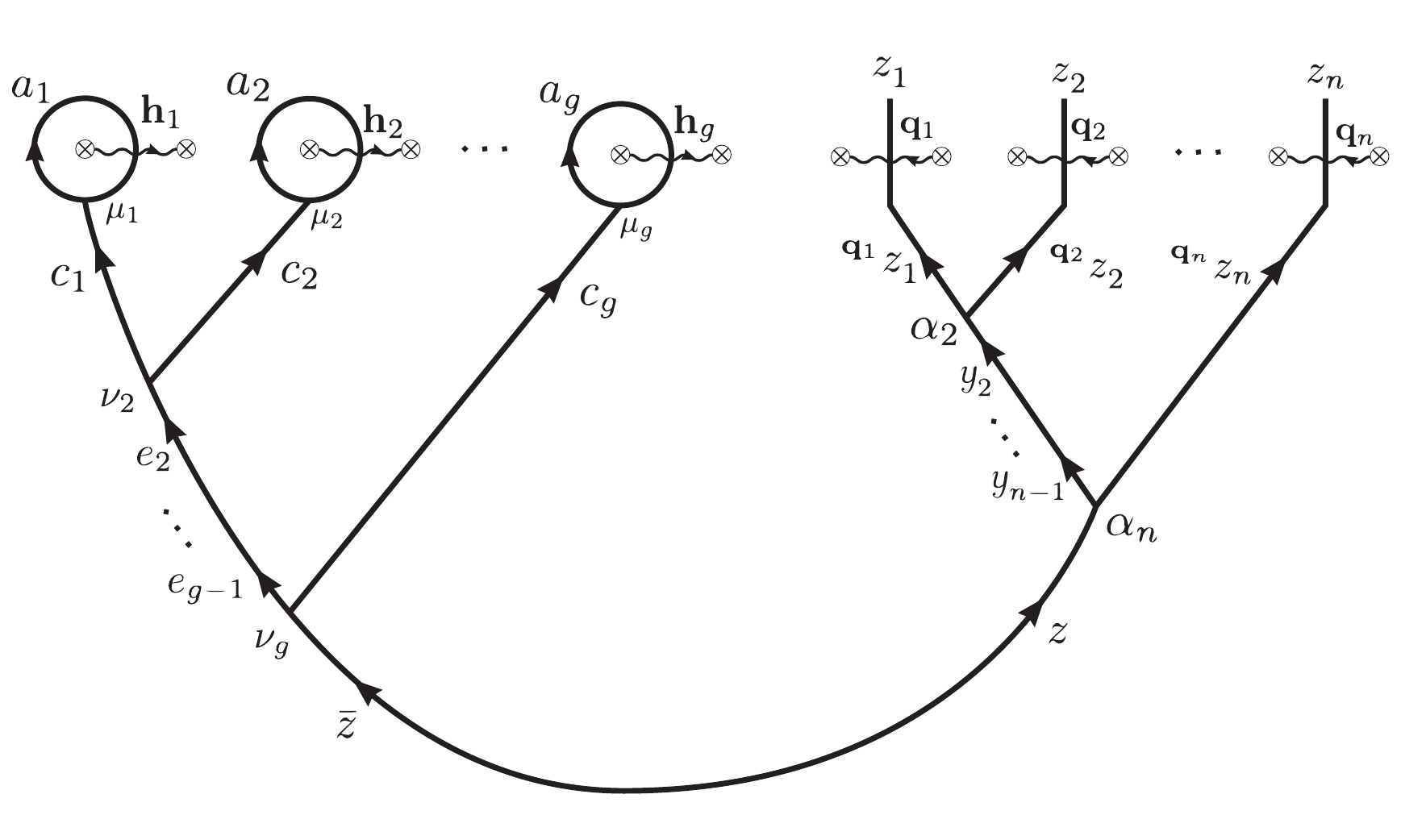}{12}
.
\label{eq:general_basis}
\end{align}

Many different choices of basis are possible, corresponding to different choices of configuration of defect branch lines, generating pairs of cycles for the genus handles, and paths connecting to boundaries.
These bases are all related by unitary transformations that can be determined the topological data of the $G$-crossed UMTC, such as the Dehn twists, the $\mathcal{S}$ transformations on each genus handle, but also relations more generally expressed in terms of the basic data ($F$-, $R$-, $U$-, and $\eta$-symbols) of the $G$-crossed UMTC.

Finally, in order to represent the mapping class group on a general surface $\Sigma_{g,n}$, we note that the mapping class group can be generated by Dehn twists around a finite set of nontrivial cycles on the surface together with braiding exchanges of boundaries (see, e.g. Ref.~\onlinecite{Bakalov01}, and references therein for more details).
The braiding exchanges of boundaries can be directly translated into braiding operators applied to the diagrammatic states, which can then be evaluated using the $F$- and $R$-symbols, as usual.
In order to evaluate the Dehn twist around an arbitrary cycle $c$ on the surface, we first apply a change of basis corresponding to change in configuration that yields only one segment of ribbon operator intersecting the cycle $c$.
The change of basis can be evaluated using the basic data of the $G$-crossed UMTC by identifying a sequence of moves that relates the bases.
In this basis, one can directly apply the Dehn twist operation, corresponding to
\begin{align}
\mathcal{T}^{({\bf g,h})}_{a_{\bf g},b_{\bf g}} = \eta_{a_{\bf g}}({\bf g,h}) \theta_{a_{\bf g}} \delta_{a_{\bf g} b_{\bf g}}
,
\label{eq:Dehn}
\end{align} 
where ${\bf h}$ is the label of the branch line around cycle $c$ and $a_{\bf g}$ is the topological charge of the ribbon operator intersecting the cycle, as shown in Fig.~\ref{fig:annulus_twist}.
We note that the Dehn twist operation can generally be applied to any annular/cylindrical segment of a general surface, and was not specific to the torus.
We also observe that defining the Dehn twist operation with the twisting on the other side of the ${\bf h}$-branch line would yield the same expression in Eq.~\eqref{eq:Dehn}, since Eq.~\eqref{eqn:action_on_twist} ensures that $\eta_{a_{\bf g}}({\bf h, \bar{h}gh }) \theta_{^{\bf \bar{h}}a_{\bf g}} = \eta_{a_{\bf g}}({\bf g,h}) \theta_{a_{\bf g}}$.

\section{Mapping Class Transformation Operators}
\label{sec:MCG_operators}

Similar to the methods used in Ref.~\onlinecite{Bakalov01} for MTCs, it is useful to diagrammatically define operators corresponding to the action of the mapping class transformations on the state space of the torus with a boundary.
This allows us to simultaneously treat the subspaces with different values of boundary charge $z$ and use diagrammatic methods to more efficiently demonstrate that the transformations (projectively) represent the mapping class group.

We define the operators $\boldsymbol{S}$ and $\boldsymbol{T}$ for a particular defect sector, though with all possible boundary charge values $z$ for that sector, to be
\begin{align}
\boldsymbol{S}^{({\bf g,h})} &\equiv 
\sum_{a_{\bf g},b_{\bf h}} \frac{\sqrt{d_{a_{\bf g}} d_{b_{\bf h}}}}{U_{\bf h}(a_{\bf g}, \overline{a_{\bf g}};0) \mathcal{D}_{\bf 0} }
\gineq{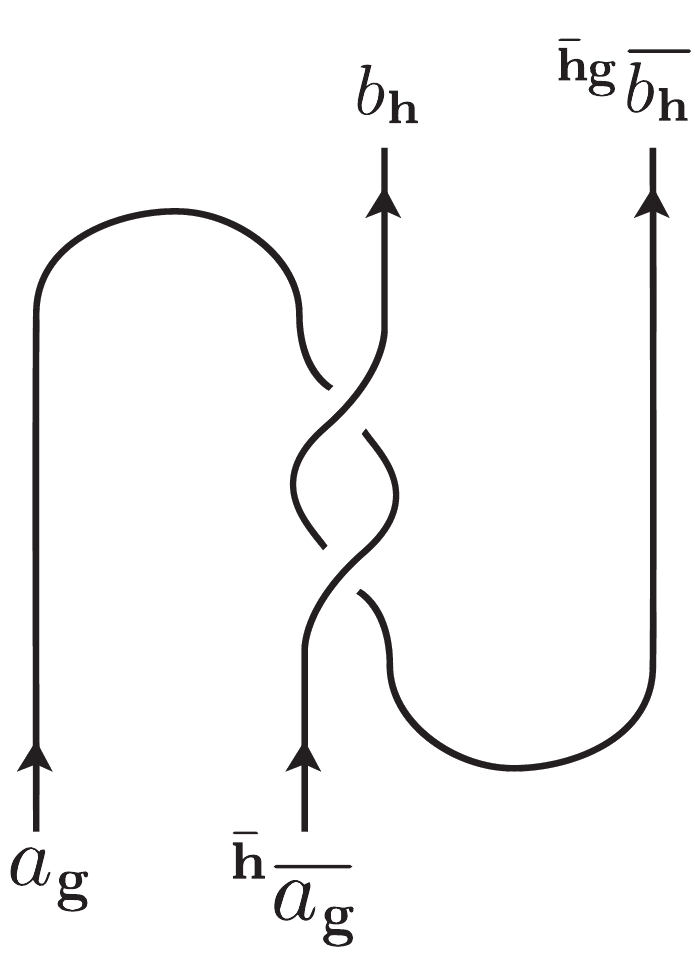}{2.5} 
\, =\,  \sum_{\substack{a_{\bf g},b_{\bf h} \\ z_{\bf \bar{h}gh\bar{g}} \\ \mu,\nu}} 
\Sgh \sqrt{\frac{d_z}{d_a d_b}} 
\gineq{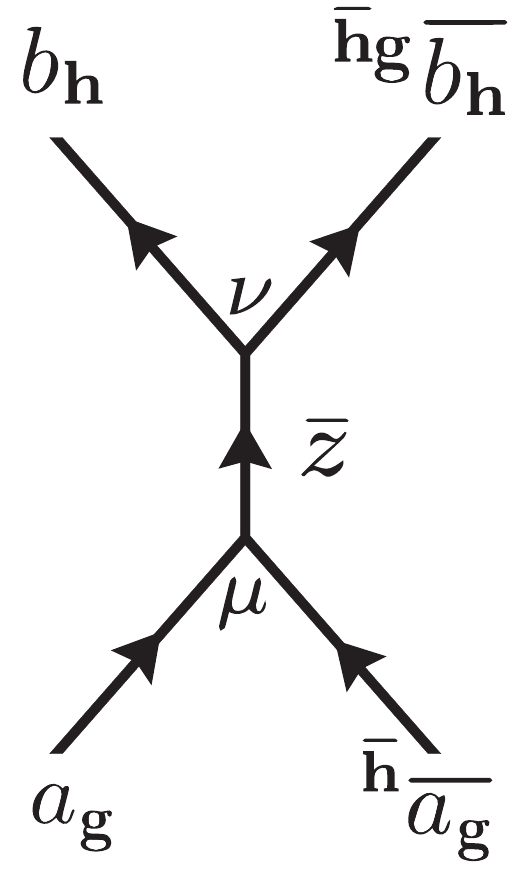}{1.5}
,
\label{eq:modular_S}
\end{align}
\begin{align}
\boldsymbol{T}^{({\bf g,h})} &\equiv 
\sum_{a_{\bf g}} \eta_{a_{\bf g}}({\bf g,h})
\gineq{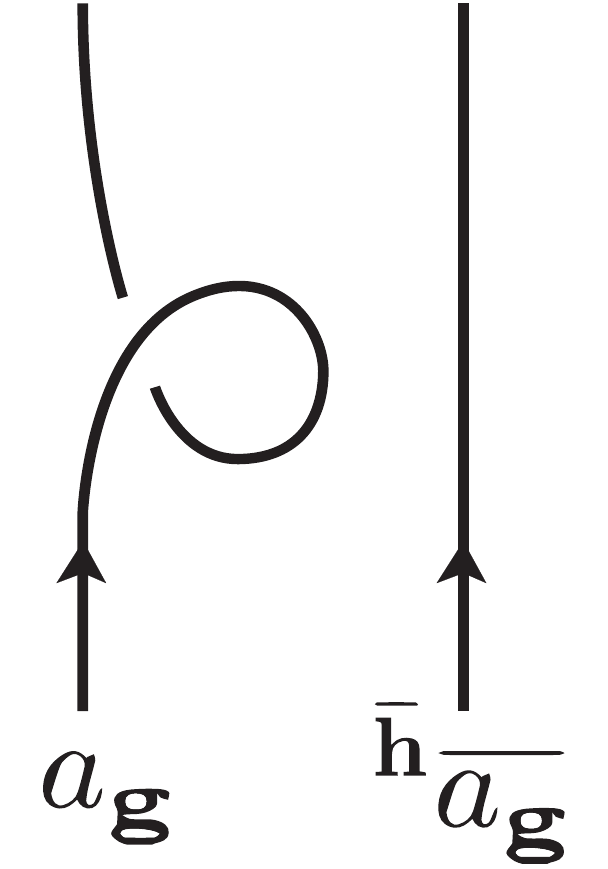}{1.75} \,
= \, \sum_{\substack{a_{\bf g},b_{\bf g} \\ z_{\bf \bar{h}gh\bar{g}} \\ \mu,\nu}} \Tgh \sqrt{\frac{d_z}{d_a d_b}} \gineq{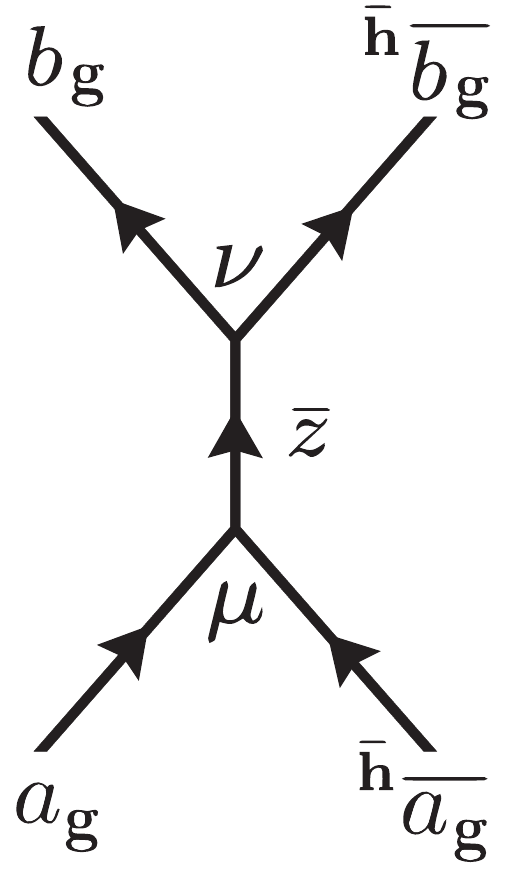}{1.5}
.
\label{eq:modular_T}
\end{align}
The matrix elements for these operators are given in Eqs.~(\ref{eq:Sz}) and (\ref{eq:Tz}).
We use the methodology (and normalization factors) of Ref.~\onlinecite{Bonderson2017} for applying these operators to the topological ground states on surfaces with nontrivial genus using the diagrammatic representation.
For example, to apply $\boldsymbol{S}^{({\bf g,h})}$, we imagine cutting across cycle $m$, breaking open the diagrammatic $a_{\bf g}$-loop to give $\ket{(a_{\bf g})^{\bf h} ; \overline{z_{\bf \bar{h}gh\bar{g}}}, \mu} \rightarrow \ket{a_{\bf g} , \,^{\bf \bar{h}}\overline{a_{\bf g}}; \overline{z_{\bf \bar{h}gh\bar{g}}}, \mu}$;
then we apply the operator
\begin{align}
\boldsymbol{S}^{({\bf g,h})}  \ket{a_{\bf g} , \,^{\bf \bar{h}}\overline{a_{\bf g}}; \overline{z_{\bf \bar{h}gh\bar{g}}}, \mu}
= \sum_{b_{\bf h},\nu} \Sgh
\ket{b_{\bf h} , \,^{\bf \bar{h}g}\overline{b_{\bf h}}; \overline{z_{\bf \bar{h}gh\bar{g}}}, \mu}
,
\end{align}
which corresponds to diagrammatically stacking the operator on the trivalent vertex representing the state, and evaluating the resulting diagram;
and finally we close up the diagram, sewing along the $-l$ cycle, to form a $b_{\bf h}$-loop winding around the $m$ cycle, giving $\ket{b_{\bf h} , \,^{\bf \bar{h}g}\overline{b_{\bf h}}; \overline{z_{\bf \bar{h}gh\bar{g}}}, \mu} \rightarrow  \ket{(b_{\bf h})^{\bf \bar{h}\bar{g}h} ; \overline{z_{\bf \bar{h}gh\bar{g}}}, \nu}$.
Applying the operator $\boldsymbol{S}^{({\bf g,h})}$ to states on surfaces with genus in this way produces the corresponding basis changes of Eq.~\eqref{eq:modular_trans_Sz}, and similarly for other operators representing mapping class transformations.
In this way, a sequence of mapping class transformation is obtained by multiplying the corresponding operators, and the resulting coefficients can be computed by stacking their diagrammatic representations.
We note that the order the operators are applied to states (or multiplied with each other) will be the reverse of the corresponding order in which the basis changes occur.

We additionally define the operators
\begin{align}
\boldsymbol{C}^{({\bf g,h})} &\equiv \sum_{a_{\bf g}} 
\eta_{a_{\bf g}}({\bf h, \bar{h} \bar{g}h}) 
\eta_{a_{\bf g}}({\bf \bar{g}h, \bar{h}g \bar{h} \bar{g}h}) 
U_{\bf \bar{h}g\bar{h}\bar{g}h}(\,^{\bf \bar{h}}\overline{a_{\bf g}}, \,^{\bf \bar{h}}a_{\bf g}; 0)
R^{^{\bf \bar{h}}a_{\bf g} \, ^{\bf \bar{h}}\overline{a_{\bf g}}}_{0}
\,\, \gineq{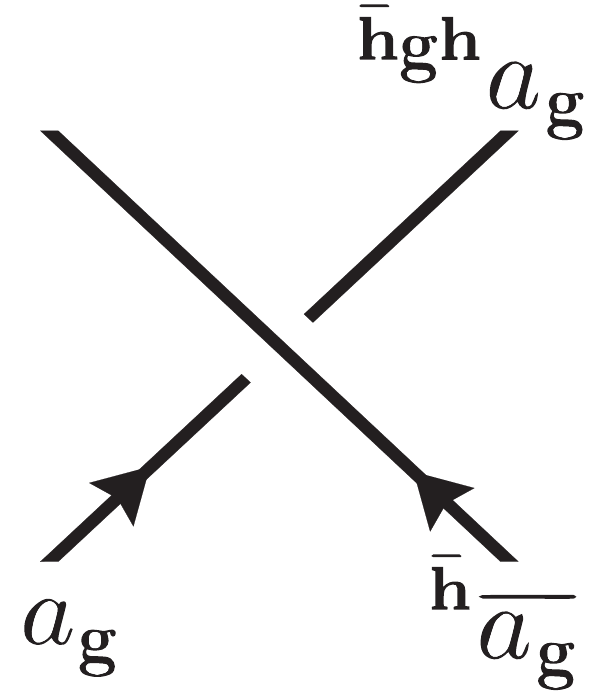}{1.75}
\notag \\
&= \sum_{\substack{a_{\bf g}, b_{\bf \bar{h}\bar{g}h}  \\ z_{\bf \bar{h}gh\bar{g}} \\ \mu,\nu}} \Cgh \sqrt{\frac{d_z}{d_a d_b}} \gineq{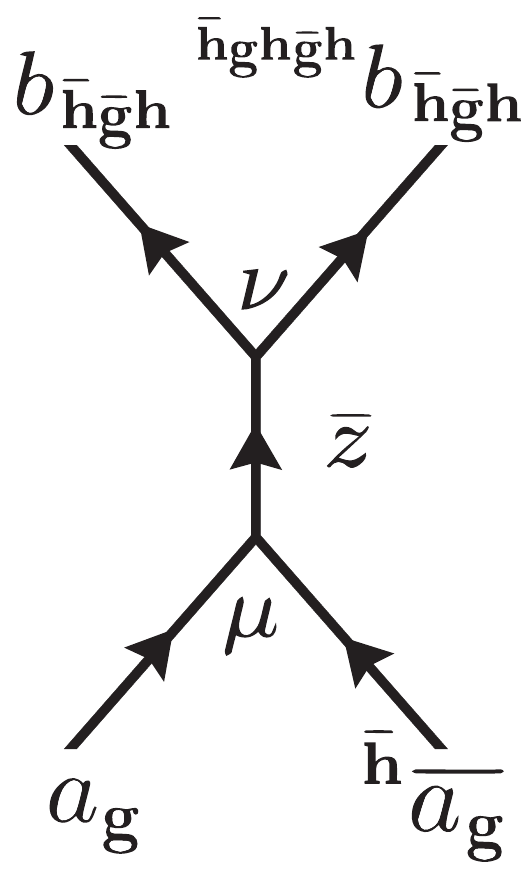}{1.5}
,
\label{eq:modular_C}
\end{align}
\begin{align}
\boldsymbol{Q}^{({\bf g,h})} &\equiv \sum_{a_{\bf g}} 
\frac{\eta_{^{\bf g\bar{h}\bar{g}h}a_{\bf g}}({\bf g\bar{h}\bar{g}h,h }) 
\eta_{^{\bf g\bar{h}\bar{g}h}a_{\bf g}}({\bf g,\bar{h} \bar{g}h })
\eta_{^{\bf g\bar{h}}\overline{a_{\bf g}}}({\bf g,\bar{h} \bar{g}h })
}
{\eta_{^{\bf g\bar{h}\bar{g}h}a_{\bf g}}({\bf g\bar{h}\bar{g}h g h \bar{g}  , g\bar{h}\bar{g}h })
U_{\bf g\bar{h}\bar{g}h}(^{\bf g\bar{h}}a_{\bf g},^{\bf g\bar{h}}\overline{a_{\bf g}};0)
}
\gineq{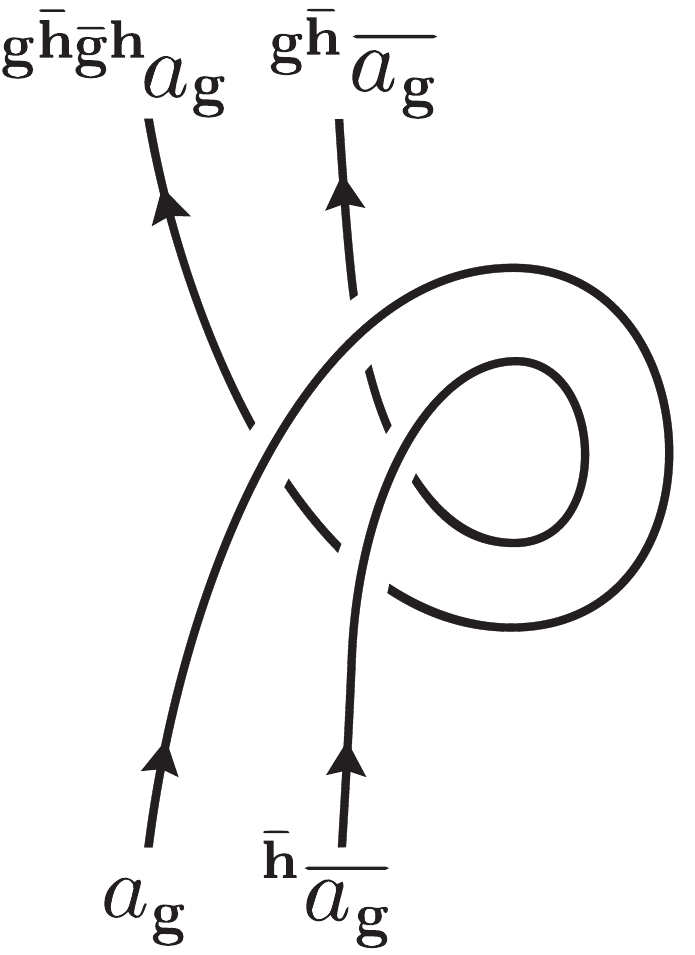}{2} 
\notag \\
&=  \sum_{\substack{a_{\bf g}, b_{\bf g\bar{h}\bar{g}h g \bar{h}gh\bar{g}}  \\ z_{\bf \bar{h}gh\bar{g}} \\ \mu,\nu}}
Q^{({\bf g,h};z)}_{(a_{\bf g}, \mu)(b_{\bf g\bar{h}\bar{g}h g \bar{h}gh\bar{g}},\nu)} \sqrt{\frac{d_z}{d_a d_b}} \gineq{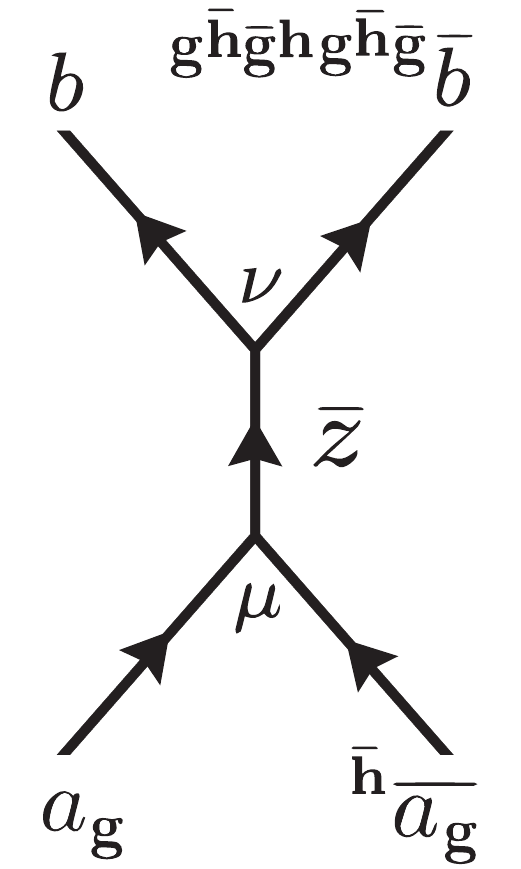}{1.5}
.
\label{eq:modular_Q}
\end{align}
The coefficients in these definitions, which only depend on ${\bf g}$, ${\bf h}$, and $a_{\bf g}$, are chosen to yield the desired mapping class relations, as we will show.
Computing the matrix elements, we find
\begin{align}
\Cgh = \delta_{\,^{\bf \bar{h}}\overline{a_{\bf g}} , b_{\bf \bar{h}\bar{g}h}} \, 
\eta_{a_{\bf g}}({\bf h, \bar{h} \bar{g}h}) 
\eta_{a_{\bf g}}({\bf \bar{g}h, \bar{h}g \bar{h} \bar{g}h}) 
U_{\bf \bar{h}g\bar{h}\bar{g}h}(\,^{\bf \bar{h}}\overline{a_{\bf g}}, \,^{\bf \bar{h}}a_{\bf g}; 0) 
R^{^{\bf \bar{h}}a_{\bf g} \, ^{\bf \bar{h}}\overline{a_{\bf g}}}_{0}
[ (R^{a_{\bf g} \, ^{\bf \bar{h}}\overline{a_{\bf g}}}_{\overline{z}})^{-1}]_{\mu \nu}
,
\label{eq:mathcal_C}
\end{align}
\begin{align}
\Qgh &=  \delta_{^{\bf g\bar{h}\bar{g}h}a_{\bf g} , b_{\bf g\bar{h}\bar{g}hg\bar{h}gh\bar{g}}} \,
\frac{\eta_{^{\bf g\bar{h}\bar{g}h}a_{\bf g}}({\bf g\bar{h}\bar{g}h,h }) 
[U_{\bf g\bar{h}\bar{g}h}(^{\bf g\bar{h}\bar{g}h} a_{\bf g} , ^{\bf g\bar{h}}\overline{a_{\bf g}} ; \bar{z})]_{\mu \nu}
}
{\eta_{^{\bf g\bar{h}\bar{g}h}a_{\bf g}}({\bf g\bar{h}\bar{g}h g h \bar{g}  , g\bar{h}\bar{g}h })
U_{\bf g\bar{h}\bar{g}h}(^{\bf g\bar{h}}a_{\bf g},^{\bf g\bar{h}}\overline{a_{\bf g}};0)
}
\, \theta_{\bar{z}}
.
\end{align}
We note that for the case $z=0$, these expressions reduce to Eqs.~\eqref{eq:C_z0} and \eqref{eq:Q_z0}, by using the various consistency relations.
When $z \in \mathcal{B}_{\bf 0}$, or equivalently when ${\bf gh}={\bf hg}$, it is straightforward to see that $Q^{({\bf g,h};z_{\bf 0})}_{(a_{\bf g}, \mu)(b_{\bf g},\nu)} = \theta_{z_{\bf 0}} \delta_{a_{\bf g} b_{\bf g}} \delta_{\mu \nu}$.

We can define the respective operators acting on all defect sectors by taking the direct sum of the group labels, that is $\boldsymbol{S} = \bigoplus\limits_{{\bf g,h}} \boldsymbol{S}^{({\bf g,h})}$, $\boldsymbol{T}= \bigoplus\limits_{{\bf g,h}} \boldsymbol{T}^{({\bf g,h})}$,  $\boldsymbol{C}= \bigoplus\limits_{{\bf g,h}} \boldsymbol{C}^{({\bf g,h})}$, and $\boldsymbol{Q}= \bigoplus\limits_{{\bf g,h}} \boldsymbol{Q}^{({\bf g,h})}$.
With these definitions, we will show for any $G$-crossed UBTC $\mathcal{B}_{G}^{\times}$ (not necessarily modular) that these operators satisfy the following relations
\begin{align}
\boldsymbol{TSTST} &= \Theta_0 \boldsymbol{S} ,
\label{eq:modrel_1}
\\
\boldsymbol{S} &= \boldsymbol{S}^{\dagger} \boldsymbol{C},
\label{eq:modrel_2}
\\
\boldsymbol{CS} &= \boldsymbol{SC} ,
\label{eq:modrel_3}
\\
\label{eq:modrel_4}
\boldsymbol{CT} &= \boldsymbol{TC} ,
\\
\label{eq:modrel_5}
\boldsymbol{C}^2 &= \boldsymbol{Q}^{-1}
.
\end{align}
By definition, the operators $\boldsymbol{T}$, $\boldsymbol{C}$, and $\boldsymbol{Q}$ are unitary when $\mathcal{B}_{G}^{\times}$ is a $G$-crossed UBTC.
On the other hand, unitarity of $\boldsymbol{S}$ is an additional condition that needs to be imposed for these operators to represent the mapping class transformations.
($\boldsymbol{S}$ is not necessarily even invertible for a general $G$-crossed UBTC.)
When $\boldsymbol{S}$ is unitary, we can rewrite Eq.~\eqref{eq:modrel_2} as
\begin{align}
\boldsymbol{S}^2 &= \boldsymbol{C}
,
\label{eq:modrel_S2C}
\end{align}
and $\boldsymbol{S}$ and $\boldsymbol{T}$ provide a projective representation of MCG($\Sigma_{1,1}$), corresponding to the generators $\mathfrak{s}$ and $\mathfrak{t}$, respectively.
In this case, we also see that $\boldsymbol{C}$ appropriately represents $\mathfrak{c}$, the generator of the center of MCG($\Sigma_{1,1}$).
Moreover, the $z=0$ sectors of these operators provide a projective representation of MCG($\Sigma_{1,0}$).

In light of this, it is natural to define a $G$-crossed UBTC to be $G$-crossed modular when it satisfies the additional conditions that there are a finite number of topological charges (simple objects) in each sector $\mathcal{B}_{\bf g}$ and the operator $\boldsymbol{S}$ is unitary.
In Sec.~\ref{sec:S_unitarity}, we will show that these conditions are implied when $\mathcal{B}_{G}^{\times}$ is a faithful $G$-crossed extension of a UMTC $\mathcal{B}_{\bf 0}$.
Faithful here means $\mathcal{B}_{\bf g} \neq \varnothing$ for all ${\bf g}\in G$.
In other words, a $G$-crossed UBTC $\mathcal{B}_{G}^{\times}$ is $G$-crossed modular if and only if it is a faithful $G$-crossed extension of a modular UBTC $\mathcal{B}_{\bf 0}$.
(The ``only if'' direction of this statement follows simply by restricting to the ${\bf 0}$-sector of $\mathcal{B}_{G}^{\times}$.)

Since Refs.~\onlinecite{Turaev2000,turaev2010,Kirillov2004,Barkeshli2019} use slightly different definitions of modularity for $G$-crossed theories, it is worth comparing them and clarifying our result with respect to these papers.
They all require $\lvert\mathcal{B}_{\bf g}\lvert$ to be finite for all ${\bf g}$.
Refs.~\onlinecite{Turaev2000,turaev2010} define a (ribbon) $G$-crossed category to be modular when the topological $S$-matrix [Eq.~\eqref{eqn:topoSmatrix}] of the ${\bf 0}$-sector is invertible, i.e. when $\mathcal{B}_{\bf 0}$ is a MTC.
Ref.~\onlinecite{Kirillov2004} defines a $G$-crossed (fusion) category to be modular when the $\tilde{s}$ operator on the extended Verlinde algebra is invertible.
(The Verlinde algebra maps to the state space on $\Sigma_{1,0}$ for modular theories, i.e. the boundary charge $z=0$ sector.)
$({\bf g},{\bf h})$ sectors of the extended Verlinde algebra are non-empty when $\mathcal{B}_{\bf g}^{\bf h} \neq \varnothing$ (even if $\mathcal{B}_{\bf h} = \varnothing$).
For this definition, Ref.~\onlinecite{Kirillov2004} observed that modularity of $\mathcal{B}_{\bf 0}$ does not necessarily imply modularity of $\mathcal{B}_{G}^{\times}$, since the $\tilde{s}$ operator maps between nonempty and empty vector spaces whenever there is a ${\bf h}$ with $\mathcal{B}_{\bf h} = \varnothing$.
Ref.~\onlinecite{Barkeshli2019} defines a $G$-crossed UBTC $\mathcal{B}_{G}^{\times}$ to be $G$-crossed modular when its topological $S$-matrix is $G$-graded unitary, meaning for each pair $({\bf g,h})$, the matrix defined by $S_{a_{\bf g} b_{\bf h}}$ with indices $a_{\bf g} \in \mathcal{B}_{\bf g}^{\bf h}$ and $b_{\bf h} \in \mathcal{B}_{\bf h}^{\bf g}$ is unitary.
This is equivalent to unitarity of the restriction of $\boldsymbol{S}$ to the $z=0$ sectors.
This definition is equivalent to that of Ref.~\onlinecite{Kirillov2004} (for unitary theories).
The results in our paper show that all four of these definitions are equivalent under the condition that the $G$-crossed UBTC $\mathcal{B}_{G}^{\times}$ is a faithful $G$-crossed extension.~\footnote{We note that our focus is on unitary category theories in this paper, while Refs.~\onlinecite{Turaev2000,turaev2010,Kirillov2004} do not require unitarity of the categories.
Much of our discussion and results can be adapted (with some care) to non-unitary theories, for which modularity is defined using invertiblity of the $S$-matrix or $\boldsymbol{S}$ operator, rather than unitarity.
However, non-unitary theories do not correspond to the low-energy effective theory describing a topological or SET phase of matter.}
This indicates that the counterexamples to modular $\mathcal{B}_{\bf 0}$ implying $G$-crossed modular $\mathcal{B}_{G}^{\times}$ only arise when the $G$-crossed extensions are unfaithful.
We emphasize that our results show that unitarity of the $z=0$ sector of $\boldsymbol{S}$ implies unitarity for all $z$ sectors, i.e. modularity automatically extends from the torus without boundary to the torus with boundary carrying possibly nontrivial topological charge $z$.

\section{Proofs of Mapping Class Operator Relations}
\label{sec:MCG_relations}

In this section, we prove that the relations in Eqs.~\eqref{eq:modrel_1}-\eqref{eq:modrel_4} hold for any $G$-crossed UBTC $\mathcal{B}_{G}^{\times}$.
For this, we do not assume modularity of either $\mathcal{B}_{G}^{\times}$ or $\mathcal{B}_{\bf 0}$.

\subsection{$\boldsymbol{TSTST} = \Theta_0 \boldsymbol{S}$}

Starting from the definition of $\boldsymbol{TSTST}$ acting on the $({\bf g,h})$-sector, we have
\begin{align}
&\boldsymbol{T}^{({\bf h, \bar{h}\bar{h}\bar{g}h})}
\boldsymbol{S}^{({\bf gh, h})}
\boldsymbol{T}^{({\bf gh, \bar{h}\bar{g}h})}
\boldsymbol{S}^{({\bf g, gh})}
\boldsymbol{T}^{({\bf g, h})}
\notag \\
& = \sum_{a_{\bf g} , c_{\bf gh}, b_{\bf h}} \frac{\sqrt{d_a d_c^2 d_b}}{\mathcal{D}_{\bf 0}^2}  
\frac{\eta_a({\bf g , h}) \theta_a \,
\eta_c({\bf gh , \bar{h}\bar{g}h}) \theta_c \,
\eta_b({\bf h , \bar{h}\bar{h}\bar{g}h}) \theta_b
}
{U_{\bf gh}(a, \bar{a};0) U_{\bf h}(c, \bar{c};0)} 
\,\, \gineq{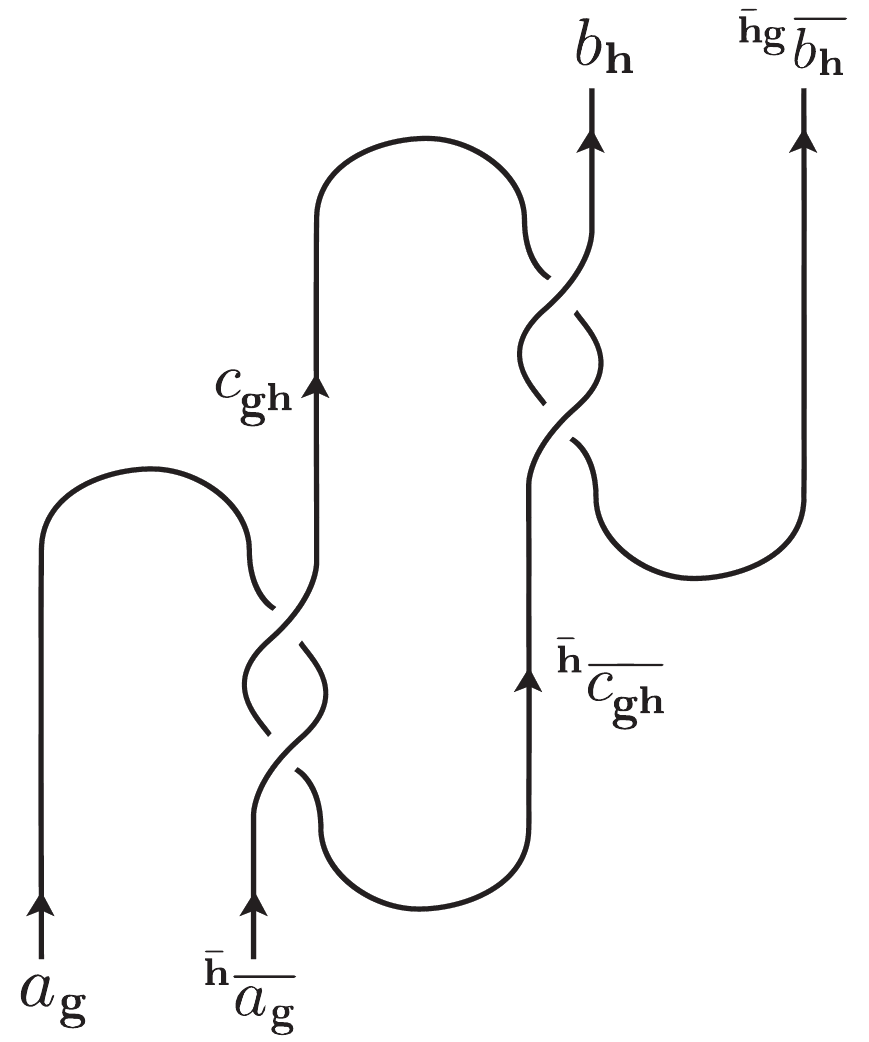}{3}
\notag \\ 
& = \sum_{a_{\bf g}, b_{\bf h}} 
\frac{\sqrt{d_a d_b}}{\mathcal{D}_{\bf 0}^2}  
\theta_a \theta_b  R^{^{\bf \bar{h}}a \, ^{\bf \bar{h}}\bar{a}}_{0} \, 
\frac{\eta_a({\bf g , h})
\eta_a({\bf gh, \bar{h}\bar{g}h}) 
\eta_{^{\bf \bar{h}gh}a}({\bf \bar{h}ghh, \bar{h}\bar{h}\bar{g}h})}
{\eta_{a}({\bf \bar{h}\bar{g}h,\bar{h}ghh })}
\eta_b({\bf h,\bar{h}\bar{h}\bar{g}h})
\notag \\
& \qquad \qquad  \times \sum_{c_{\bf gh}} d_c \theta_c  \frac{U_{\bf \bar{h}\bar{g}h}(c, \bar{c};0)}{U_{\bf h}(c,\bar{c};0)} \eta_c({\bf gh , \bar{h}\bar{g}h}) \eta_{\bar{c}}({\bf \bar{h}\bar{g}h ,\bar{h}gh }) \,\,
\gineq{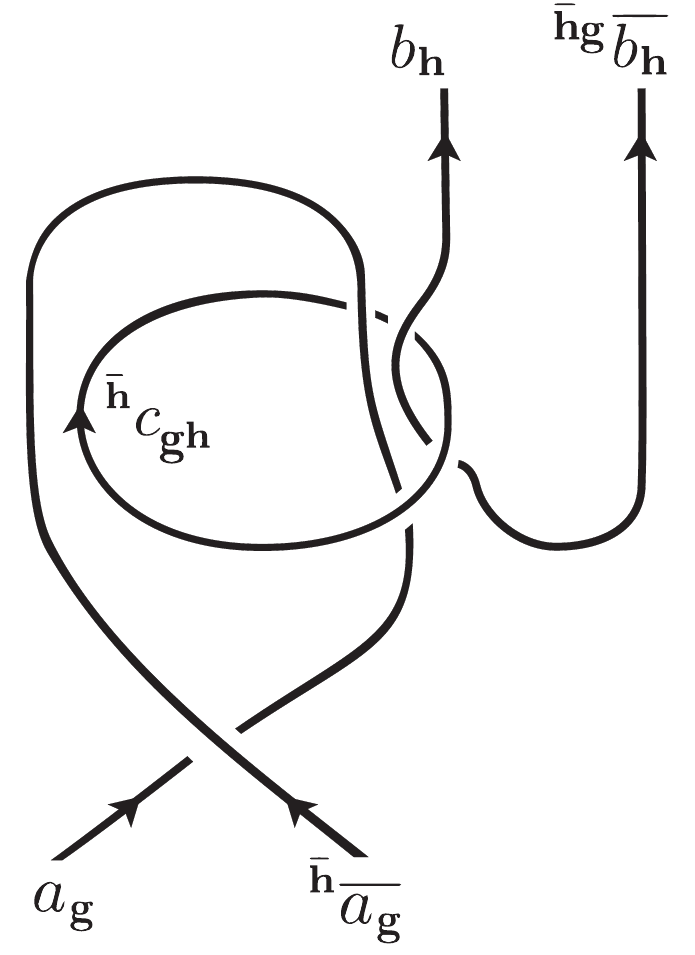}{2.25} \, .
\notag \\ 
& = \sum_{a_{\bf g}, b_{\bf h}} 
\frac{\sqrt{d_a d_b}}{\mathcal{D}_{\bf 0}^2}  
\theta_a \theta_b  R^{^{\bf \bar{h}}a \, ^{\bf \bar{h}}\bar{a}}_{0} \, 
\frac{\eta_a({\bf g , h})
\eta_a({\bf gh, \bar{h}\bar{g}h}) 
\eta_{^{\bf \bar{h}gh}a}({\bf \bar{h}ghh, \bar{h}\bar{h}\bar{g}h})}
{\eta_{a}({\bf \bar{h}\bar{g}h,\bar{h}ghh })}
\eta_b({\bf h,\bar{h}\bar{h}\bar{g}h})
\notag \\
& \qquad \qquad  \times \sum_{c_{\bf gh}} d_{^{\bf \bar{h}}c} \theta_{^{\bf \bar{h}}c}
\frac{\eta_{^{\bf \bar{h}}\bar{c}}({\bf \bar{h}gh, h})}
{U_{\bf \bar{h}ghh}(^{\bf \bar{h}}\bar{c}, \,^{\bf \bar{h}}{c};0)}
\,\,
\gineq{Equations/ST3/STcubed_eq2_low.pdf}{2.25}
\, .
\label{eq:ST3_1}
\end{align}
The second equality follows from the sequence of relations shown diagrammatically in Fig.~\ref{fig:ST3_1}, together with collecting all factors that depend directly on $c$ (not $a$ or $b$) on the second line.
The third equality rewrites the $c$ dependent terms using Eq.~\eqref{eq:U_eta_consistency}, two applications of Eq.~\eqref{eq:eta_consistency}, Eq.~\eqref{eq:U_k_relation}, and Eq.~\eqref{eqn:action_on_twist}.


We will use the relation
\begin{align}
& \sum_{c_{\bf gh}} d_{^{\bf \bar{h}}c} \theta_{^{\bf \bar{h}}c}
\frac{\eta_{^{\bf \bar{h}}\bar{c}}({\bf \bar{h}gh, h})}
{U_{\bf \bar{h}ghh}(^{\bf \bar{h}}\bar{c}, \,^{\bf \bar{h}}{c};0)}
\,\,
\gineq{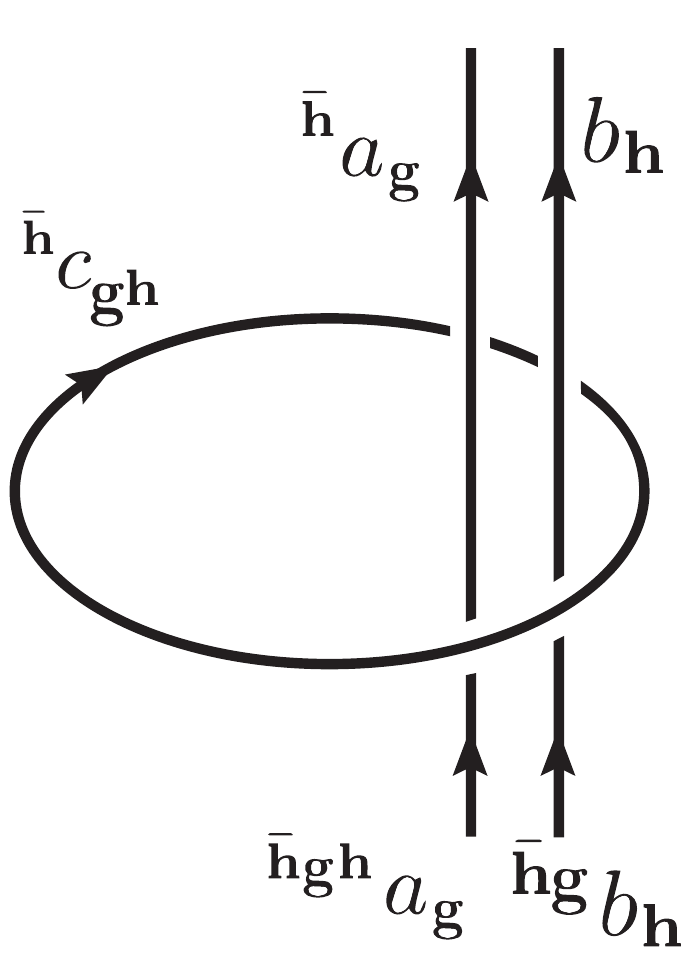}{2}
=\sum_{c_{\bf gh}} \sum_{\substack{x_{\bf gh} \\ \mu,\nu}} 
\sqrt{\frac{d_x}{d_a d_b}}
d_{^{\bf \bar{h}}c} \theta_{^{\bf \bar{h}}c}
\frac{[U_{\bf \bar{h}\bar{h}\bar{g}h}(\,^{\bf \bar{h}}a,b; \,^{\bf \bar{h}}x )]_{\mu \nu}}
{U_{\bf \bar{h}ghh}(^{\bf \bar{h}}\bar{c}, \,^{\bf \bar{h}}{c};0)}
\,\,
\gineq{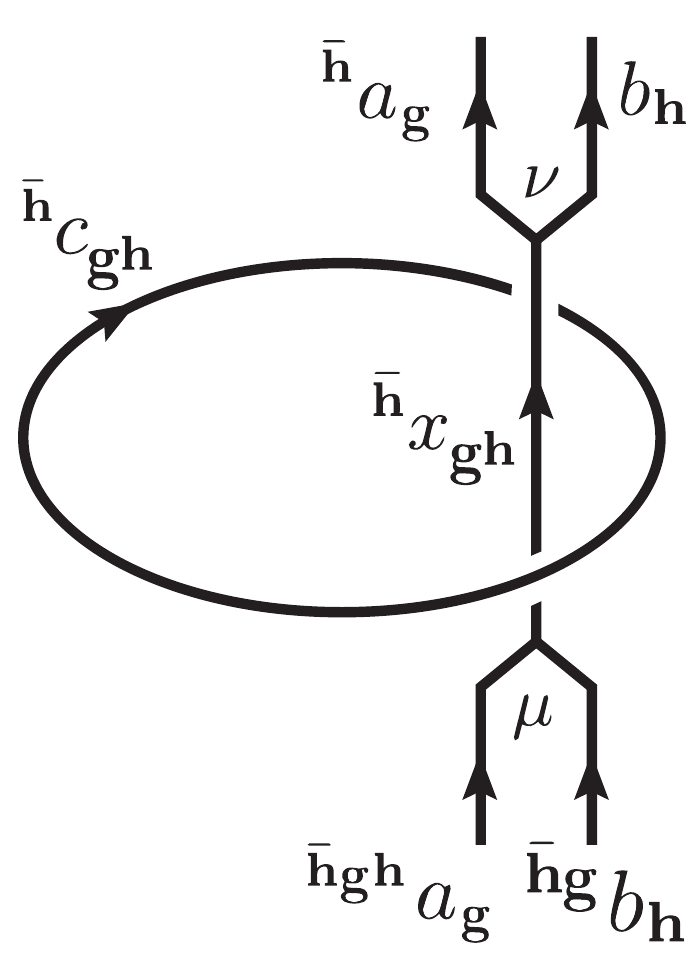}{2}
\notag \\
& \qquad \qquad=
\sum_{\substack{x_{\bf gh} \\ \mu,\nu}} 
\sqrt{\frac{d_x}{d_a d_b}}
[U_{\bf \bar{h}\bar{h}\bar{g}h}(\,^{\bf \bar{h}}a,b; \,^{\bf \bar{h}}x )]_{\mu \nu}
\frac{\mathcal{D}_{\bf 0} \Theta_{\bf 0}}
{\eta_{^{\bf \bar{h}}x}({\bf \bar{h}ghh, \bar{h}\bar{h}\bar{g}h}) \theta_{^{\bf \bar{h}}x}}
\,\,
\gineq{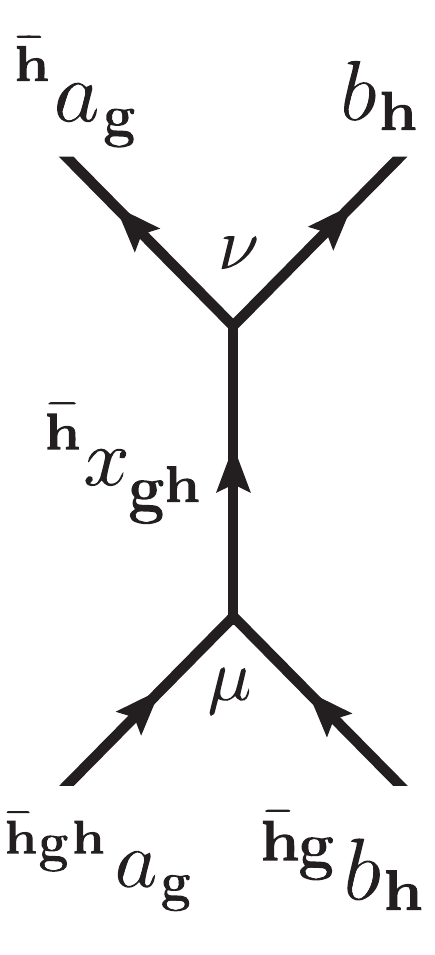}{1.25}
\notag \\
& \qquad \qquad=
\mathcal{D}_{\bf 0} \Theta_{\bf 0}
\frac{1}
{\theta_{^{\bf \bar{h}gh}a} 
\eta_{^{\bf \bar{h}gh}a} ({\bf \bar{h}ghg\bar{h}\bar{g}h, \bar{h}gh\bar{g}h})
\eta_{^{\bf \bar{h}gh}a} ({\bf \bar{h}ghh, \bar{h}\bar{h}\bar{g}h})
}
\notag \\
& \qquad \qquad \qquad \qquad \qquad \qquad \times 
\frac{1}
{\theta_{^{\bf \bar{h}g}b}
\eta_{^{\bf \bar{h}g}b} ({\bf \bar{h}gh\bar{g}h, \bar{h}gh})
\eta_{^{\bf \bar{h}g}b} ({\bf \bar{h}ghh, \bar{h}\bar{h}\bar{g}h})
}
\,\,
\gineq{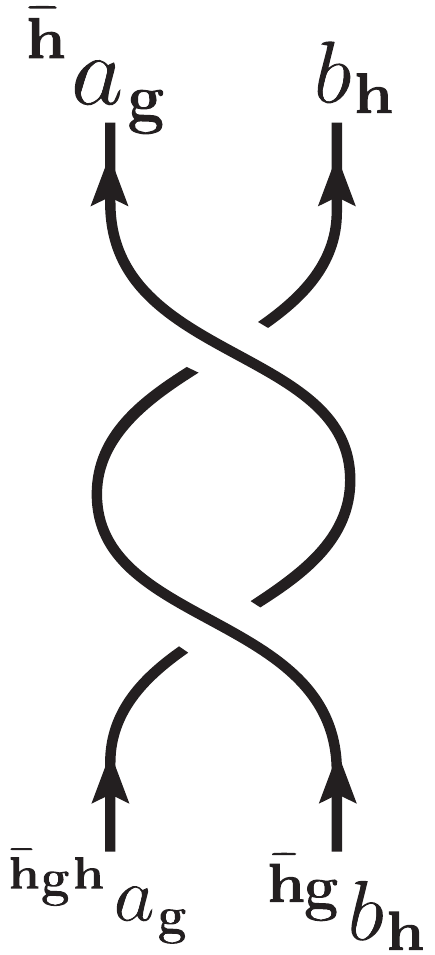}{1.1}
\, .
\label{eq:ST3_2}
\end{align}
The second equality is obtained using Eq.~\eqref{eq:loop_twist}.
The third equality is obtained using Eq.~\eqref{eq:U&eta_consistency_2} and the inverse of Eq.~\eqref{eq:G-crossed_ribbon}.

Combining Eqs.~\eqref{eq:ST3_1} and \eqref {eq:ST3_2}, we find
\begin{align}
&\boldsymbol{T}^{({\bf h, \bar{h}\bar{h}\bar{g}h})}
\boldsymbol{S}^{({\bf gh, h})}
\boldsymbol{T}^{({\bf gh, \bar{h}\bar{g}h})}
\boldsymbol{S}^{({\bf g, gh})}
\boldsymbol{T}^{({\bf g, h})}
\notag \\
& \qquad =  \Theta_{\bf 0} \sum_{a_{\bf g}, b_{\bf h}} 
\frac{\sqrt{d_a d_b}}{\mathcal{D}_{\bf 0}}  
R^{^{\bf \bar{h}}a \, ^{\bf \bar{h}}\bar{a}}_{0}
\,
\frac{\theta_a  
\eta_a({\bf g , h})
\eta_a({\bf gh, \bar{h}\bar{g}h}) 
}
{\theta_{^{\bf \bar{h}gh}a}
\eta_{a}({\bf \bar{h}\bar{g}h,\bar{h}ghh })
\eta_{^{\bf \bar{h}gh}a} ({\bf \bar{h}ghg\bar{h}\bar{g}h,\bar{h}gh\bar{g}h})
}
\notag \\
&  \qquad \qquad \qquad \qquad \times 
\frac{\theta_b \eta_b({\bf h,\bar{h}\bar{h}\bar{g}h})}
{\theta_{^{\bf \bar{h}g}b}
\eta_{^{\bf \bar{h}g}b} ({\bf \bar{h}gh\bar{g}h, \bar{h}gh})
\eta_{^{\bf \bar{h}g}b} ({\bf \bar{h}ghh, \bar{h}\bar{h}\bar{g}h})
}
\,\,
\gineq{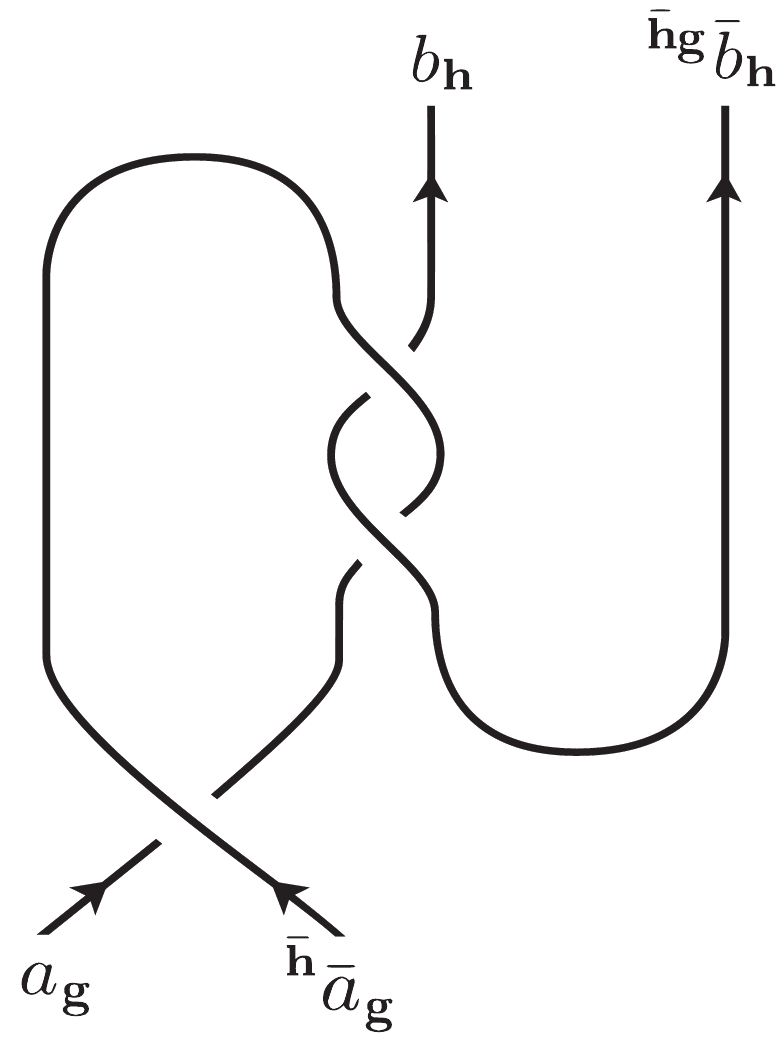}{2}
\notag \\
&  \qquad =  \Theta_{\bf 0} \sum_{a_{\bf g}, b_{\bf h}} 
\frac{\sqrt{d_a d_b}}{\mathcal{D}_{\bf 0}}  
R^{^{\bf \bar{h}}a \, ^{\bf \bar{h}}\bar{a}}_{0} 
\,
\frac{\eta_a({\bf h, \bar{h}\bar{g}h})}
{\eta_{a}({\bf \bar{h}\bar{g}h,\bar{h}gh\bar{g}h})
\eta_b({\bf \bar{h}\bar{g}h,\bar{h}gh})
}
\,\,
\gineq{Equations/ST3/STcubed_eq4_low.pdf}{2}
\notag \\
&  \qquad = \Theta_{\bf 0} \sum_{a_{\bf g}, b_{\bf h}} 
\frac{\sqrt{d_a d_b}}{\mathcal{D}_{\bf 0}}  
\frac{1}{U_{\bf h}(a, \bar{a};0)}
\gineq{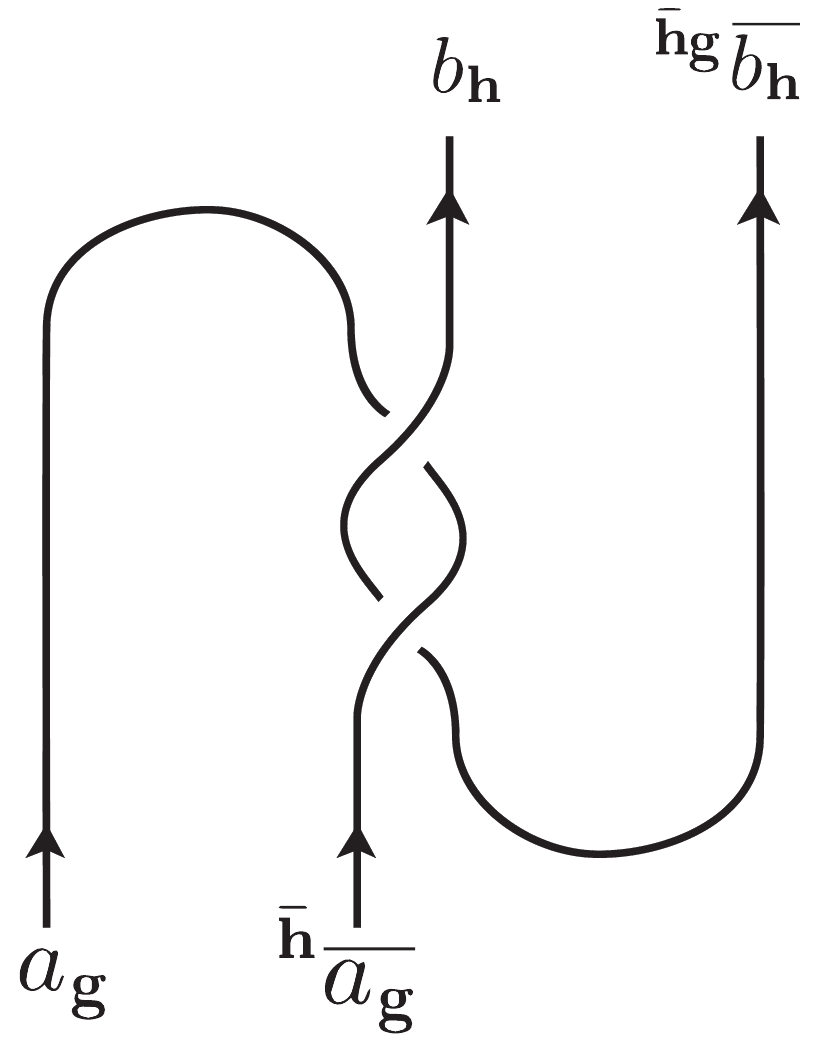}{2} 
\,\,
= \Theta_{\bf 0}  \boldsymbol{S}^{({\bf g ,h})} \, .
\label{eq:ST3_final}
\end{align}
The second equality is obtained using two applications of Eq.~\eqref{eqn:action_on_twist} and several applications of Eq.~\eqref{eq:eta_consistency}.
The third equality is obtained using the sequence of relations shown diagrammatically in Fig. \ref{fig:ST3_2}.
This proves the claimed relation $\boldsymbol{T}\boldsymbol{S}\boldsymbol{T}\boldsymbol{S}\boldsymbol{T}= \Theta_{\bf 0} \boldsymbol{S}$.


\subsection{$\boldsymbol{S} = \boldsymbol{S}^{\dagger} \boldsymbol{C} $}

Starting from the definition, we find
\begin{align}
\boldsymbol{S}^{({\bf{g,h}})} &= \sum_{a_{\bf g}, b_{\bf h}} \frac{\sqrt{d_a d_b}}{\mathcal{D}_{\bf 0}} \frac{1}{U_{\bf h}(a,\bar{a};0)}
    \gineq{Equations/ST3/STcubed_final_low.pdf}{2}
\notag \\
=& \sum_{a_{\bf g}, b_{\bf h}} \frac{\sqrt{d_a d_b}}{\mathcal{D}_{\bf 0}} 
U_{\bf \bar{h}\bar{g}h}(b,\bar{b};0)
\eta_a({\bf h,\bar{h} })\, 
\frac{\eta_{^{\bf \bar{h}}a}({\bf \bar{h}\bar{g}h, \bar{h}g\bar{h}\bar{g}h})}
{\eta_{^{\bf \bar{h}}a}({\bf \bar{h}, \bar{h}\bar{g}h})} 
R^{^{\bf \bar{h}}a \, ^{\bf \bar{h}}\bar{a}}_{0}
U_{\bf \bar{h}g\bar{h}\bar{g}h}(^{\bf \bar{h}}\bar{a} , ^{\bf \bar{h}}a ; 0)
\gineq{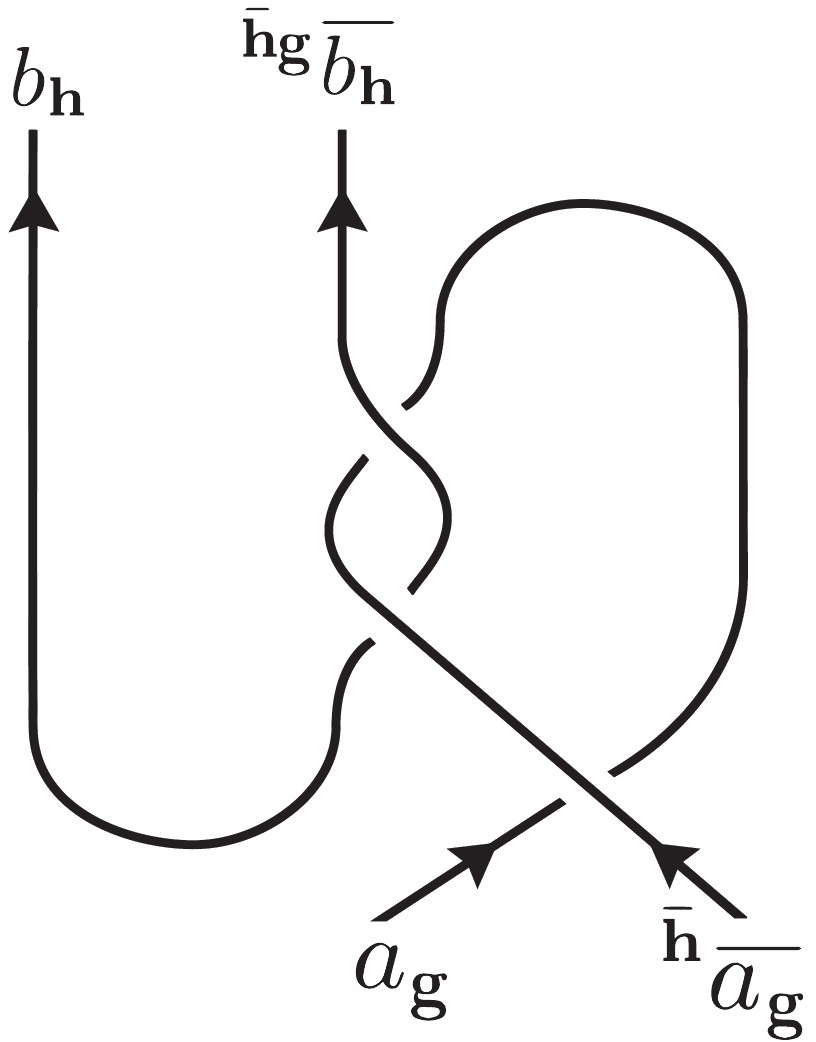}{2}
\notag \\
=& \sum_{a_{\bf g}, b_{\bf h}} 
\eta_a({\bf h, \bar{h}\bar{g}h}) 
\eta_a({\bf \bar{g}h, \bar{h}g\bar{h}\bar{g}h}) 
U_{\bf \bar{h}g\bar{h}\bar{g}h}(^{\bf \bar{h}}\bar{a},^{\bf \bar{h}}a;0)
R^{^{\bf \bar{h}}a \, ^{\bf \bar{h}}\bar{a}}_{0} \,
\frac{\sqrt{d_a d_b}}{\mathcal{D}_{\bf 0}}
U_{\bf \bar{h}\bar{g}h}(b,\bar{b};0)
\gineq{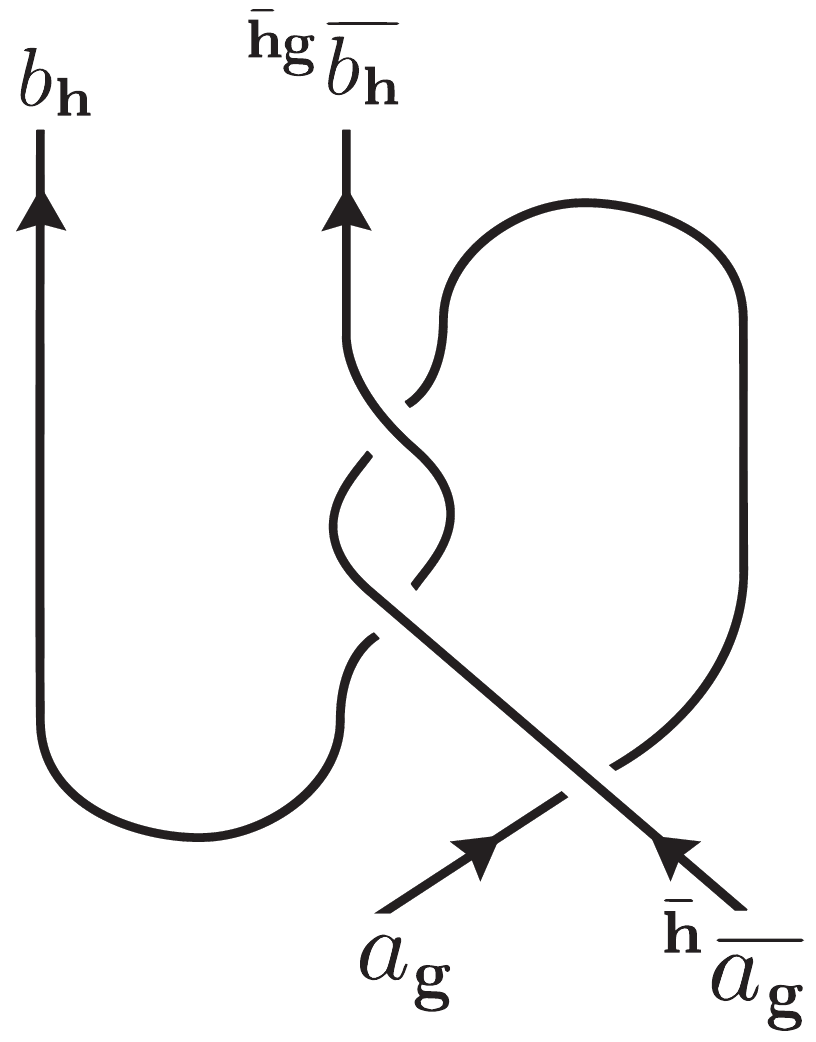}{2} 
\notag \\
=& \boldsymbol{S}^{({\bf h, \bar{h}\bar{g}h}) \, \dagger} \, \boldsymbol{C}^{({\bf g,h})} 
\label{eq:SCS_1}
\end{align}
The second equality is obtained using the sequence of relations shown diagrammatically in Fig.~\ref{fig:SCS}.
The third equality is obtained by two applications of Eq.~\eqref{eq:eta_consistency}.
The find equality follows from the definition of $\boldsymbol{C}^{({\bf g,h})}$, together with
\begin{equation}
\boldsymbol{S}^{({\bf g,h}) \, \dagger} = \sum_{a_{\bf g}, b_{\bf h}} 
\frac{\sqrt{d_a d_b}}{\mathcal{D}_{\bf 0}} U_{\bf h}(a,\bar{a};0)
\gineq{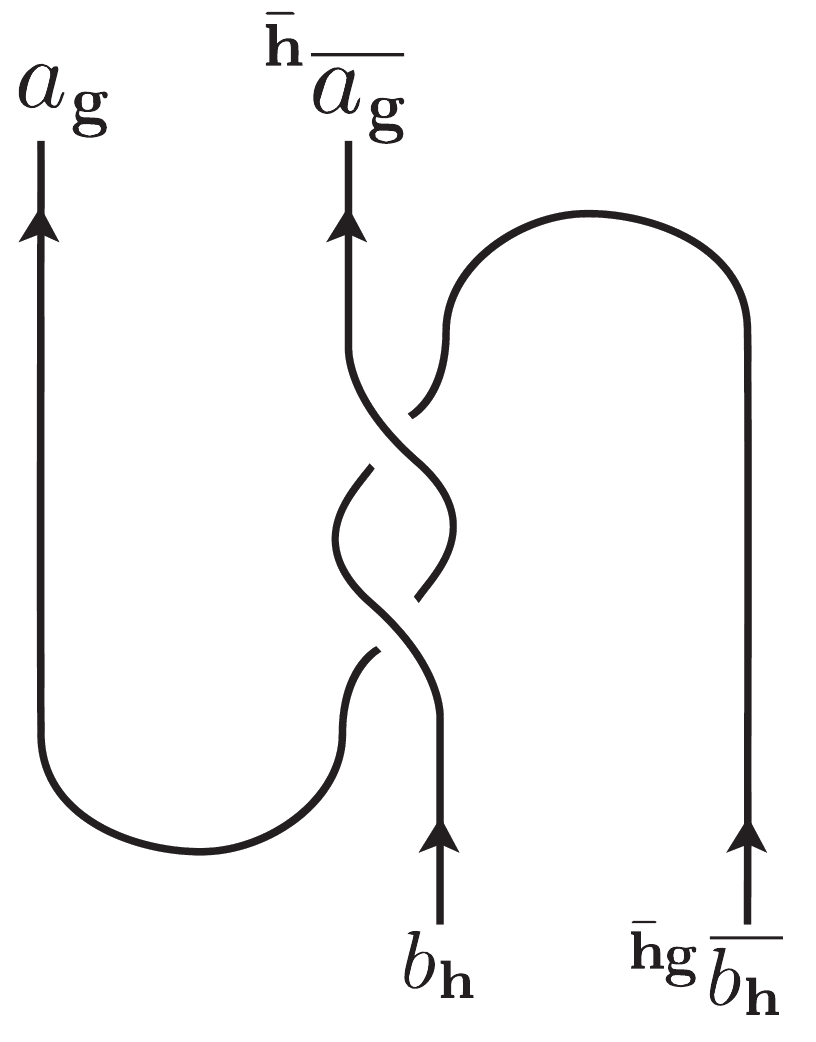}{2}
\, .
\end{equation}
This proves the claimed relation $\boldsymbol{S}=\boldsymbol{S}^{\dagger} \boldsymbol{C}$.


\subsection{$\boldsymbol{CS} = \boldsymbol{SC}$}


Starting from the definitions, we find
\begin{align}
\boldsymbol{C}^{({\bf h, \bar{h}\bar{g}h})} \boldsymbol{S}^{({\bf g,h })}
&= \sum_{a_{\bf g}, b_{\bf h}} \frac{\sqrt{d_a d_b}}{\mathcal{D}_{\bf 0}} 
\frac{1}{U_{\bf h}(a,\bar{a};0)}
\eta_b({\bf \bar{h}\bar{g}h, \bar{h}g\bar{h}\bar{g}h})
\eta_b({\bf \bar{h}\bar{h}\bar{g}h, \bar{h}ghg\bar{h}\bar{g}h})
\notag\\
& \qquad \qquad \times
U_{\bf \bar{h}ghg\bar{h}\bar{g}h}(^{\bf \bar{h}g}\bar{b}, \,^{\bf \bar{h}g}b ;0) \,
R^{^{\bf \bar{h}g}b \, ^{\bf \bar{h}g}\overline{b}}_0
\,\,\gineq{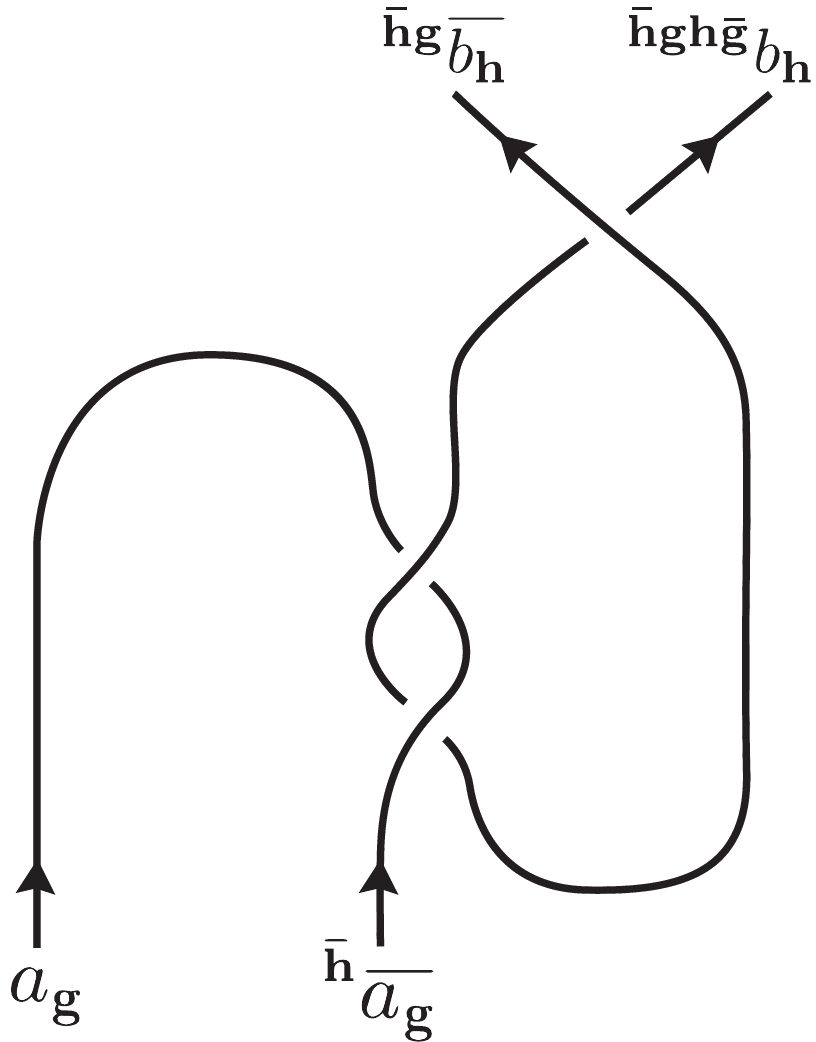}{2.25}
\notag \\
&=\sum_{a_{\bf g}, b_{\bf h}} \frac{\sqrt{d_a d_b}}{\mathcal{D}_{\bf 0}} 
\frac{\eta_b({\bf \bar{h}\bar{g}h, \bar{h}g\bar{h}\bar{g}h})
\eta_b({\bf \bar{h}\bar{h}\bar{g}h, \bar{h}ghg\bar{h}\bar{g}h})}
{\eta_b({\bf \bar{h}\bar{g}h, \bar{h}g\bar{h}})}
\frac{\eta_{^{\bf \bar{h}g}b}({\bf \bar{h}gh, \bar{h}g\bar{h}\bar{g}h})}
{\eta_{^{\bf \bar{h}g}b}({\bf \bar{h}g\bar{h}\bar{g}h, \bar{h}ghg\bar{h}\bar{g}h})}
\notag\\
& \qquad \qquad \times
\frac{\eta_a({\bf h, \bar{h}\bar{g}h})}
{\eta_a({\bf \bar{h}\bar{g}h, \bar{h}gh\bar{g}h })}
\eta_{^{\bf \bar{h}gh}a}({\bf \bar{h}gh\bar{g}h, \bar{h}g\bar{h}\bar{g}h })
R^{^{\bf \bar{h}}a \,^{\bf \bar{h}}\overline{a}}_0
\,\, \gineq{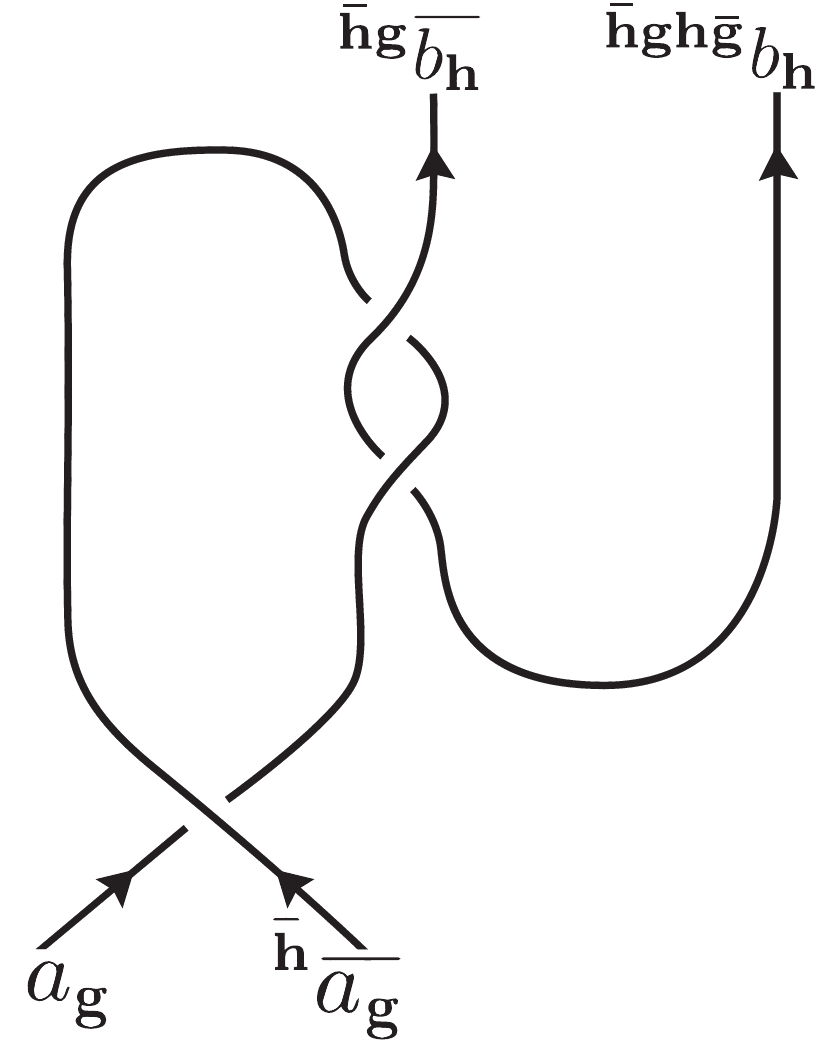}{2.25}
\notag \\
&= \sum_{a_{\bf g}, b_{\bf h}} \frac{\sqrt{d_a d_b}}{\mathcal{D}_{\bf 0}} 
\eta_a({\bf h, \bar{h}\bar{g}h}) 
\eta_a({\bf \bar{g}h, \bar{h}g\bar{h}\bar{g}h })
R^{^{\bf \bar{h}}a \,^{\bf \bar{h}}\overline{a}}_0 
\,\, \gineq{Equations/SCCS/SCCS_eq2_low.pdf}{2.25}
\notag \\
&= \boldsymbol{S}^{({\bf \bar{h}\bar{g}h , \bar{h}g\bar{h}\bar{g}h })}
\boldsymbol{C}^{({\bf g,h})}
.
\label{eq:SCCS}
\end{align}
The second equality is obtained using the sequence of relations shown diagrammatically in Fig.~\ref{fig:SCCS}.
The third equality is obtained by three applications of Eq.~\eqref{eq:eta_consistency}.
This proves the claimed relation $\boldsymbol{CS} = \boldsymbol{SC}$.

\subsection{$\boldsymbol{CT} = \boldsymbol{TC}$}

Starting from the definitions, we find
\begin{align}
\boldsymbol{C}^{({\bf g,gh})} \boldsymbol{T}^{({\bf g,h})} 
&= \sum_{a_{\bf g}} 
\eta_a({\bf g,h}) \theta_{a}
\eta_a({\bf gh, \bar{h}\bar{g}h})
\eta_a({\bf h, \bar{h}\bar{h}\bar{g}h})
U_{\bf \bar{h}\bar{h}\bar{g}h} (^{\bf \bar{h}}\bar{a}, \,^{\bf \bar{h}}a;0) 
R^{^{\bf \bar{h}}a \, ^{\bf \bar{h}}\bar{a}}_{0}
\,
\gineq{Equations/Defs/Cgh_1_low.pdf}{1.5}
\\
&=
\sum_{a_{\bf g}} 
\frac{\eta_{a}({\bf h, \bar{h}gh}) 
\eta_a({\bf gh, \bar{h}\bar{g}h})
\eta_a({\bf h, \bar{h}\bar{h}\bar{g}h})
}
{\eta_{^{\bf \bar{h}}a}({\bf \bar{h}gh, \bar{h}\bar{h}\bar{g}h})}
\frac{\eta_{^{\bf \bar{h}}\overline{a}}({\bf \bar{h}\bar{g}h , \bar{h}gh})}
{\eta_{^{\bf \bar{h}}\overline{a}}({\bf \bar{h}gh, \bar{h}\bar{h}\bar{g}h})
}
\notag \\
& \qquad \qquad  \qquad \times U_{\bf \bar{h}g\bar{h}\bar{g}h}\left( ^{\bf \bar{h}}\bar{a}, \,^{\bf \bar{h}}a ;0 \right)
\theta_{^{\bf \bar{h}}\bar{a}}
R^{^{\bf \bar{h}}a \, ^{\bf \bar{h}}\bar{a}}_{0}
\,
\gineq{Equations/Defs/Cgh_1_low.pdf}{1.5}
\notag \\
&= \sum_{a_{\bf g}} 
\eta_a({\bf h, \bar{h}\bar{g}h}) 
\eta_a({\bf \bar{g}h, \bar{h}g\bar{h}\bar{g}h}) 
U_{\bf \bar{h}g\bar{h}\bar{g}h}( ^{\bf \bar{h}}\bar{a}, \,^{\bf \bar{h}}a;0)
R^{^{\bf \bar{h}}a \, ^{\bf \bar{h}}\bar{a}}_{0}
\notag \\
& \qquad \qquad  \qquad \qquad \times \eta_{^{\bf \bar{h}}\bar{a}}({\bf \bar{h}\bar{g}h, \bar{h}g\bar{h}\bar{g}h }) 
\theta_{^{\bf \bar{h}}\bar{a}}
\,\,
\gineq{Equations/Defs/Cgh_1_low.pdf}{1.5}
\notag \\
&=\boldsymbol{T}^{({\bf \bar{h}\bar{g}h,\bar{h}g\bar{h}\bar{g}h})} \boldsymbol{C}^{({\bf g,h})} \,
.
\label{eq:TCCT}
\end{align}
The second equality uses the relations
\begin{align}
\theta_{a_{\bf g}} &= \frac{\eta_{a}({\bf h, \bar{h}gh})}{\eta_a({\bf g,h})}
\theta_{^{\bf \bar{h}}a_{\bf g}}
= \frac{\eta_{a}({\bf h, \bar{h}gh})}
{\eta_a({\bf g,h})}
\eta_{^{\bf \bar{h}}\overline{a}}({\bf \bar{h}\bar{g}h , \bar{h}gh})
U_{\bf \bar{h}gh}(^{\bf \bar{h}}\bar{a},^{\bf \bar{h}}a;0)
\theta_{^{\bf \bar{h}}\overline{a_{\bf g}}}
,
\end{align}
which follow from Eqs.~\eqref{eqn:action_on_twist} and \eqref{eq:lemmatwist}, together with the relation
\begin{align}
U_{\bf \bar{h}gh} \left(^{\bf \bar{h}}\overline{a_{\bf g}},^{\bf \bar{h}}a_{\bf g};0 \right)
U_{\bf \bar{h}\bar{h}\bar{g}h} \left(^{\bf \bar{h}}\overline{a_{\bf g}}, \,^{\bf \bar{h}}a_{\bf g};0 \right)
&= \frac{U_{\bf \bar{h}g\bar{h}\bar{g}h}\left( ^{\bf \bar{h}}\overline{a_{\bf g}}, \,^{\bf \bar{h}}a_{\bf g};0 \right)}
{\eta_{^{\bf \bar{h}}\overline{a}}({\bf \bar{h}gh, \bar{h}\bar{h}\bar{g}h})
\eta_{^{\bf \bar{h}}a}({\bf \bar{h}gh, \bar{h}\bar{h}\bar{g}h})
}
,
\end{align}
which is an application of Eq.~\eqref{eq:U&eta_consistency_2}.
The third equality in Eq.~\eqref{eq:TCCT} is obtained using four applications of Eq.~\eqref{eq:eta_consistency}, while the last equality follows from the definitions.
This proves the claimed relaion $\boldsymbol{CT} = \boldsymbol{TC}$.

\subsection{$\boldsymbol{C}^2 =  \boldsymbol{Q}^{-1}$}

Starting from the definition, we have
\begin{align}
\boldsymbol{C}^{({\bf \bar{h}\bar{g}h, \bar{h}g\bar{h}\bar{g}h})}
\boldsymbol{C}^{({\bf g,h})}
&= \sum_{a_{\bf g}} 
\eta_{a}({\bf h, \bar{h}\bar{g}h})
\eta_{a}({\bf \bar{g}h, \bar{h}g\bar{h}\bar{g}h})
U_{\bf \bar{h}g\bar{h}\bar{g}h}(^{\bf \bar{h}}\bar{a}, \,^{\bf \bar{h}}a;0)
R^{^{\bf \bar{h}}a \, ^{\bf \bar{h}}\bar{a}}_0
\notag \\
& \qquad \qquad \times
\eta_{^{\bf \bar{h}}\bar{a}}({\bf \bar{h}g\bar{h}\bar{g}h, \bar{h}ghg\bar{h}\bar{g}h}) 
\eta_{^{\bf \bar{h}}\bar{a}}({\bf \bar{h}gg\bar{h}\bar{g}h, \bar{h}gh\bar{g}hg\bar{h}\bar{g}h})
\notag \\
&\qquad \qquad \times 
U_{\bf \bar{h}gh\bar{g}hg\bar{h}\bar{g}h}(^{\bf \bar{h}gh}a,\,^{\bf \bar{h}gh}\bar{a};0)
R^{^{\bf \bar{h}gh}\bar{a} \, ^{\bf \bar{h}gh}a}_0
    \gineq{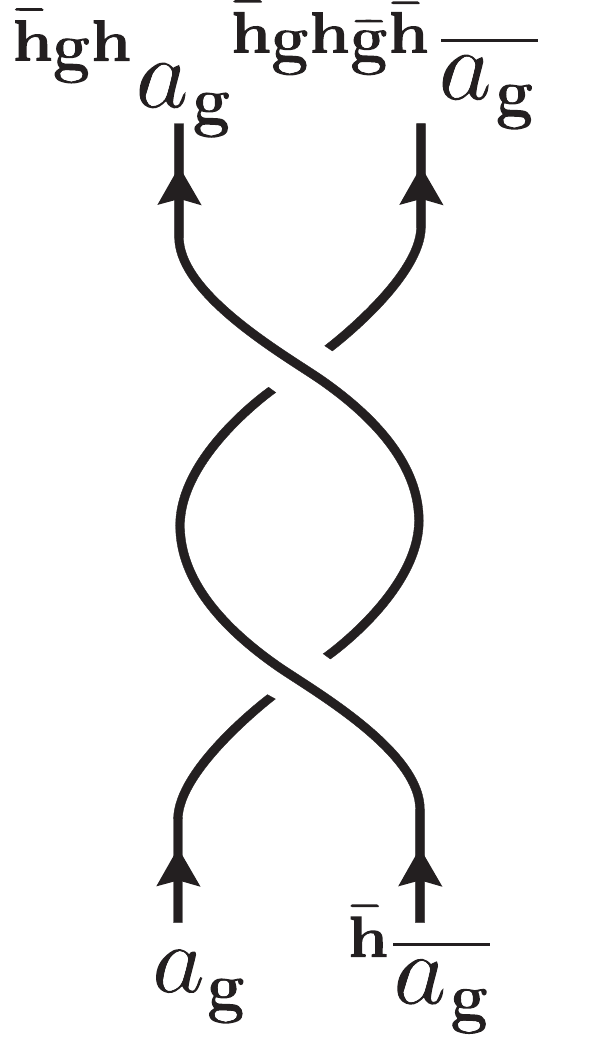}{1.75}
\notag \\
&= \sum_{\substack{a_{\bf g} \\ z_{\bf \bar{h}gh\bar{g}} \\ \mu,\nu}}
\frac{\eta_{a}({\bf h, g\bar{h}\bar{g}h }) 
}
{\eta_{a}({\bf g\bar{h}\bar{g}h, \bar{h}gh\bar{g}hg\bar{h}\bar{g}h })
}
U_{\bf g\bar{h}\bar{g}h}(^{\bf \bar{h}}a,\,^{\bf \bar{h}}\bar{a};0)
\notag \\
& \qquad \qquad \times [U_{\bf g\bar{h}\bar{g}h}(a , \,^{\bf \bar{h}}\bar{a} ; \bar{z})^{-1}]_{\mu \nu}
\, \theta_{\bar{z}}^{-1}
\frac{\sqrt{d_{z}}}{d_{a}}
\gineq{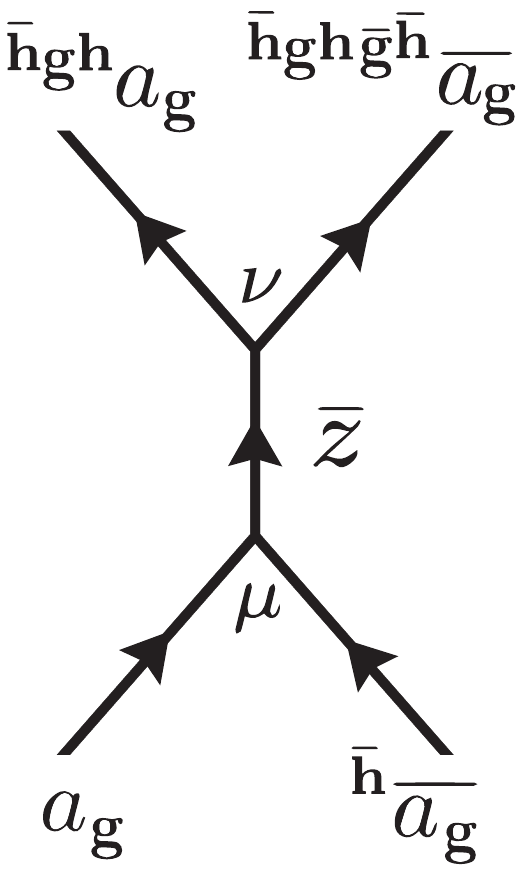}{1.5}
\notag \\
&= \boldsymbol{Q}^{({\bf \bar{h}gh g\bar{h}\bar{g}h ,\bar{h}gh\bar{g} h g\bar{h}\bar{g}h})\, -1}
.
\label{eq:C_sqrd}
\end{align}
The second equality follows by applying Eqs.~\eqref{eq:R_invariant}, \eqref{eq:G-crossed_ribbon} twice, \eqref{eq:U_eta_consistency}, \eqref{eqn:action_on_twist} twice, and \eqref{eq:eta_consistency} seven times.
The third equality follows from the definition of $\boldsymbol{Q}$, noting that
\begin{align}
\gineq{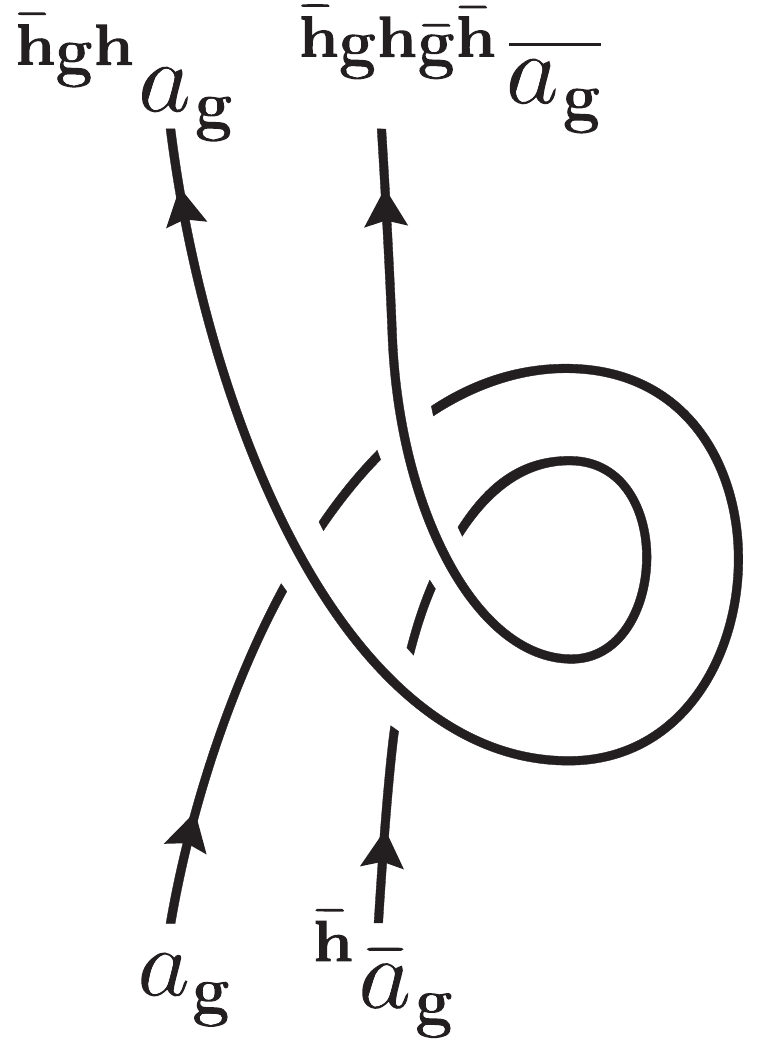}{2}
&= \sum_{\substack{z_{\bf \bar{h}gh\bar{g}} \\ \mu,\nu}}
\eta_a({\bf \bar{h}gh g\bar{h}\bar{g}h, \bar{h}gh\bar{g} \bar{h}\bar{g} h g\bar{h}\bar{g}h})
\eta_{^{\bf \bar{h}}a}({\bf \bar{h}gh g\bar{h}\bar{g}h, \bar{h}gh\bar{g} \bar{h}\bar{g} h g\bar{h}\bar{g}h})
\notag \\
& \qquad \qquad \qquad \times [U_{\bf g\bar{h}\bar{g}h}(a , \,^{\bf \bar{h}}\bar{a} ; \bar{z})^{-1}]_{\mu \nu}
\, \theta_{\bar{z}}^{-1}
\frac{\sqrt{d_{z}}}{d_{a}}
\gineq{Equations/Defs/Qinv_1_low.pdf}{1.5} .
\end{align}
This proves the claimed relation $\boldsymbol{C}^2 = \boldsymbol{Q}^{-1}$.

\section{Faithful $G$-Crossed Extensions $\mathcal{B}_{G}^{\times}$ of Modular $\mathcal{B}_{\bf 0}$ are $G$-Crossed Modular}
\label{sec:S_unitarity}

We have shown that the relations in Eqs.~(\ref{eq:modrel_1})-(\ref{eq:modrel_5}) hold for any $G$-crossed UBTC, without requiring modularity.
However, these relations are not quite sufficient to establish that $\boldsymbol{S}$ and $\boldsymbol{T}$ provide a unitary projective representation of the mapping class group of the torus, since $\boldsymbol{S}$ generally need not be unitary (or even invertible).
For this, one needs to restrict to $G$-crossed UBTCs for which $\boldsymbol{S}$ is unitary.
With this additional condition, it is clear that $\boldsymbol{S}^2 = \boldsymbol{C}$, and the operators $\boldsymbol{S}$ and $\boldsymbol{T}$ constitute a unitary projective representation of the mapping class groups of the torus, with or without a boundary that may carry nontrivial topological charge.
As such, it is natural to define $G$-crossed modularity for $G$-crossed UBTCs by the conditions that there are a finite number of topological charges (simple objects) in each sector $\mathcal{B}_{\bf g}$ and $\boldsymbol{S}$ is unitary. 
However, it is useful to determine the most basic conditions that are equivalent to this notion of $G$-crossed modularity.

In this section, we prove that unitarity of the topological $S$-matrix ($z=0$ sector) of a UBTC $\mathcal{B}_{\bf 0}$ implies the unitarity of $\boldsymbol{S}$ for any faithful $G$-crossed extension $\mathcal{B}_{G}^{\times}$ of $\mathcal{B}_{\bf 0}$.
Faithful here means $\mathcal{B}_{\bf g} \neq \varnothing$ for all ${\bf g}\in G$.
In other words, $\mathcal{B}_{\bf 0}$ modular implies that a faithful $G$-crossed extension $\mathcal{B}_{G}^{\times}$ of it is necessarily $G$-crossed modular.
We emphasize that our results show that the unitarity of the $z\neq 0$ sectors of $\boldsymbol{S}$ follows from the unitarity of the $z=0$ sector, so additional conditions do not need to be imposed for this.

The strategy we use is to prove that $\boldsymbol{S}^2 = \boldsymbol{C}$ holds whenever $\mathcal{B}_{\bf 0}$ is a UMTC and $\mathcal{B}_{\bf g} \neq \varnothing$ for all ${\bf g}\in G$.
Combining Eqs.~\eqref{eq:modrel_2}, \eqref{eq:modrel_3}, and the invertibility of $\boldsymbol{C}$, all of which are automatically true for any $G$-crossed UBTC, together with $\boldsymbol{S}^2 = \boldsymbol{C}$ then implies $\boldsymbol{S}^{-1} = \boldsymbol{S}^{\dagger}$.

Ref.~\onlinecite{Barkeshli2019} showed that modularity of $\mathcal{B}_{\bf 0}$ implies
\begin{equation}
\label{eq:g0_orthog_1}
\sum_{x_{\bf 0} \in \mathcal{B}_{\bf 0}^{\bf g}} S_{a_{\bf g} x_{\bf 0}} S_{b_{\bf g} x_{\bf 0}}^{\ast}
=\sum_{x_{\bf 0} \in \mathcal{B}_{\bf 0}^{\bf g}} S_{x_{\bf 0} a_{\bf g}} S_{x_{\bf 0} b_{\bf g} }^{\ast}
= \delta_{a_{\bf g} b_{\bf g}}
\end{equation}
and the $G$-crossed Verlinde formula
\begin{align}
N_{a_{\bf g} b_{\bf h} }^{c_{\bf gh}} &= \sum_{x_{\bf 0} \in \mathcal{C}_{\bf 0}^{\bf g,h}} \frac{S_{a_{\bf g} x_{\bf 0}} S_{b_{\bf h} x_{\bf 0}} S^{\ast}_{c_{\bf gh} x_{\bf 0}} }{S_{0 x_{\bf 0}}} \eta_{x_{\bf 0}}({\bf \bar{h}, \bar{g} })
.
\label{eq:Gverlinde}
\end{align}

Using this, we see that for $x_{\bf 0} \in \mathcal{B}_{\bf 0}^{\bf g}$ we have
\begin{align}
\delta_{0 x_{\bf 0}}
&= \sum_{b_{\bf 0}} \frac{d_{b_{\bf 0}}}{ \mathcal{D}_{\bf 0}} S_{b_{\bf 0} x_{\bf 0}}^{\ast} 
=\frac{1 }{d_{a_{\bf g}} }
\sum_{b_{\bf 0},c_{\bf g}}  N_{a_{\bf g}b_{\bf 0}}^{c_{\bf g}} \frac{d_{c_{\bf g}} }{ \mathcal{D}_{\bf 0}}  S_{b_{\bf 0} x_{\bf 0}}^{\ast}
\notag \\
&= \frac{1 }{d_{a_{\bf g}} } \sum_{b_{\bf 0},c_{\bf g}} \sum_{y_{\bf 0} \in \mathcal{B}_{\bf 0}^{\bf g}} \frac{S_{a_{\bf g} y_{\bf 0}}  S_{b_{\bf 0} y_{\bf 0}} S_{c_{\bf g} y_{\bf 0}}^{\ast}}{S_{0 y_{\bf 0}}}
\frac{d_{c_{\bf g}} }{ \mathcal{D}_{\bf 0}} S_{b_{\bf 0} x_{\bf 0}}^{\ast}
= \frac{S_{a_{\bf g} x_{\bf 0}}}{d_{a_{\bf g}}  S_{0 x_{\bf 0}}} \sum_{c_{\bf g}}   \frac{d_{c_{\bf g}}}{\mathcal{D}_{\bf 0}} S_{c_{\bf g} x_{\bf 0}}^{\ast}
.
\end{align}
The first and fourth equalities follow from modularity of $\mathcal{B}_{\bf 0}$, the second follows from Eq.~\eqref{qd_rel}, and the third utilizes Eq.~\eqref{eq:Gverlinde}.
When $x_{\bf 0} = 0$, $\frac{S_{a_{\bf g} x_{\bf 0}}}{d_{a_{\bf g}}  S_{0 x_{\bf 0}}} =1$.
When $x_{\bf 0} \neq 0$ and there exists an $a_{\bf g}$ such that $S_{a_{\bf g} x_{\bf 0}} \neq 0$, we can divide this equation by $\frac{S_{a_{\bf g} x_{\bf 0}}}{d_{a_{\bf g}}  S_{0 x_{\bf 0}}}$ to obtain Eq.~\eqref{eq:g0_orthog_2}.
Thus, we find that
\begin{equation}
\label{eq:g0_orthog_2}
\sum_{c_{\bf g}} \frac{d_{c_{\bf g}}}{\mathcal{D}_{\bf 0}} S_{c_{\bf g} x_{\bf 0}}
= \delta_{0 x_{\bf 0}}
\end{equation}
for $x_{\bf 0} \in \mathcal{B}_{\bf 0}^{\bf g}$, when $\mathcal{B}_{\bf 0}$ is modular.
(For $x_{\bf 0}$ with $S_{a_{\bf g} x_{\bf 0}} = 0$ for all $a_{\bf g}$, Eq.~\eqref{eq:g0_orthog_2} is trivially true.)

Next, we use
\begin{equation}
S_{c_{\bf g} b_{\bf 0}}^{\ast} = \frac{S_{c_{\bf g} \overline{b_{\bf 0}}}}{U_{\bf \bar{g}}(b,\bar{b};0)} 
\end{equation}
with Eqs.~\eqref{eq:G_pre_Verlinde} and \eqref{eq:g0_orthog_2} to obtain
\begin{align}
\label{eq:g0_orthog_3}
\sum_{c_{\bf g}}  S_{c_{\bf g} a_{\bf 0}} S_{c_{\bf g} b_{\bf 0}}^{\ast}
&= \sum_{c_{\bf g}}  S_{c_{\bf g} a_{\bf 0}} \frac{ S_{c_{\bf g} \overline{b_{\bf 0}}}}{U_{\bf \bar{g}}(b,\bar{b};0)}
= \sum_{c_{\bf g}}   \sum_{\substack{ e_{\bf 0} \in \mathcal{B}_{\bf 0}^{\bf g} \\ \mu}} \frac{[U_{\bf \bar{g}}(a,\bar{b};e)]_{\mu \mu} S_{c_{\bf g} e_{\bf 0}} S_{c_{\bf g} 0} }{U_{\bf \bar{g}}(b,\bar{b};0)} 
= \delta_{a_{\bf 0} b_{\bf 0}}
\end{align}
for $a_{\bf 0},b_{\bf 0} \in \mathcal{B}_{\bf 0}^{\bf g}$.

As noted in Ref.~\onlinecite{Barkeshli2019}, Eq.~\eqref{eq:g0_orthog_1} can be used to define a generalization of $\omega$-loops for ${\bf g}$-defects given by
\begin{equation}
\label{eq:omega_ag}
\gineq{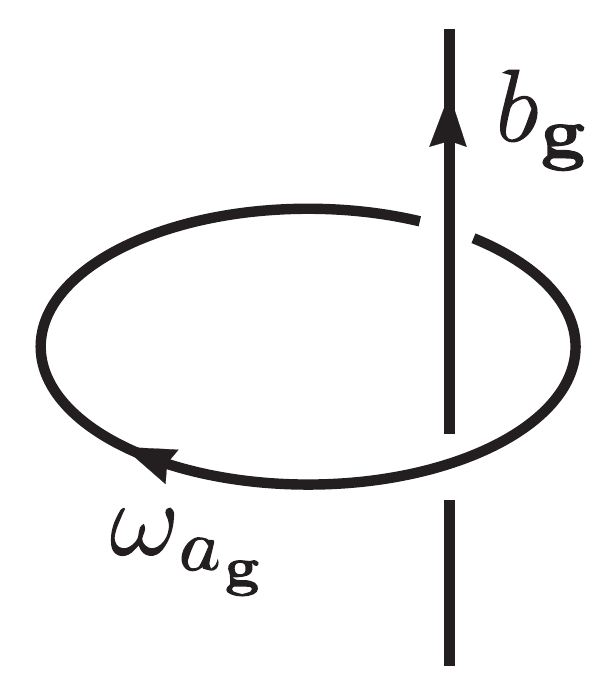}{1.5}
= \sum_{x_{\bf 0} \in \mathcal{C}_{\bf 0}^{\bf g}} S_{0 a_{\bf g}} S^{\ast}_{x_{\bf 0} a_{\bf g}}
\gineq{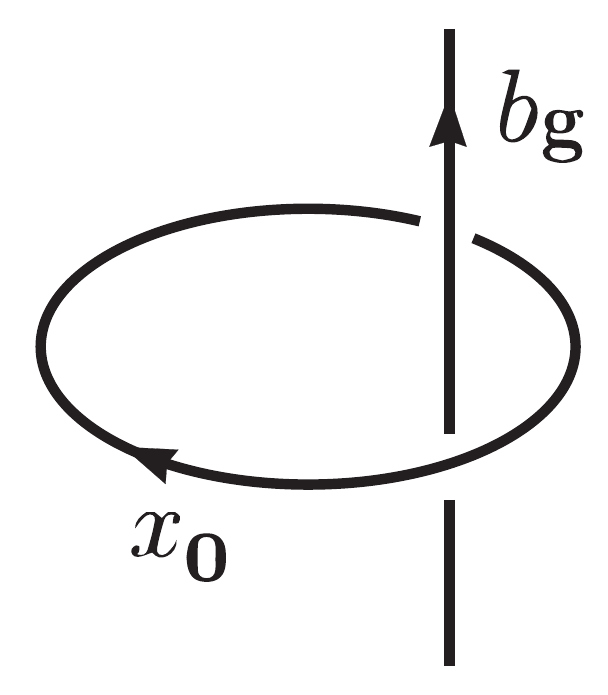}{1.5}
=\delta_{a_{\bf g} b_{\bf g}}
\gineq{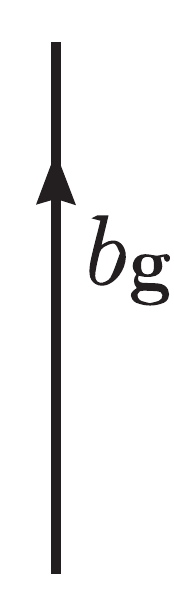}{0.45}
.
\end{equation}
Similarly, we can use Eq.~\eqref{eq:g0_orthog_3} to define generalized $\omega$-loops for ${\bf g}$-invariant quasiparticles $a_{\bf 0} \in  \mathcal{B}_{\bf 0}^{\bf g}$ by
\begin{equation}
\label{eq:omega_a0_ginv}
\gineq{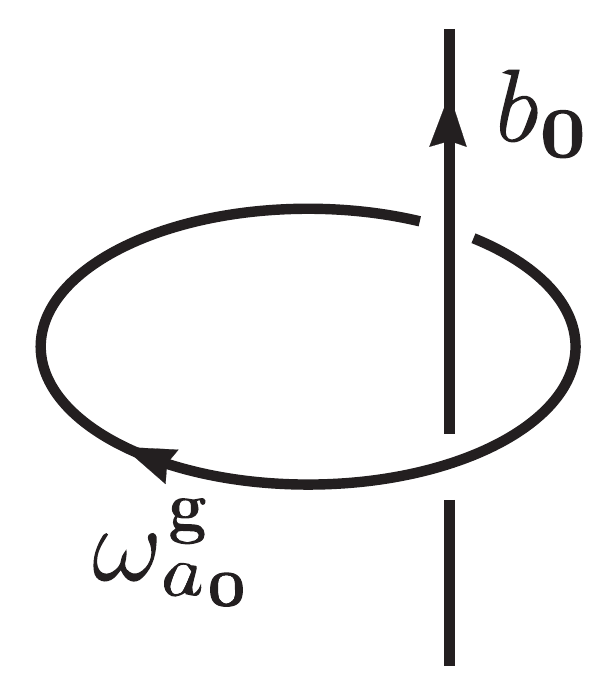}{1.5}
= \sum_{x_{\bf g}} S_{0 a_{\bf 0}} S^{\ast}_{x_{\bf g} a_{\bf 0}}
\gineq{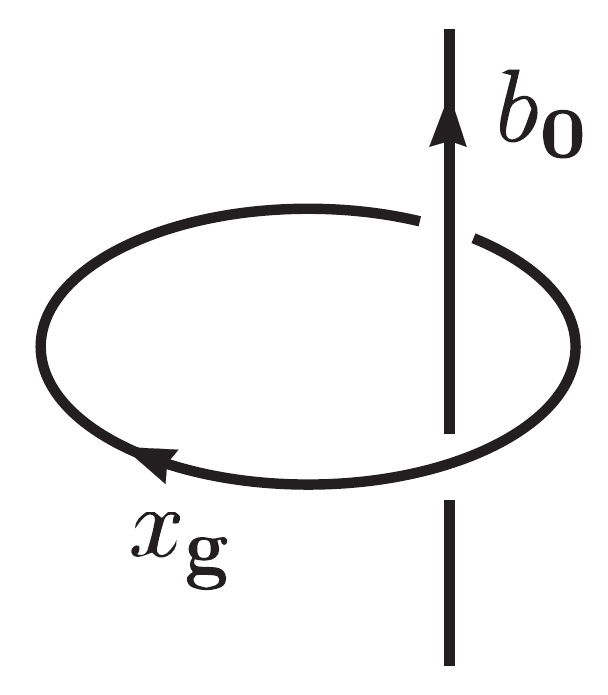}{1.5}
=\delta_{a_{\bf 0} b_{\bf 0}}
\gineq{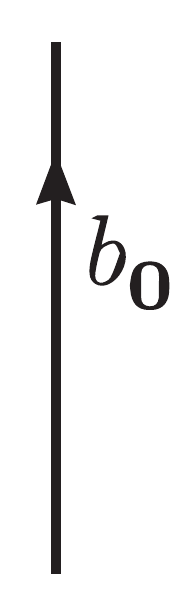}{0.45}
.
\end{equation}
We note that when $b_{\bf 0} \neq \,^{\bf g}b_{\bf 0}$, these diagrams automatically evaluate to zero.

\begin{figure}[H]
    \centering
    \includegraphics[width=13cm]{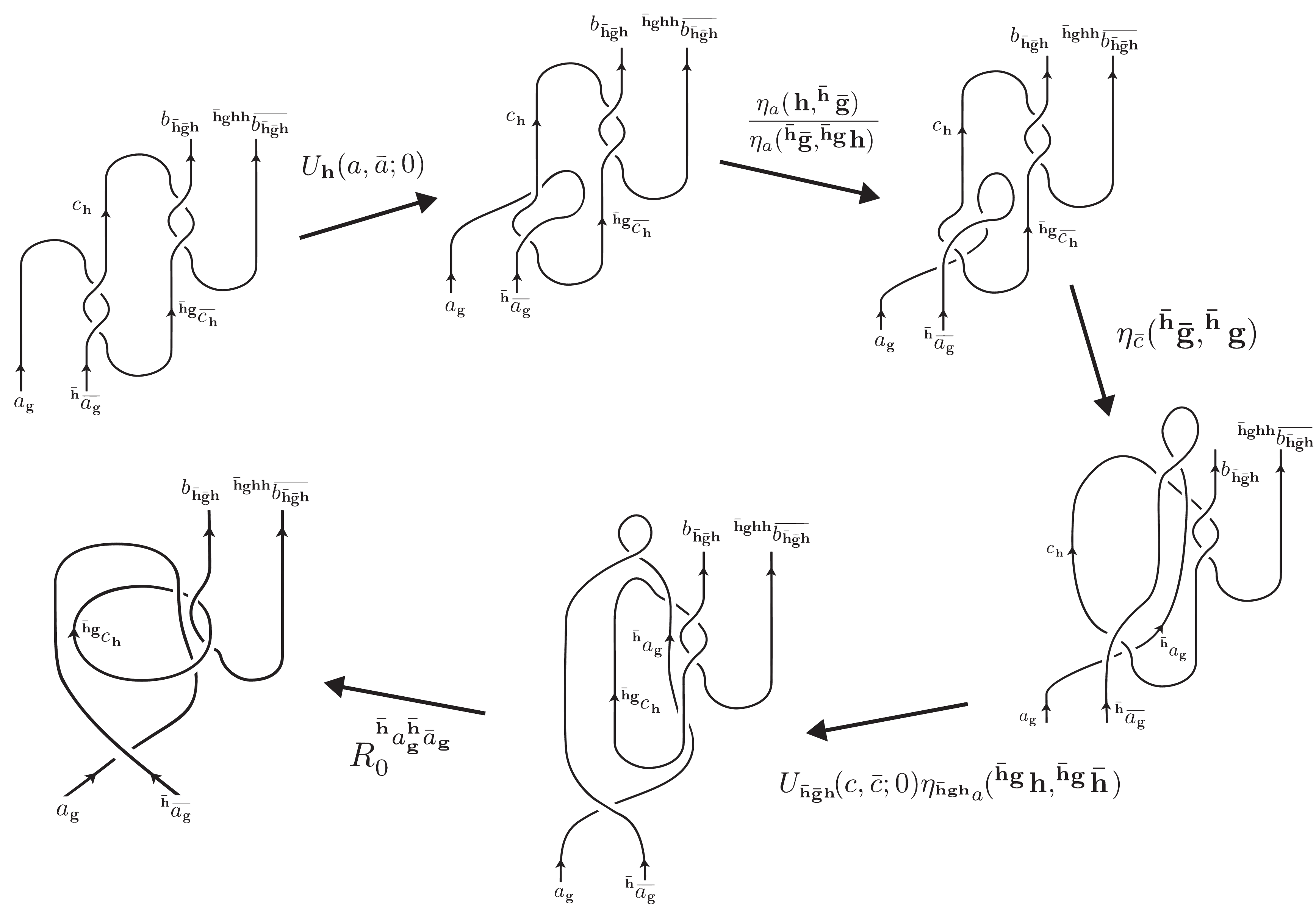}
    \caption{Diagrammatic steps used in deriving Eq.~\eqref{eq:S2C}.}
    \label{fig:S2C}
\end{figure}

We can now use Eq.~\eqref{eq:omega_a0_ginv} to prove that $\boldsymbol{S}^{({\bf{h,\bar{h}\bar{g}h}})} \boldsymbol{S}^{({\bf{g,h}})}  = \boldsymbol{C}^{({\bf{g,h}})}$.
We start by writing the expression for $\boldsymbol{S}^2$
\begin{align}
\boldsymbol{S}^{({\bf{h,\bar{h}\bar{g}h}})} \boldsymbol{S}^{({\bf{g,h}})} 
&= \sum_{a_{\bf g}, c_{\bf h}, b_{\bf \bar{h}\bar{g}h}} 
\frac{\sqrt{d_a d_c^2 d_b}}{\mathcal{D}_{\bf 0}^2} 
\frac{1}{U_{\bf h}(a,\bar{a};0) U_{\bf \bar{h}\bar{g}h}(c,\bar{c};0)}
\,\, \gineq{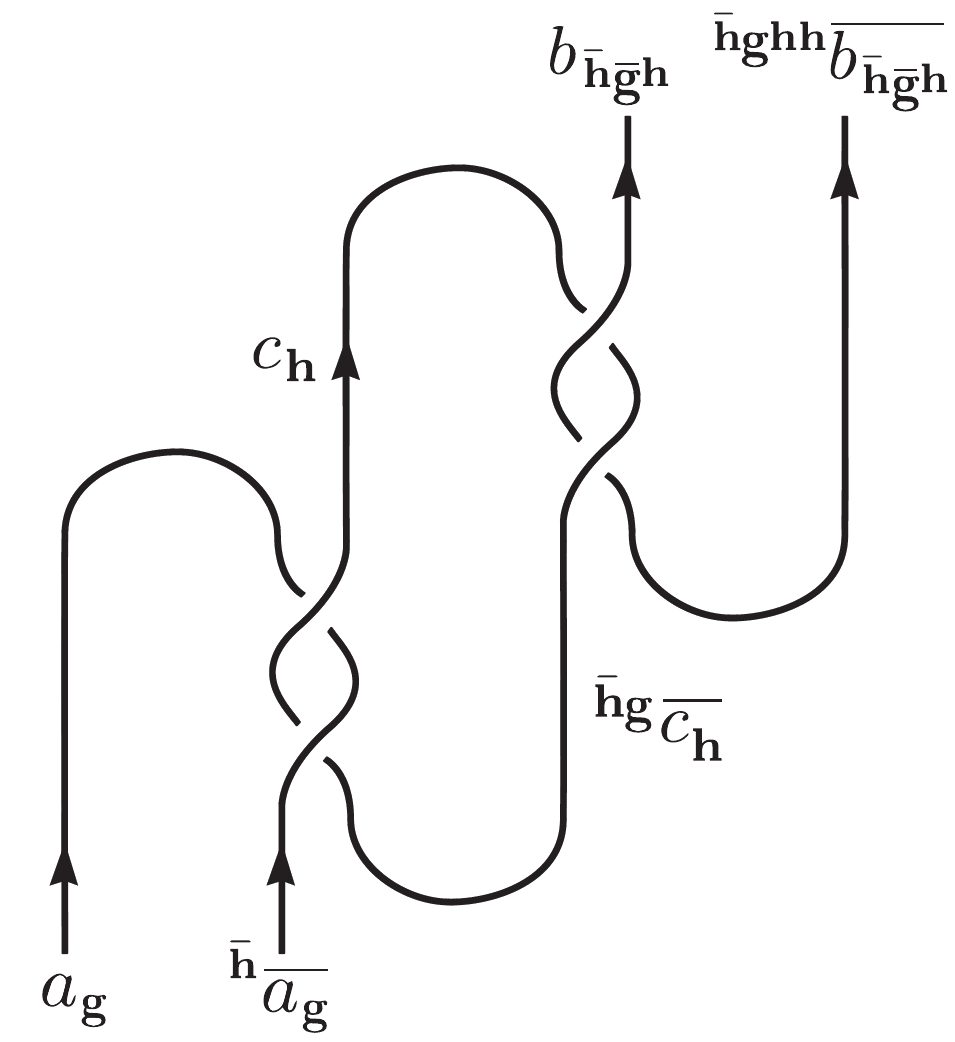}{3}
\notag \\
&=\sum_{a_{\bf g},  b_{\bf \bar{h}\bar{g}h}} 
\sqrt{d_a d_b}
\frac{\eta_a({\bf h,\bar{h}\bar{g}h})}
{\eta_a({\bf \bar{h}\bar{g}h, \bar{h}gh \bar{g}h})}
\eta_{^{\bf \bar{h}gh}a}({\bf \bar{h}gh \bar{g}h, \bar{h}g\bar{h}\bar{g}h}) 
R^{^{\bf \bar{h}}a_{\bf g} \,  ^{\bf \bar{h}}\overline{a_{\bf g}}}_{0}
\notag \\
& \qquad \qquad \times \sum_{c_{\bf h}} \frac{ d_c}{\mathcal{D}_{\bf 0}^2}
\eta_c({\bf \bar{h}\bar{g}h, \bar{h}gh})
\gineq{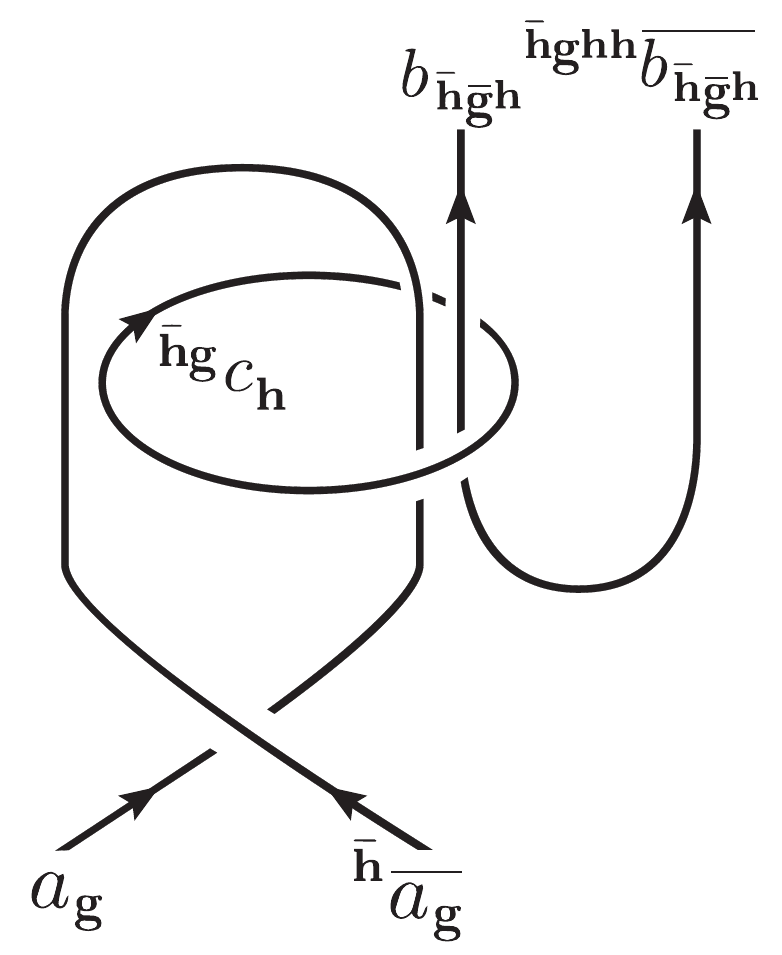}{2.5}
\notag \\
&=\sum_{\substack{a_{\bf g},  b_{\bf \bar{h}\bar{g}h} \\ x_{\bf 0}, \mu, \nu}} 
\sqrt{d_x}
\frac{\eta_a({\bf h,\bar{h}\bar{g}h})}
{\eta_a({\bf \bar{h}\bar{g}h, \bar{h}gh \bar{g}h})}
\eta_{^{\bf \bar{h}gh}a}({\bf \bar{h}gh \bar{g}h, \bar{h}g\bar{h}\bar{g}h}) 
R^{^{\bf \bar{h}}a_{\bf g} \,  ^{\bf \bar{h}}\overline{a_{\bf g}}}_{0}
\notag \\
& \qquad \qquad \times \left[U_{\bf \bar{h}g\bar{h}\bar{g}h}(^{\bf \bar{h}}a, b;x)\right]_{\mu\nu}
\sum_{c_{\bf h}} \frac{ d_c}{\mathcal{D}_{\bf 0}^2}
\,\, \gineq{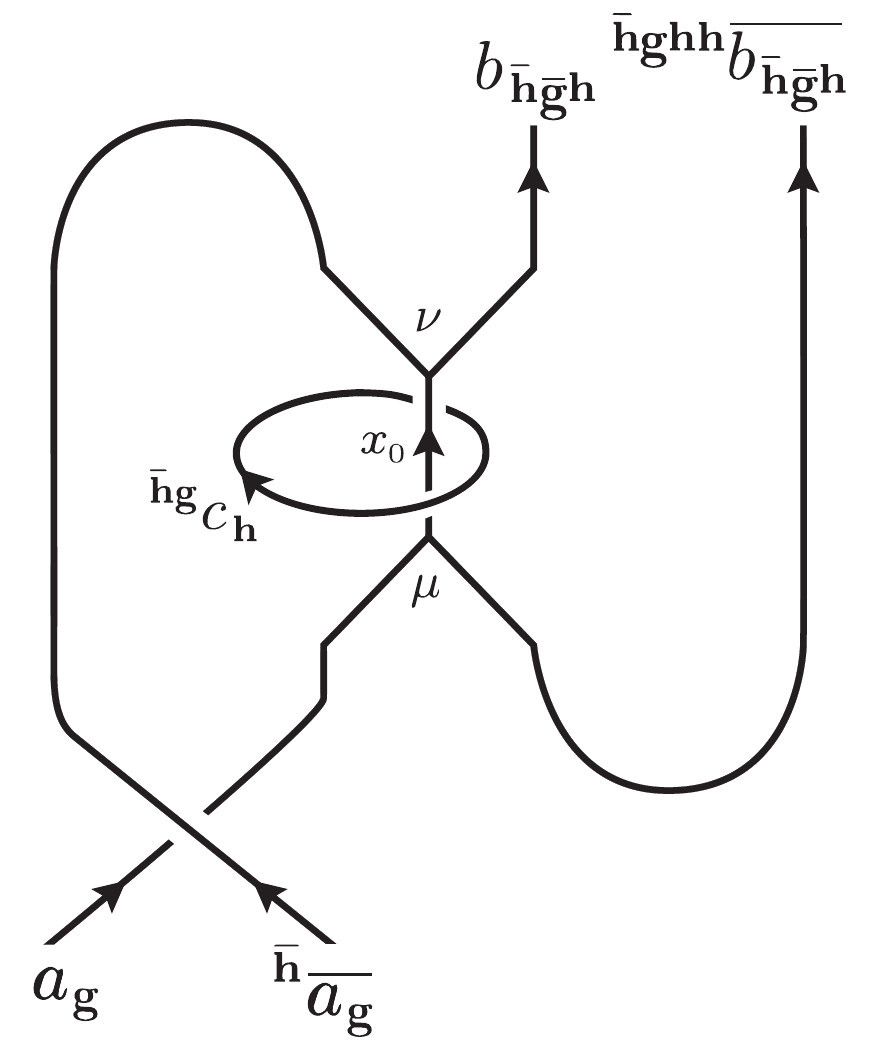}{3}
\notag \\
&=\sum_{a_{\bf g}} 
\frac{\eta_a({\bf h,\bar{h}\bar{g}h})}
{\eta_a({\bf \bar{h}\bar{g}h, \bar{h}gh \bar{g}h})}
\eta_{^{\bf \bar{h}gh}a}({\bf \bar{h}gh \bar{g}h, \bar{h}g\bar{h}\bar{g}h}) 
R^{^{\bf \bar{h}}a_{\bf g} \,  ^{\bf \bar{h}}\overline{a_{\bf g}}}_{0}
\notag \\
& \qquad \qquad \times U_{\bf \bar{h}g\bar{h}\bar{g}h}(^{\bf \bar{h}}a, ^{\bf \bar{h}}\bar{a};0)
\,\, \gineq{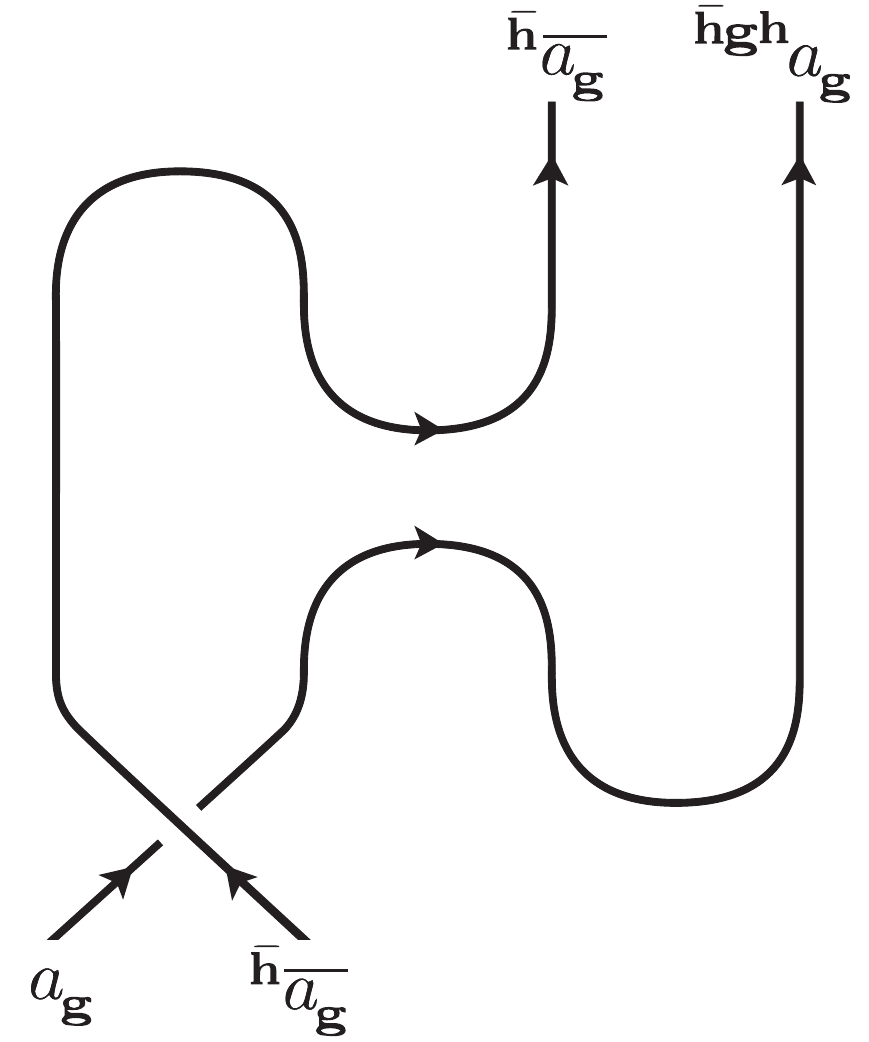}{3} 
\notag \\
&= \sum_{a_{\bf g}} 
\eta_a({\bf h,\bar{h}\bar{g}h})
\eta_a({\bf \bar{g}h, \bar{h}g \bar{h} \bar{g}h})
U_{\bf \bar{h}g\bar{h}\bar{g}h}(^{\bf \bar{h}}\bar{a}, ^{\bf \bar{h}}a;0)
R^{^{\bf \bar{h}}a_{\bf g} \,  ^{\bf \bar{h}}\overline{a_{\bf g}}}_{0}
\,\, \gineq{Equations/Defs/Cgh_1_low.pdf}{1.5}
\notag \\
&= \boldsymbol{C}^{({\bf{g,h}})}
.
\label{eq:S2C}
\end{align}
The second equality follows from the sequence of relations shown diagrammatically in Fig.~\ref{fig:S2C}.
(This sequence is the same as that of Fig.~\ref{fig:ST3_1}, except with different defect sectors involved due to the particular mapping class transformations being examined.)

We used the relation $\eta_{{\bf{^{\bar{h}g}}}\bar{c}}({\bf{{\bar{h}}gh, \bar{h}\bar{g}h}}) =  \eta_{\bar{c}}({\bf{\bar{h}\bar{g}h,\bar{h}gh}})$ in obtaining the third equality.
The fourth equality utilizes Eq.~\eqref{eq:omega_a0_ginv}.
The fifth equality was obtained using the fact that $\frac{\chi_{{\bf{^{\bar{h}}}}a_{\bf{g}}}}{\chi_{^{{\bf{\bar{h}gh}}}a_{\bf{g}}}} =  \frac{U_{\bf{\bar{h}g\bar{h}\bar{g}h}}({\bf{^{\bar{h}}}}\bar{a},{\bf{^{\bar{h}}}}a; 0)}{U_{\bf{\bar{h}g\bar{h}\bar{g}h}}({\bf{^{\bar{h}}}}a,{\bf{^{\bar{h}}}}\bar{a}; 0)}$, obtained from Eq.~\eqref{eq:G-crossed_FS}. 
This proves the relation $\boldsymbol{S}^2 = \boldsymbol{C}$ and, hence, $G$-crossed modularity of $\mathcal{B}_{G}^{\times}$ follows from modularity of $\mathcal{B}_{\bf 0}$ when $\mathcal{B}_{G}^{\times}$ is a faithful $G$-crossed extension.

\section{Examples}

\subsection{$\mathbb{Z}_2$-Toric Code with $\mathbb{Z}_2^{\text{em}}$ Symmetry}

We consider the toric code UMTC, which has topological charges $\mathcal{B}_{\bf 0} = \{ I,e,m,\psi \}$ ($I$ is the vacuum charge) and fusion rules given by $\mathbb{Z}_2 \times \mathbb{Z}_2$ group multiplication.
The quantum dimensions are all $d_{a_{\bf 0}}=1$ and the topological twist factors are $\theta_{I}=\theta_{e}=\theta_{m}=1$ and $\theta_{\psi}=-1$.
We can choose a gauge in which the $F$-symbols are all trivial ($F=1$ when allowed by the fusion rules) and the $R$-symbols are all trivial except for $R^{em}=R^{e\psi}=R^{\psi m}=R^{\psi \psi}=-1$.
The modular transformations of this UMTC are given by
\begin{equation}
\mathcal{S}^{({\bf 0},{\bf 0})}=
\frac{1}{2}
\left[
	\begin{array}{rrrr}
1 & 1 & 1 & 1 \\
1 & 1 & -1 & -1 \\
1 & -1 & 1 & -1 \\
1 & -1 & -1 & 1
    \end{array}
	\right],
\qquad 
\mathcal{T}^{({\bf 0},{\bf 0})}=\left[
	\begin{array}{rrrr}
1 & 0 & 0 & 0 \\
0 & 1 & 0 & 0 \\
0 & 0 & 1 & 0 \\
0 & 0 & 0 & -1
    \end{array}
\right],
\end{equation}
where the basis states are written in the order $\ket{\Phi_{I}}^{({\bf 0,0})}$, $\ket{\Phi_{e}}^{({\bf 0,0})}$, $\ket{\Phi_{m}}^{({\bf 0,0})}$, and $\ket{\Phi_{\psi}}^{({\bf 0,0})}$.
This UMTC has a $\mathbb{Z}_2$ autoequivalence that interchanges the $e$ and $m$ charges, which we refer to as $e-m$ symmetry.

We let $G=\mathbb{Z}_2$ act on the UMTC with the $e-m$ topological symmetry, i.e. $\rho_{\bf 1}(I)=I$, $\rho_{\bf 1}(e)=m$, $\rho_{\bf 1}(m)=e$, and $\rho_{\bf 1}(\psi)=\psi$.
The $G$-crossed extensions with this symmetry action have two defect topological charges  $\mathcal{B}_{\bf 1} = \{ \sigma^{+}, \sigma^{-} \}$ and fusion rules
\begin{align}
\sigma^{\pm} &= \psi \otimes \sigma^{\pm} = e \otimes \sigma^{\mp} = m \otimes \sigma^{\mp}, \\
\sigma^{\pm} \otimes \sigma^{\pm} &= I \oplus \psi , \qquad
\sigma^{\pm} \otimes \sigma^{\mp} = e \oplus m 
.
\end{align}
The defects have quantum dimensions $d_{\sigma^{\pm}}=\sqrt{2}$.
Up to gauge equivalences, there are two distinct $\mathbb{Z}_2^{\text{em}}$-crossed extensions of the toric code UMTC, which can be distinguished by the (shared) Frobenius-Schur indicator of the defects $\kappa_{\sigma} = \pm 1$.
We will not write all the topological data here, but refer to Sec.~XI.I,2 of Ref.~\onlinecite{Barkeshli2019} for details.
In a particular gauge choice (setting $s_e =s_\psi =1$ in Ref.~\onlinecite{Barkeshli2019}), the $z=I$ mapping class transformations were found to be
\begin{equation}
\mathcal{S}^{(I)}=
\left[
	\begin{array}{c|cc|cc|cc}
	  \mathcal{S}^{({\bf 0},{\bf 0})} & 0 & 0 & 0 & 0 & 0 & 0\\
	\hline
	0 & 0 & 0 & \frac{1}{\sqrt{2}} &  \frac{1}{\sqrt{2}} & 0 & 0\\
	 0 & 0 & 0 & -\frac{1}{\sqrt{2}}  &  \frac{1}{\sqrt{2}}  & 0 & 0\\
	\hline
	0 & \frac{1}{\sqrt{2}} & -\frac{1}{\sqrt{2}}  & 0 & 0 & 0 & 0\\
	 0 & \frac{1}{\sqrt{2}}  & \frac{1}{\sqrt{2}}  & 0 & 0 & 0 & 0\\
	\hline
	0 & 0 & 0 & 0 & 0 & 0 & 1\\
	0 & 0 & 0 & 0 & 0 & 1 & 0
	\end{array}
	\right],
\qquad
\mathcal{T}^{(I)}=\left[
	\begin{array}{c|cc|cc|cc}
	\mathcal{T}^{({\bf 0},{\bf 0})} & 0 & 0 & 0 & 0 & 0 & 0\\
	\hline
	0 & 1 & 0 & 0 & 0 & 0 & 0\\
	0 & 0 & -1 & 0  & 0 & 0 & 0\\
	\hline
	0 & 0 & 0 & 0 & 0 & \theta_{\sigma^{+}} & 0\\
	0 & 0 & 0 & 0 & 0 & 0 & \theta_{\sigma^{-}}\\
	\hline
	0 & 0 & 0 & \theta_{\sigma^{+}} & 0 & 0 & 0 \\
	0 & 0 & 0 & 0 & \theta_{\sigma^{-}} & 0 & 0
	\end{array}
	\right]
,
\end{equation}
where the basis states of the defect sectors are written in the order:  
$\ket{\Phi_{I}}^{({\bf 0,1})}$, $\ket{\Phi_{\psi}}^{({\bf 0,1})}$; $\ket{\Phi_{\sigma^+}}^{({\bf 1,0})}$, $\ket{\Phi_{\sigma^-}}^{({\bf 1,0})}$; $\ket{\Phi_{\sigma^+}}^{({\bf 1,1})}$, and $\ket{\Phi_{\sigma^-}}^{({\bf 1,1})}$.
The topological twists of the defects (which are only gauge invariant up to a sign in this example) are
\begin{equation}
\theta_{\sigma^{\pm}} = \left(\sqrt{\kappa_{\sigma}}e^{-i\frac{\pi}{8}}\right)^{\pm 1}
,
\end{equation}
where $\sqrt{\kappa_{\sigma}} = 1$ and $i$ for $\kappa_{\sigma} = 1$ and $-1$, respectively.

For this example, the only other possible value of topological charge on the boundary of $\Sigma_{1,1}$ is $z=\psi$, but only for the nontrivial defect sectors, i.e. the state space of the $({\bf 0,0})$ sector is empty for $z\neq I$.
Using the topological data from Ref.~\onlinecite{Barkeshli2019} (with $s_e = s_\psi =1$), we find
\begin{equation}
\mathcal{S}^{(\psi)}=
\left[
	\begin{array}{c|cc|cc|cc}
	  - & - & - & - & - & - & -\\
	\hline
	- & 0 & 0 & \frac{i}{\sqrt{2}} &  \frac{-i}{\sqrt{2}} & 0 & 0\\
	 - & 0 & 0 & \frac{1}{\sqrt{2}}  &  \frac{1}{\sqrt{2}}  & 0 & 0\\
	\hline
	- & \frac{1}{\sqrt{2}} & \frac{i}{\sqrt{2}}  & 0 & 0 & 0 & 0\\
	 - & \frac{1}{\sqrt{2}}  & \frac{-i}{\sqrt{2}}  & 0 & 0 & 0 & 0\\
	\hline
	- & 0 & 0 & 0 & 0 & \kappa_{\sigma}e^{i\frac{\pi}{4}} & 0\\
	- & 0 & 0 & 0 & 0 & 0 & \kappa_{\sigma}e^{-i\frac{\pi}{4}}
	\end{array}
	\right],
\qquad
\mathcal{T}^{(\psi)}=\left[
	\begin{array}{c|cc|cc|cc}
	- & - & - & - & - & - & -\\
	\hline
	- & 1 & 0 & 0 & 0 & 0 & 0\\
	- & 0 & 1 & 0  & 0 & 0 & 0\\
	\hline
	- & 0 & 0 & 0 & 0 & \theta_{\sigma^{+}} & 0\\
	- & 0 & 0 & 0 & 0 & 0 & \theta_{\sigma^{-}}\\
	\hline
	- & 0 & 0 & \theta_{\sigma^{+}} & 0 & 0 & 0 \\
	- & 0 & 0 & 0 & \theta_{\sigma^{-}} & 0 & 0
	\end{array}
	\right]
,
\end{equation}
where the basis states are written in the order:  
$\ket{\Phi_{e;\psi}}^{({\bf 0,1})}$, $\ket{\Phi_{m;\psi}}^{({\bf 0,1})}$; $\ket{\Phi_{\sigma^+;\psi}}^{({\bf 1,0})}$, $\ket{\Phi_{\sigma^-;\psi}}^{({\bf 1,0})}$; $\ket{\Phi_{\sigma^+;\psi}}^{({\bf 1,1})}$, and $\ket{\Phi_{\sigma^-;\psi}}^{({\bf 1,1})}$.

The $\mathbb{Z}_2^{\text{em}}$ symmetry defects of toric code are sometimes erroneously referred to as Ising non-Abelian anyons, because these defects exhibit fusion and braiding transformations that are projectively the same as Ising anyons.
Despite sharing some similar topological properties, these defects do not qualify as anyons, as they are extrinsic and connected to each other by symmetry branch lines.
However, it is worth considering topological properties that clearly manifest this distinction, and indeed the mapping class transformations are an example of such distinct topological properties.
This can be viewed through the lens of quantum computational power, for which an unpublished construction of Bravyi and Kitaev (see e.g. Ref.~\onlinecite{Freedman06a}) found that the Ising TQFTs can generate a computationally universal gate set (Clifford gates together with the crucial $\pi/8$-phase gate).
While the Clifford gates can be produced from braiding and measurement of Ising anyons, one needs mapping class transformations on surfaces with genus to achieve the $\pi/8$-phase gate needed for universality.
In particular, by encoding a qubit with basis states that (partically) correspond to $\ket{\Phi_{\sigma ; I}}$ and $\ket{\Phi_{\sigma ; \psi}}$ associated with one of the handles of the surface, a $\pi/8$-phase gate can be effected on the qubit by applying $\mathcal{S}\mathcal{T}^{2}\mathcal{S}^{-1}$ on the handle.
In contrast, no such protocol involving mapping class transformations allow one to generate the $\pi/8$-phase gate or computational universality for the $\mathbb{Z}_2^{\text{em}}$ symmetry defects of toric code.

\subsection{$\mathbb{Z}_N$ Anyons with $N$ odd and $\mathbb{Z}_2^{\text{cc}}$ Symmetry}

We consider the $\mathbb{Z}_N^{(p)}$ UMTC with $N$ odd and $\gcd(N,p)=1$.
This UMTC has topological charges $\mathcal{B}_{\bf 0} = \{ 0,1,\ldots, N-1 \}$ and fusion rules given by $\mathbb{Z}_N$ group multiplication $a \otimes b = [a+b]_N$, where $[x]_N = x \text{ mod }N$.
The quantum dimensions are all $d_{a}=1$ and the topological twist factors are $\theta_{a}=e^{i \frac{2 \pi p}{N} a^2}$.
We can choose a gauge in which the $F$-symbols are all trivial ($F=1$ when allowed by the fusion rules) and the $R$-symbols are $R^{ab}=e^{i \frac{2 \pi p}{N} ab}$.
The modular transformations of this UMTC are given by
\begin{equation}
\mathcal{S}_{ab}= \frac{1}{\sqrt{N}} e^{i \frac{4 \pi p}{N} ab}
,
\qquad 
\mathcal{T}_{ab} = e^{i \frac{2 \pi p}{N} a^2} \delta_{ab}
.
\end{equation}
This UMTC has a $\mathbb{Z}_2$ autoequivalence that interchanges $a$ and $\bar{a} = [-a]_{N}$ for all charges, which we refer to as charge conjugation symmetry.

We let $G=\mathbb{Z}_2$ act on the UMTC with the topological charge conjugation symmetry, i.e. $\rho_{\bf 1}(a)= \bar{a}$.
The $G$-crossed extensions with this symmetry action have one defect topological charge  $\mathcal{B}_{\bf 1} = \{ \sigma \}$ and fusion rules
\begin{align}
\sigma \otimes a_{\bf 0} =  \sigma, \qquad \sigma \otimes \sigma = \bigoplus_{a_{\bf 0} \in \mathcal{B}_{\bf 0}} a_{\bf 0} 
.
\end{align}
The defect has quantum dimension $d_{\sigma}=\sqrt{N}$.

Up to gauge equivalences, there are two distinct $\mathbb{Z}_2^{\text{cc}}$-crossed extensions of the $\mathbb{Z}_N^{(p)}$ UMTCs, which can be distinguished by the Frobenius-Schur indicator of the defect $\kappa_{\sigma} = \pm 1$.
We will not write all the topological data here, but refer to Sec.~XI.G of Ref.~\onlinecite{Barkeshli2019} for details.
In a particular gauge choice (setting $r=1$ in Ref.~\onlinecite{Barkeshli2019}), the $z=I$ mapping class transformations were found to be
\begin{equation}
\mathcal{S}^{(0)}=
\left[
	\begin{array}{c|c|c|c}
	  \mathcal{S}^{({\bf 0},{\bf 0})} &  0 & 0 & 0\\
	\hline
	0 & 0 & 1 & 0 \\
	\hline
	0 & 1 & 0 & 0\\
	\hline
	0 & 0 & 0 & \Theta_{\bf 0} \theta_{\sigma}^{-2}
	\end{array}
	\right],
\qquad
\mathcal{T}^{(0)}=\left[
	\begin{array}{c|c|c|c}
	  \mathcal{T}^{({\bf 0},{\bf 0})} &  0 & 0 & 0\\
	\hline
	0 & 1 & 0 & 0 \\
	\hline
	0 & 0 & 0 & \theta_{\sigma }\\
	\hline
	0 & 0 & \theta_{\sigma } & 0
	\end{array}
	\right]
,
\end{equation}
where the basis states of the defect sectors are written in the order:  
$\ket{\Phi_{0}}^{({\bf 0,1})}$; $\ket{\Phi_{\sigma}}^{({\bf 1,0})}$; and $\ket{\Phi_{\sigma}}^{({\bf 1,1})}$.
Here, $\Theta_{\bf 0} = \frac{1}{\sqrt{N}} \sum_{a=0}^{N-1} e^{i \frac{2 \pi p}{N} a^2}$, and the topological twists of the defects are
\begin{equation}
\theta_{\sigma} = \kappa_\sigma \left(\frac{\kappa_\sigma}{\sqrt{N}} \sum_{a=0}^{N-1} (-1)^{p a} e^{-i \frac{\pi p}{N} a^2} \right)^{-\frac{1}{2}}
,
\end{equation}
which is gauge invariant up to a sign in this example.

For this example, the state space on $\Sigma_{1,1}$ is nonempty for any boundary topological charge $z \in \mathcal{B}_{\bf 0}$.
(The state space of the $(\bf 0,0)$ sector is only nonempty for $z=0$.)
Using the topological data from Ref.~\onlinecite{Barkeshli2019} (setting $r=1$), we find
\begin{equation}
\mathcal{S}^{(z)}=
\left[
	\begin{array}{c|c|c|c}
	  - &  - & - & -\\
	\hline
	- & 0 & \theta_{x}^{-1} & 0 \\
	\hline
	- & \theta_{x}^{-1} & 0 & 0\\
	\hline
	- & 0 & 0 & \Theta_{\bf 0} \theta_{\sigma}^{-2} \theta_{x}^{-1}
	\end{array}
	\right],
\qquad
\mathcal{T}^{(z)}=\left[
	\begin{array}{c|c|c|c}
	  - &  - & - & -\\
	\hline
	- & \theta_{x} & 0 & 0 \\
	\hline
	- & 0 & 0 & \theta_{\sigma }\\
	\hline
	- & 0 & \theta_{\sigma } & 0
	\end{array}
	\right]
,
\end{equation}
where $x$ is related to $z$ by $[-2x]_{N}=z$, and the basis states of the defect sectors are written in the order:  
$\ket{\Phi_{x;z}}^{({\bf 0,1})}$; $\ket{\Phi_{\sigma;z}}^{({\bf 1,0})}$; and $\ket{\Phi_{\sigma;z}}^{({\bf 1,1})}$.

\subsection{Three-Fermion Model with $S_3$ Symmetry}

In order to have a nonempty state space when the boundary carries the topological charge of a nontrivial symmetry defect, one must have ${\bf \bar{h}gh\bar{h}} \neq {\bf 0}$, i.e. ${\bf g}$ and ${\bf h}$ must not commute.
The smallest non-Abelian group is $S_3$ and the smallest UMTC for which has $S_3$ topological symmetry is the three-fermion model, so we will consider this UMTC with $G=S_3$.
We will not compute the mapping class transformations here, as we do not have sufficient topological data for the defects, but we will instead focus on the topological state spaces on the torus with boundary to demonstrate the ability to have boundaries that carry defect charge.
(For more details, we refer to Sec.~XI.M of Ref.~\onlinecite{Barkeshli2019}, where much of the topological data was computed.)

The three-fermion model has topological charges $\mathcal{B}_{\bf 0} = \{I,\psi_1, \psi_2, \psi_3 \}$ and fusion rules given by $\mathbb{Z}_2 \times \mathbb{Z}_2$ group multiplication.
The quantum dimensions are all $d_{a_{\bf 0}}=1$ and the topological twist factors are $\theta_{I}=1$ and $\theta_{\psi_j}=-1$ (i.e. the three nontrivial charges are all fermions).
We can choose a gauge in which the $F$-symbols are all trivial ($F=1$ when allowed by the fusion rules) and the $R$-symbols are all trivial except for $R^{\psi_j \psi_j}=R^{\psi_1 \psi_2}=R^{\psi_2 \psi_3}=R^{\psi_3 \psi_1}=-1$.
This UMTC has topological symmetry group $S_3$ corresponding to the permutations of the three fermions.

The defects sectors of the $S_3$-crossed extensions of the three fermion model with this symmetry can be written as $\mathcal{B}_{(12)} = \{Z^{+}, Z^{-} \}$, $\mathcal{B}_{(23)} = \{X^{+}, X^{-} \}$, $\mathcal{B}_{(13)} = \{Y^{+}, Y^{-} \}$, $\mathcal{B}_{(123)} = \{ W \}$, and $\mathcal{B}_{(132)} = \{\overline{W} \}$.
The fusion rules of the defects are given by
\begin{eqnarray}
\psi_{1} \otimes X^{\pm} &=& \psi_{2} \otimes X^{\mp} = \psi_{3} \otimes X^{\mp} = X^{\pm}, \\
X^{\pm} \otimes X^{\pm} &=& I \oplus \psi_{1} , \qquad X^{\pm} \otimes X^{\mp} = \psi_{2} \oplus \psi_{3} ,
\end{eqnarray}
\begin{eqnarray}
\psi_{2} \otimes Y^{\pm} &=& \psi_{1} \otimes Y^{\mp} = \psi_{3} \otimes Y^{\mp} = Y^{\pm}, \\
Y^{\pm} \otimes Y^{\pm} &=& I\oplus \psi_{2} , \qquad Y^{\pm} \otimes Y^{\mp} = \psi_{1} \oplus \psi_{3} ,
\end{eqnarray}
\begin{eqnarray}
\psi_{3} \otimes Z^{\pm}  &=& \psi_{1} \otimes Z^{\mp} = \psi_{2} \otimes Z^{\mp} = Z^{\pm}, \\
Z^{\pm} \otimes Z^{\pm} &=& I \oplus \psi_{3} , \qquad Z^{\pm} \otimes Z^{\mp} = \psi_{1}\oplus \psi_{2} ,
\end{eqnarray}
\begin{eqnarray}
a_{\bf 0} \otimes W &=& W, \qquad  a_{\bf 0} \otimes \overline{W} = \overline{W}, \\
W \otimes \overline{W} &=& I \oplus \psi_1 \oplus \psi_2 \oplus \psi_3, \\
W\otimes W &=& 2\overline{W}, \qquad \overline{W}\otimes\overline{W} = 2W,
\label{eq:Wfusion}
\end{eqnarray}
\begin{eqnarray}
X^r \otimes Y^s &=& Y^r \otimes Z^s = Z^r \otimes X^s = {W}, \\
Y^r \otimes X^s &=& Z^r \otimes Y^s = X^r \otimes Z^s = \overline{W},
\end{eqnarray}
\begin{eqnarray}
Z^r \otimes W &=& \overline{W} \otimes Z^r = Y^r \otimes \overline{W} = W \otimes Y^r = X^+ \oplus X^-,\\
X^r \otimes W &=& \overline{W} \times X^r= Z^r \otimes \overline{W} = W \otimes Z^r = Y^+ \oplus Y^- ,\\
Y^r \otimes W &=& \overline{W} \otimes Y^r=  X^r \otimes \overline{W} = W \otimes X^r = Z^+ \oplus Z^-, 
\end{eqnarray}

With this, we can write the state space for the torus with boundary in the presence of defect branch lines.
The topological charge values $a_{\bf g}$ labeling basis states $\ket{\Phi_{a_{\bf g};I}}^{({\bf g,h})}$ of the $({\bf g,h})$ sectors for boundary charge $z=I$ are given by
\begin{equation}
z=I: \qquad 
 \begin{tabular}{c|c|c|c|c|c|c|}
${\bf g}$\textbackslash ${\bf h}$ & ${\bf 0}$ & $(12)$ & $(23)$ & $(13)$ & $(123)$ & $(132)$\\
\hline
${\bf 0}$ & $I,\psi_1,\psi_2,\psi_3$ & $I,\psi_3$ & $I,\psi_1$ & $I,\psi_2$ & $I$ & $I$\\
\hline
$(12)$ & $Z^{+},Z^{-}$ & $Z^{+},Z^{-}$ & $-$ & $-$ & $-$ & $-$\\
\hline
$(23)$ & $X^{+},X^{-}$ & $-$ & $X^{+},X^{-}$ & $-$ & $-$ & $-$\\
\hline
$(13)$ & $Y^{+},Y^{-}$ & $-$ & $-$ & $Y^{+},Y^{-}$ & $-$ & $-$\\
\hline
$(123)$ & $W$ & $-$ & $-$ & $-$ & $W$ & $-$\\
\hline
$(132)$ & $\overline{W}$ & $-$ & $-$ & $-$ & $-$ & $\overline{W}$\\
\hline
	\end{tabular}
 	\label{tab:I_sectors}
\end{equation}
The topological charge values $a_{\bf g}$ labeling basis states $\ket{\Phi_{a_{\bf g};\psi_1}}^{({\bf g,h})}$ of the $({\bf g,h})$ sectors for boundary charge $z=\psi_1$ are given by
\begin{equation}
z=\psi_1 : \qquad
	\begin{tabular}{c|c|c|c|c|c|c|}
${\bf g}$\textbackslash ${\bf h}$ & ${\bf 0}$ & $(12)$ & $(23)$ & $(13)$ & $(123)$ & $(132)$\\
\hline
${\bf 0}$ & $-$ & $-$ & $\psi_2,\psi_3$ & $-$ & $\psi_3$ & $\psi_2$\\
\hline
$(12)$ & $-$ & $-$ & $-$ & $-$ & $-$ & $-$\\
\hline
$(23)$ & $X^{+},X^{-}$ & $-$ & $X^{+},X^{-}$ & $-$ & $-$ & $-$\\
\hline
$(13)$ & $-$ & $-$ & $-$ & $-$ & $-$ & $-$\\
\hline
$(123)$ & $W$ & $-$ & $-$ & $-$ & $W$ & $-$\\
\hline
$(132)$ & $\overline{W}$ & $-$ & $-$ & $-$ & $-$ & $\overline{W}$\\
\hline
	\end{tabular}
	\label{tab:psi1_sectors}
\end{equation}
The topological charge values $a_{\bf g}$ labeling basis states $\ket{\Phi_{a_{\bf g};\psi_2}}^{({\bf g,h})}$ of the $({\bf g,h})$ sectors for boundary charge $z=\psi_2$ are given by
\begin{equation}
z=\psi_2 : \qquad
 \begin{tabular}{c|c|c|c|c|c|c|}
${\bf g}$\textbackslash ${\bf h}$ & ${\bf 0}$ & $(12)$ & $(23)$ & $(13)$ & $(123)$ & $(132)$\\
\hline
${\bf 0}$ & $-$ & $-$ & $-$ & $\psi_1,\psi_3$ & $\psi_1$ & $\psi_3$\\
\hline
$(12)$ & $-$ & $-$ & $-$ & $-$ & $-$ & $-$\\
\hline
$(23)$ & $-$ & $-$ & $-$ & $-$ & $-$ & $-$\\
\hline
$(13)$ & $Y^{+},Y^{-}$ & $-$ & $-$ & $Y^{+},Y^{-}$ & $-$ & $-$\\
\hline
$(123)$ & $W$ & $-$ & $-$ & $-$ & $W$ & $-$\\
\hline
$(132)$ & $\overline{W}$ & $-$ & $-$ & $-$ & $-$ & $\overline{W}$\\
\hline
	\end{tabular}
	\label{tab:psi2_sectors}
\end{equation}
The topological charge values $a_{\bf g}$ labeling basis states $\ket{\Phi_{a_{\bf g};\psi_3}}^{({\bf g,h})}$ of the $({\bf g,h})$ sectors for boundary charge $z=\psi_3$ are given by
\begin{equation}
z=\psi_3 : \qquad
	\begin{tabular}{c|c|c|c|c|c|c|}
${\bf g}$\textbackslash ${\bf h}$ & ${\bf 0}$ & $(12)$ & $(23)$ & $(13)$ & $(123)$ & $(132)$\\
\hline
${\bf 0}$ & $-$ & $\psi_1,\psi_2$ & $-$ & $-$ & $\psi_2$ & $\psi_1$\\
\hline
$(12)$ & $Z^{+},Z^{-}$ & $Z^{+},Z^{-}$ & $-$ & $-$ & $-$ & $-$\\
\hline
$(23)$ & $-$ & $-$ & $-$ & $-$ & $-$ & $-$\\
\hline
$(13)$ & $-$ & $-$ & $-$ & $-$ & $-$ & $-$\\
\hline
$(123)$ & $W$ & $-$ & $-$ & $-$ & $W$ & $-$\\
\hline
$(132)$ & $\overline{W}$ & $-$ & $-$ & $-$ & $-$ & $\overline{W}$\\
\hline
	\end{tabular}
	\label{tab:psi3_sectors}
\end{equation}
The topological charge and fusion vertex values $(a_{\bf g};\mu)$ labeling basis states $\ket{\Phi_{a_{\bf g};W,\mu}}^{({\bf g,h})}$ of the $({\bf g,h})$ sectors for boundary charge $z=W$ are given by
\begin{equation}
z=W : \qquad 
	\begin{tabular}{c|c|c|c|c|c|c|}
${\bf g}$\textbackslash ${\bf h}$ & ${\bf 0}$ & $(12)$ & $(23)$ & $(13)$ & $(123)$ & $(132)$\\
\hline
${\bf 0}$ & $-$ & $-$ & $-$ & $-$ & $-$ & $-$\\
\hline
$(12)$ & $-$ & $-$ & $Z^{+},Z^{-}$ & $-$ & $Z^{+},Z^{-}$ & $-$\\
\hline
$(23)$ & $-$ & $-$ & $-$ & $X^{+},X^{-}$ & $X^{+},X^{-}$ & $-$\\
\hline
$(13)$ & $-$ & $Y^{+},Y^{-}$ & $-$ & $-$ & $Y^{+},Y^{-}$ & $-$\\
\hline
$(123)$ & $-$ & $(W;1),(W;2)$ & $(W;1),(W;2)$ & $(W;1),(W;2)$ & $-$ & $-$\\
\hline
$(132)$ & $-$ & $-$ & $-$ & $-$ & $-$ & $-$\\
\hline
	\end{tabular}
	\label{tab:W_sectors}
\end{equation}
The topological charge and fusion vertex values $(a_{\bf g};\mu)$ labeling basis states $\ket{\Phi_{a_{\bf g};\overline{W},\mu}}^{({\bf g,h})}$ of the $({\bf g,h})$ sectors for boundary charge $z=\overline{W}$ are given by
\begin{equation}
z=\overline{W}: \qquad
	\begin{tabular}{c|c|c|c|c|c|c|}
${\bf g}$\textbackslash ${\bf h}$ & ${\bf 0}$ & $(12)$ & $(23)$ & $(13)$ & $(123)$ & $(132)$\\
\hline
${\bf 0}$ & $-$ & $-$ & $-$ & $-$ & $-$ & $-$\\
\hline
$(12)$ & $-$ & $-$ & $-$ & $Z^{+},Z^{-}$ & $-$ & $Z^{+},Z^{-}$\\
\hline
$(23)$ & $-$ & $X^{+},X^{-}$ & $-$ & $-$ & $-$ & $X^{+},X^{-}$\\
\hline
$(13)$ & $-$ & $-$ & $Y^{+},Y^{-}$ & $-$ & $-$ & $Y^{+},Y^{-}$\\
\hline
$(123)$ & $-$ & $-$ & $-$ & $-$ & $-$ & $-$\\
\hline
$(132)$ & $-$ & $(\overline{W};1),(\overline{W};2)$ & $(\overline{W};1),(\overline{W};2)$ & $(\overline{W};1),(\overline{W};2)$ & $-$ & $-$\\
\hline
	\end{tabular}
	\label{tab:barW_sectors}
\end{equation}
We see that for the cases where the boundary carries a symmetry defect ($z=W$ or $\overline{W}$), the fusion multiplicities from Eq.~\eqref{eq:Wfusion} arise in the basis state labels for $a_{\bf g} = W$ and $\overline{W}$.

\acknowledgements

We thank Z. Wang for useful discussions.
This work was performed in part at the Aspen Center for Physics, which is supported by National Science Foundation grant PHY-1607611.


\appendix

\section{Review of $G$-Crossed UBTC}
\label{sec:review}

In this appendix, we provide a brief review of $G$-crossed UBTCs (working with skeletonizations) following Ref.~\onlinecite{Barkeshli2019}, to which we direct the reader for more details and derivations.

The fusion structure of a $G$-crossed UBTC $\mathcal{B}_{G}^{\times}$ (i.e. ignore the braiding), is described by a $G$-graded unitary fusion tensor category (UFTC)
\begin{equation}
\mathcal{B}_{G} = \bigoplus_{{\bf g} \in G} \mathcal{B}_{\bf g} 
.
\end{equation}
This is a UFTC for which the fusion rules are additionally required to be $G$-graded.
In more detail, each topological charge (simple object) is assigned a particular element of the symmetry group $G$ and the fusion rules must respect the group multiplication of $G$.
We denote the identity element of $G$ as ${\bf 0}$ and inverses as ${\bf \bar{g}} ={\bf g}^{-1}$
Topological charges assigned the group element ${\bf g}$ correspond to the distinct types of ${\bf g}$-defects, and we recognize ${\bf 0}$-defects as the quasiparticles of the topological phase.
In this way, we write topological charges corresponding to ${\bf g}$-defects as $a_{\bf g} \in \mathcal{B}_{\bf g}$~\footnote{In a slight abuse of notation, we will use the same symbol in referencing both the category and its set of topological charges.} and the associative fusion algebra takes the form
\begin{equation}
a_{\bf{g}} \otimes b_{\bf{h}} = \sum_{c \in \mathcal{B}_{\bf{gh}}} N_{a_{\bf g} b_{\bf h}}^{c_{\bf gh}} c_{\bf{gh}}
.
\end{equation}
The fusion multiplicities $N_{a b}^c$ are non-negative integers indicating how topological charges can combine or split.
We require a unique vacuum charge, which we denote as $0$ for which $N_{a 0}^{c} = N_{0 a}^{c} \delta_{a c}$.
For each charge $a_{\bf{g}}$, we require a unique conjugate charge $\overline{a_{\bf{g}}} \in \mathcal{B}_{\bf \bar{g}}$ for which $N_{a b}^{0} = \delta_{b \bar{a}}$.

Each fusion product has an associated fusion and splitting vector space, for which we write the basis states as
\begin{equation}
\left( d_{c} / d_{a}d_{b} \right) ^{1/4}
\gineq{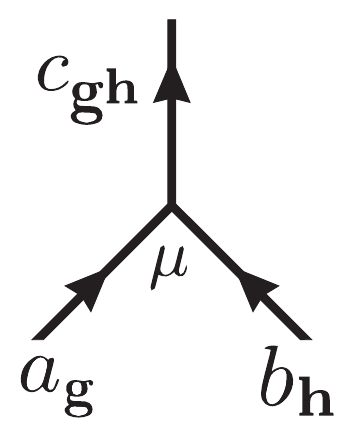}{1.35}
=\left\langle a_{\bf g},b_{\bf h};c_{\bf gh},\mu \right\lvert \in
V_{ab}^{c} ,
\label{eq:bra}
\end{equation}
\begin{equation}
\left( d_{c} / d_{a}d_{b}\right) ^{1/4}
\gineq{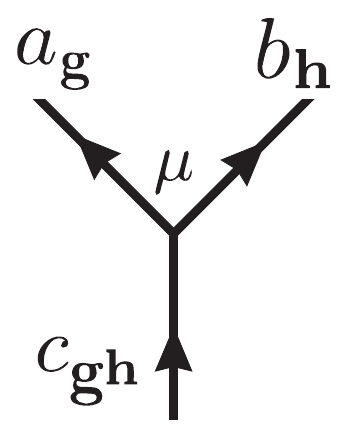}{1.35}
=\left\lvert a_{\bf g},b_{\bf h};c_{\bf gh},\mu \right\rangle \in
V_{c}^{ab},
\label{eq:ket}
\end{equation}
where $\text{dim} V_{ab}^{c} = \text{dim} V^{ab}_{c} = N_{ab}^{c}$, and $\mu = 1,\ldots , N_{ab}^{c}$.
The normalization factors in translating between diagrammatic and bra/ket notation are given in terms of the quantum dimensions $d_{a}$ of the respective charges, which are chosen to make bending lines unitary transformations.
More general states and operators can be formed diagrammatically by stacking together trivalent vertices such that lines glued together have the same topological charge.

Diagrams can be reduced using the inner product
\begin{equation}
\label{eq:inner_product}
\langle a_{\bf g},b_{\bf h};c'_{\bf gh},\mu'  \,\lvert\,  a_{\bf g},b_{\bf h};c_{\bf gh},\mu \rangle =
\sqrt{\frac{d_{c}}{d_{a}d_{b}}} \,
\gineq{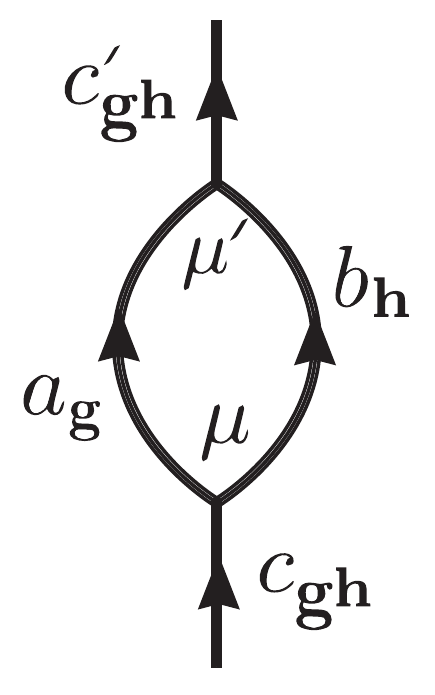}{1.5}
\, = \delta_{c c'} \delta_{\mu \mu'} \gineq{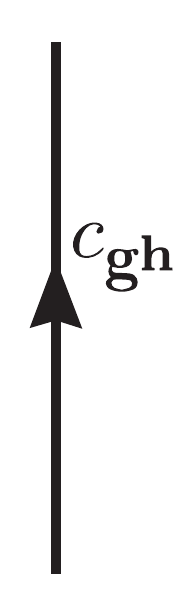}{0.7}
,
\end{equation}
which relates fusion and splitting spaces as duals.
We require the inner product to be Hermitian, which requires that we choose the quantum dimensions to be equal to the Frobenius-Perron dimensions, i.e. $d_{a}$ equals the largest (positive) eigenvalue of the matrix $\mathbf{N}_{a}$ defined by  $[\mathbf{N}_{a}]_{bc} = N_{a b}^c$.
This Hermitian inner product provides the fusion category with a pivotal structure, i.e. it enables us to bend the the lines, with appropriate unitary transformations.
It also makes it spherical, i.e. $d_a = d_{\bar{a}}$.
The quantum dimensions obey the relation
\begin{equation}
\label{qd_rel}
d_{a_{\bf{g}}} d_{b_{\bf{h}}} = \sum_{c_{\bf gh}} N_{a b}^c d_{c_{\bf{gh}}}
.
\end{equation}

We also define the total quantum dimension of a ${\bf g}$-sector $\mathcal{B}_{\bf g}$ to be
\begin{equation}
\mathcal{D}_{\bf{g}} = \sqrt{\sum_{a_{\bf{g}}} d_{a_{\bf{g}}}^2}
.
\end{equation}
For any ${\bf{g}} \in G$ with nonempty $\mathcal{B}_{\bf{g}} \neq \emptyset$, we have $\mathcal{D}_{\bf{g}} = \mathcal{D}_{\bf{0}}$.

The inner product allows us to write the partition of identity for a pair of charges $a_{\bf g}$ and $b_{\bf h}$ as
\begin{equation}
\label{eq:Id}
\openone_{ab} =
\gineq{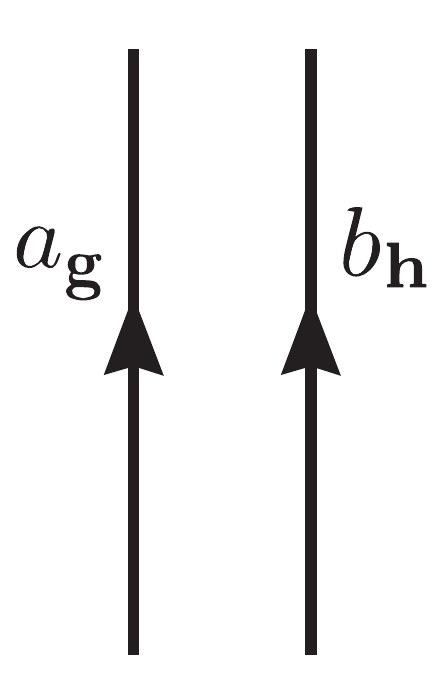}{1.25}
 = \sum\limits_{c_{\bf gh},\mu }
\sqrt{\frac{d_{c}}{d_{a}d_{b}}} \;
\gineq{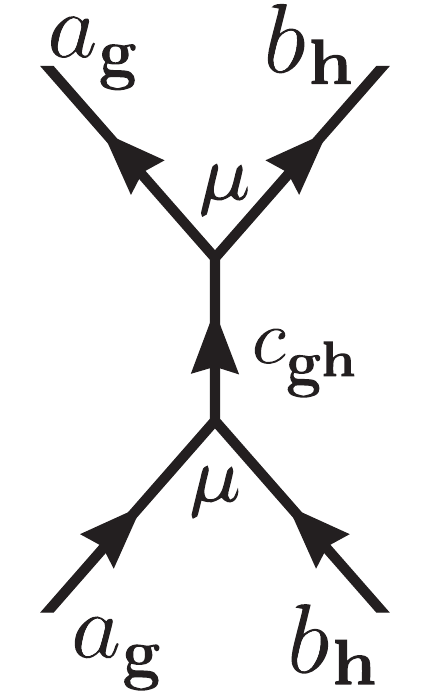}{1.25}
.
\end{equation}

The notion of associativity of fusion on the state space is encoded in the $F$-moves
\begin{equation}
\gineq{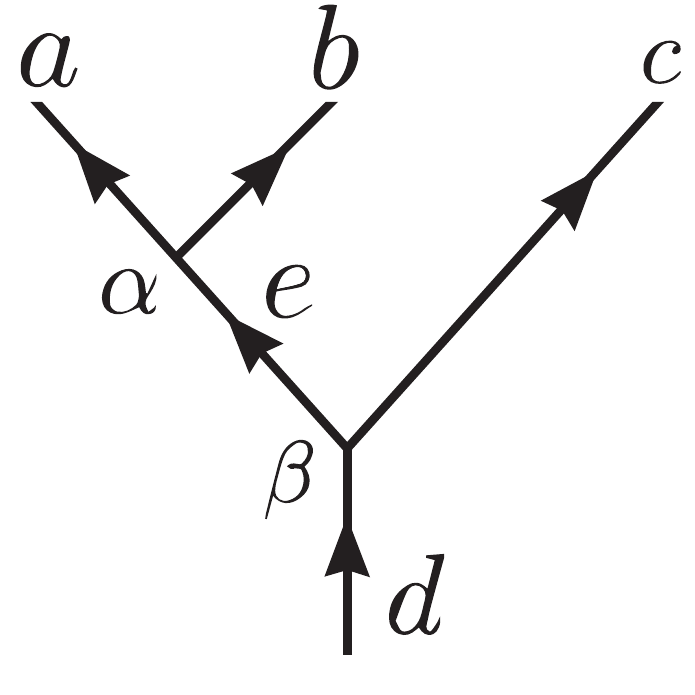}{1.5}
= \sum_{f,\mu,\nu} \left[F_d^{abc}\right]_{(e,\alpha,\beta)(f,\mu,\nu)}
\gineq{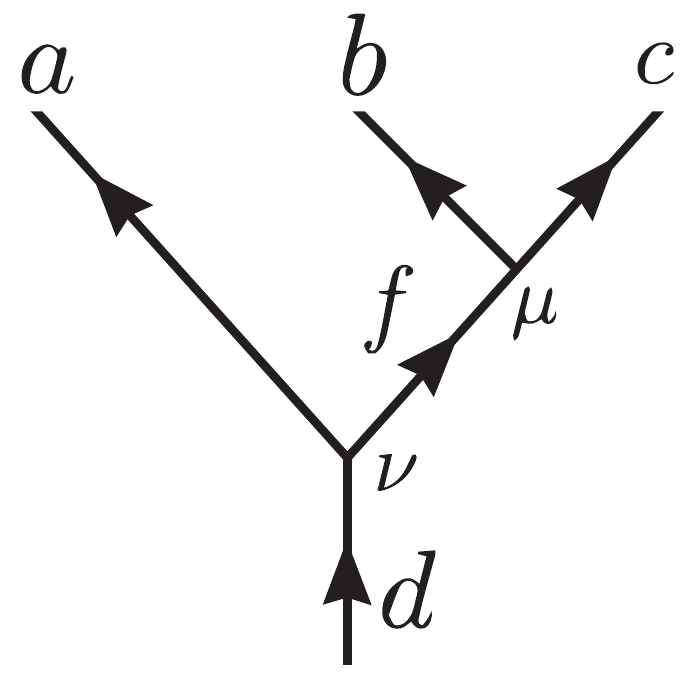}{1.5}
.
\end{equation}
These may be viewed as changes of bases, and are required to satisfy consistency conditions known as the pentagon equations.
We make a canonical gauge choice such that $F^{abc}_{d} = \openone$ whenever any of $a,b,c$ is the vacuum charge $0$.
For a unitary FTC, the $F$-moves are required to be unitary transformations, i.e. the $F$-symbols satisfy
\begin{eqnarray}
\left[ \left( F_{d}^{abc}\right) ^{-1}\right] _{\left( f,\mu
,\nu \right) \left( e,\alpha ,\beta \right) }
&=& \left[ \left( F_{d}^{abc}\right) ^{\dagger }\right]
_{\left( f,\mu ,\nu \right) \left( e,\alpha ,\beta \right) }
= \left[ F_{d}^{abc}\right] _{\left( e,\alpha ,\beta \right) \left( f,\mu
,\nu \right) }^{\ast }
.
\end{eqnarray}

Line bending can be related to $F$-moves through the relation
\begin{equation}
\gineq{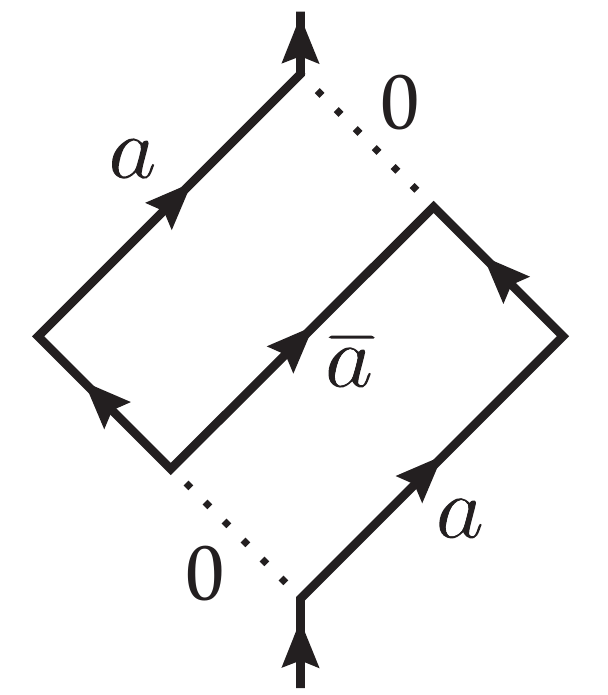}{2.5}
= \varkappa_{a}
\gineq{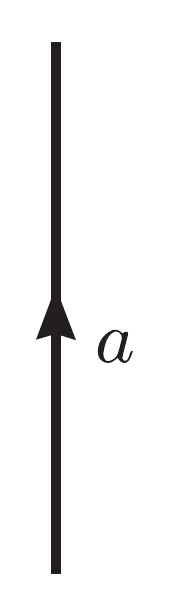}{0.8}
,
\end{equation}
from which we see that
\begin{equation}
\label{eq:FSInd}
[F^{a \bar{a} a}_a]_{00} = \frac{\varkappa_{a}}{d_a}
.
\end{equation}
Here, $\varkappa_{a} = \varkappa_{\bar{a}}^{-1}$ is a phase, not necessarily equal to one.
When $a=\bar{a}$, $\varkappa_{a} = \pm 1$ is a gauge invariant quantity known as the Frobenius-Shur indicator.

$G$-crossed braiding is a generalization of regular braiding that incorporates symmetry action and fractionalization of the group $G$.
Unlike regular braiding, the topological charges of the braided objects need not remain fixed, and compatibility of braiding with fusion does not simply require that sliding lines over or under fusion vertices be trivial.
In order to incorporate $G$-crossed braiding, we select a group action $\rho : G \mapsto \text{Aut}(\mathcal{B}_{G}^{\times})$, where the details of what it means to be an element of $\text{Aut}(\mathcal{B}_{G}^{\times})$ can be imposed as additional consistency conditions, which we will state below.
We write a shorthand for the symmetry action on topological charges as
\begin{equation}
\,^{\bf k}a_{\bf g} = \rho_{\bf k}(a_{\bf g})  \in \mathcal{B}_{\bf kg\bar{k}}
.
\end{equation}
At the level of fusion rules, these permutations of topological charge must satisfy
\begin{equation}
N_{ \,^{\bf k}a_{\bf g} \,^{\bf k}b_{\bf h}}^{\,^{\bf k}c_{\bf gh}} = N_{a_{\bf g} b_{\bf h}}^{c_{\bf gh}}
,
\end{equation}
which also implies
\begin{equation}
d_{a_{\bf g}} = d_{\,^{\bf k}a_{\bf g} }
.
\end{equation}

Next, we define the $G$-crossed braiding operation as 
\begin{eqnarray}
R^{a_{\bf g} b_{\bf h}}&=&
\gineq{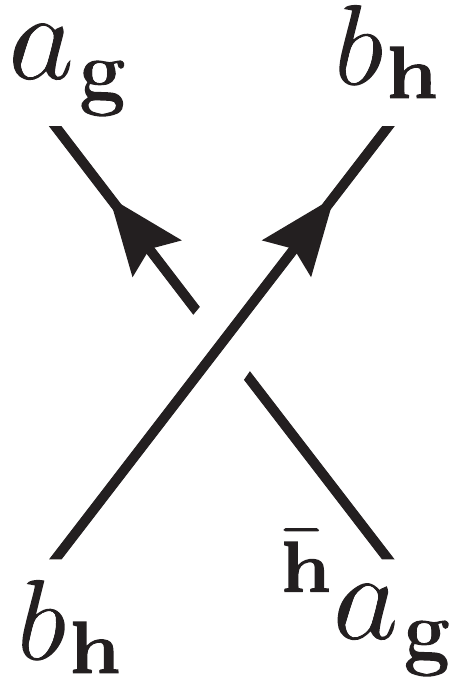}{1.25}
=\sum\limits_{c,\mu ,\nu }\sqrt{\frac{d_{c}}{d_{a}d_{b}}}\left[
R_{c_{\bf gh}}^{a_{\bf g} b_{\bf h}}\right] _{\mu \nu }
\gineq{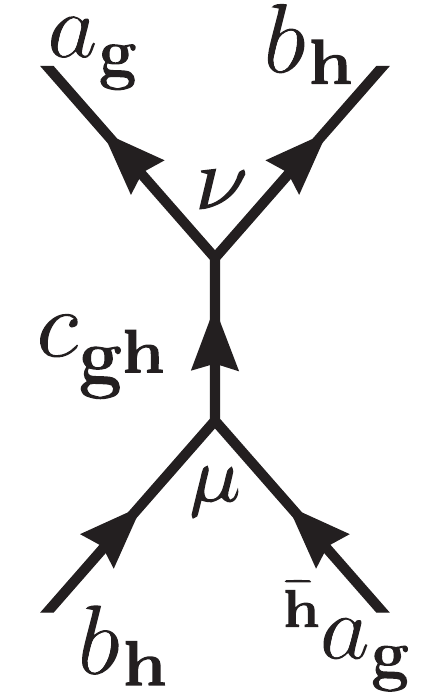}{1.25}
,
\label{eq:G-crossed_R}
\end{eqnarray}%
which yield maps between splitting/fusion state spaces $R^{ab}_{c} : V^{b_{\bf h} \,^{\bf \bar h}a_{\bf g}}_{c_{\bf gh}} \rightarrow V^{a_{\bf g} b_{\bf h}}_{c_{\bf gh}}$ that result from exchanging the two defects in a counterclockwise manner.
This operator incorporates the symmetry action of the over-crossing line's group label on the under-crossing line's topological charge, which corresponds to a convention where the symmetry defect branch-sheets go into the page from the defect world-lines.
Similarly, the clockwise $G$-crossed braiding exchange operator is
\begin{eqnarray}
&& \left(R^{ a_{\bf g} b_{\bf h}}\right)^{-1} =
\gineq{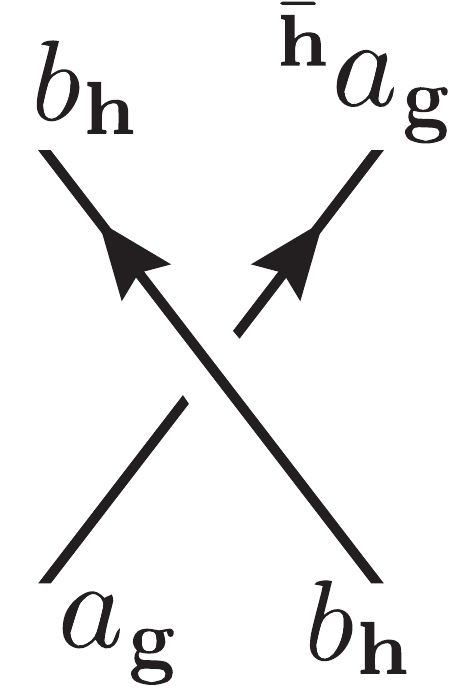}{1.25}
\,\,
=\sum\limits_{c,\mu ,\nu }\sqrt{\frac{d_{c}}{d_{a}d_{b}}}\left[\left(
R_{c_{\bf gh}}^{a_{\bf g} b_{\bf h}}\right)^{-1}\right] _{\mu \nu }
 \gineq{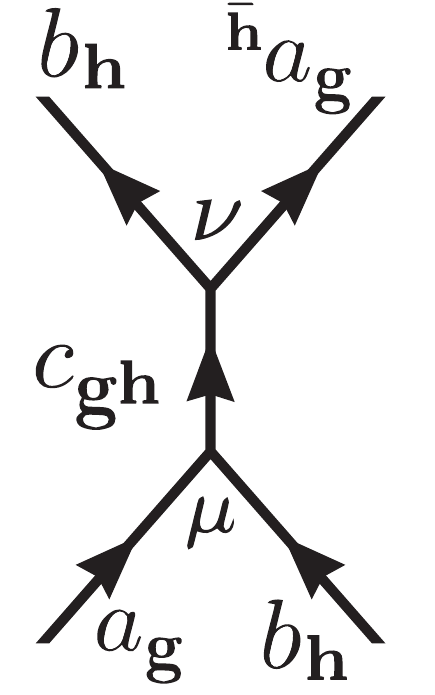}{1.25}
.
\end{eqnarray}
The $G$-crossed braiding can equivalently be specified in terms of the application of these operators to the state space, acting on the trivalent vertices as
\begin{equation}
\gineq{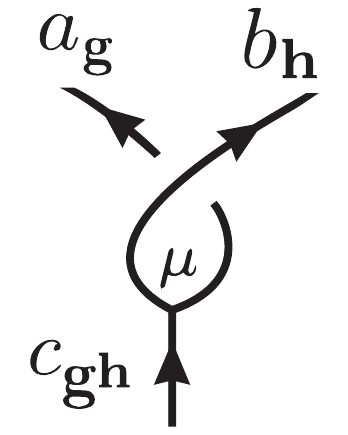}{1.25}
=\sum\limits_{\nu }\left[R_{c_{\bf gh}}^{a_{\bf g} b_{\bf h}}\right] _{\mu \nu }
\gineq{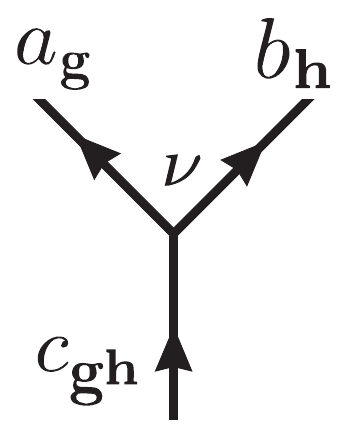}{1.25}
.
\end{equation}

The symmetry action on the topological state space is incorporated by sliding a defect line over a trivalent vertex, that is
\begin{eqnarray}
\label{eq:GcrossedU}
\gineq{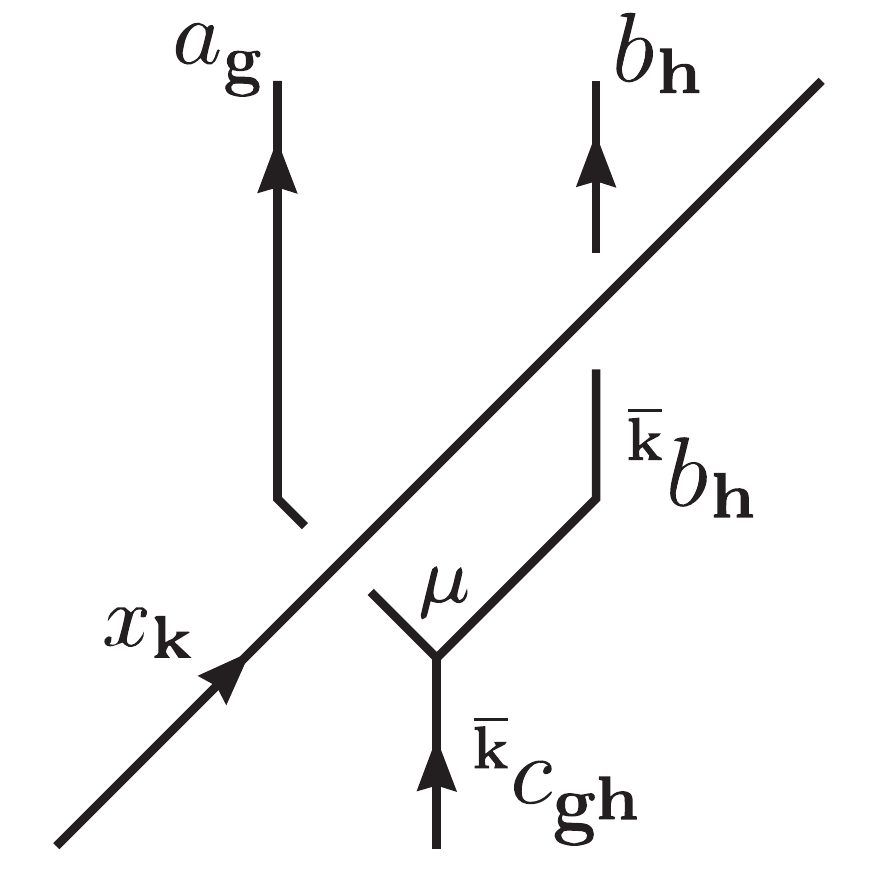}{2}
&=&
\sum_{\nu} \left[ U_{\bf k}\left(a_{\bf g}, b_{\bf h}; c_{\bf gh} \right)\right]_{\mu \nu}
\gineq{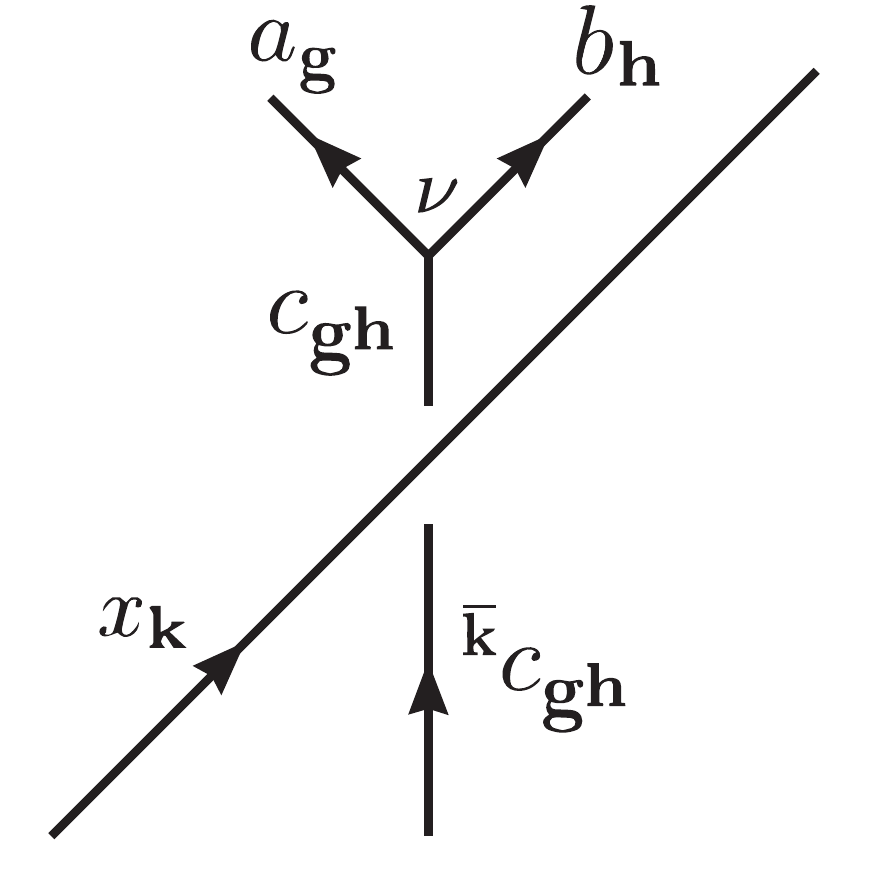}{2}
\,\, ,
\end{eqnarray}
while symmetry fractionalization is incorporated by sliding a line under a trivalent vertex, that is
\begin{eqnarray}
\label{eq:Gcrossed_eta}
\gineq{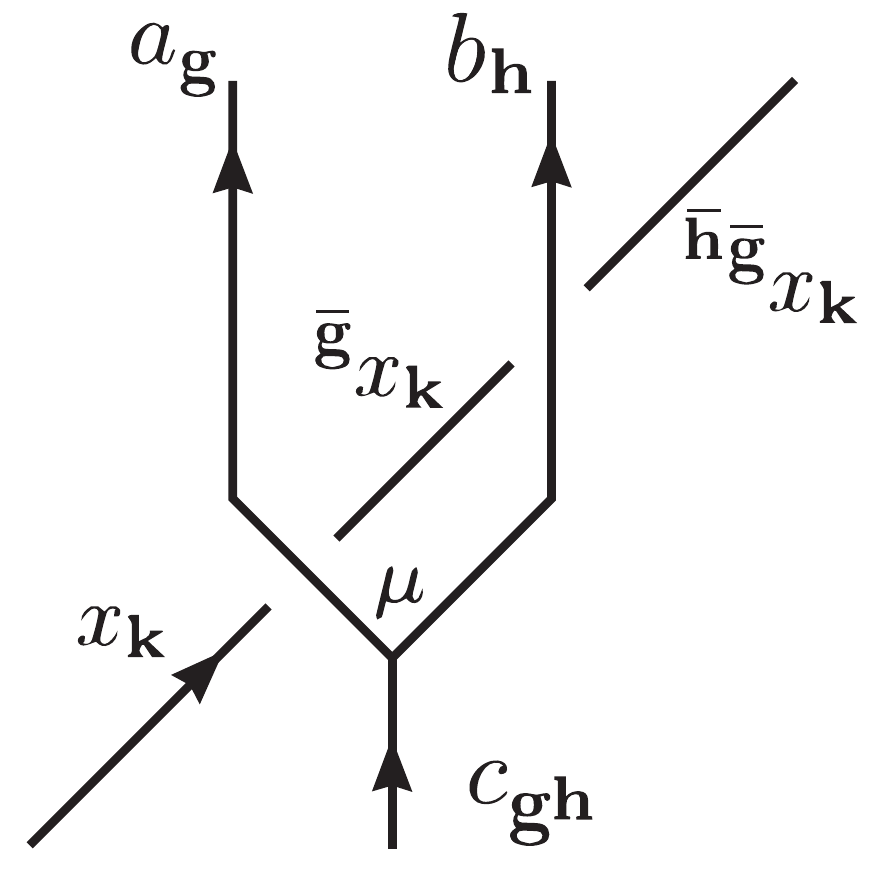}{2}
&=&
\eta_{x_{\bf k}}\left({\bf g},{\bf h}\right)
\gineq{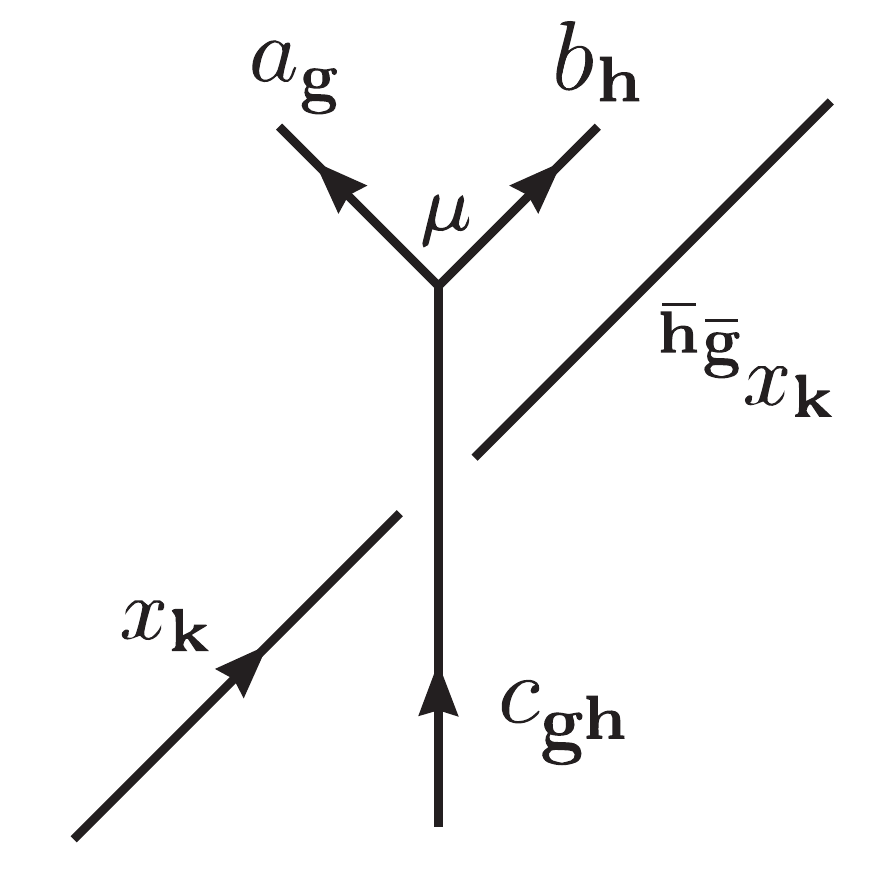}{2}
\,\, .
\end{eqnarray}
The $N$-, $F$-, $R$-, $U$-, and $\eta$-symbols, collectively referred to as the ``basic data,'' define a $G$-crossed BTC, and (in addition to the pentagon equation on the $F$-symbols) they are are required to satisfy consistency conditions known as the heptagon equations.  
Canonical gauge choices allow the expected quantities to be trivial, namely
\begin{align}
R^{a_{\bf g} 0}_{a_{\bf g}} &= R^{0 b_{\bf h}}_{b_{\bf h}} = 1 , 
\\
U_{\bf 0}\left(a_{\bf g},b_{\bf h};c_{\bf gh}\right) &= \openone ,
\\
U_{\bf k}\left(a_{\bf g},0;a_{\bf g}\right) &= U_{\bf k}\left(0,b_{\bf h};b_{\bf h}\right) = 1 ,
\\
\eta_{0}\left({\bf g},{\bf h}\right) &= \eta_{x_{\bf k}}\left({\bf g},{\bf 0}\right) = \eta_{x_{\bf k}}\left({\bf 0},{\bf h}\right) = 1
.
\end{align}
For a unitary $G$-crossed BTC, the $F$-, $R$-, $U$-, and $\eta$-symbols are all required to be unitary transformations.

The basic data satisfies consistency conditions that allows us to interpret the $U$-symbols as corresponding to the symmetry action and the $\eta$-symbols as corresponding to symmetry fractionalization.
Compatibility of the $F$-moves with sliding lines over and under fusion vertices yields
\begin{eqnarray}
&&\sum_{\alpha',\beta',\mu'\nu'} \left[U_{\bf k}(\,^{\bf k}a,\,^{\bf k}b ;\,^{\bf k}e )\right]_{\alpha \alpha'} \left[U_{\bf k}(\,^{\bf k}e, \,^{\bf k}c ;\,^{\bf k}d)\right]_{\beta \beta'} \left[F_{\,^{\bf k}d}^{ \,^{\bf k}a \,^{\bf k}b \,^{\bf k}c}\right]_{(^{\bf k}e,\alpha',\beta')(^{\bf k}f,\mu',\nu')} \notag \\
&& \qquad \qquad \qquad \times \left[U_{\bf k}(\,^{\bf k}b,\,^{\bf k}c ;\,^{\bf k}f)^{-1}\right]_{\mu' \mu} \left[U_{\bf k}(\,^{\bf k}a,\,^{\bf k}f ;\,^{\bf k}d)^{-1}\right]_{\nu' \nu} = \left[F_{d}^{abc}\right]_{(e,\alpha,\beta)(f,\mu,\nu)}
,
\label{eq:G-crossed_F_consistency}
\end{eqnarray}
and
\begin{equation}
\eta_{^{\bf \bar{g}}x}\left({\bf h}, {\bf k} \right) \eta_{x}\left({\bf g}, {\bf hk} \right) = \eta_{x}\left({\bf g}, {\bf h} \right) \eta_{x}\left({\bf gh}, {\bf k} \right)
.
\label{eq:eta_consistency}
\end{equation}
Compatibility of the $R$-moves with sliding lines over and under fusion vertices yields the $G$-crossed Yang-Baxter equation
\begin{equation}
\gineq{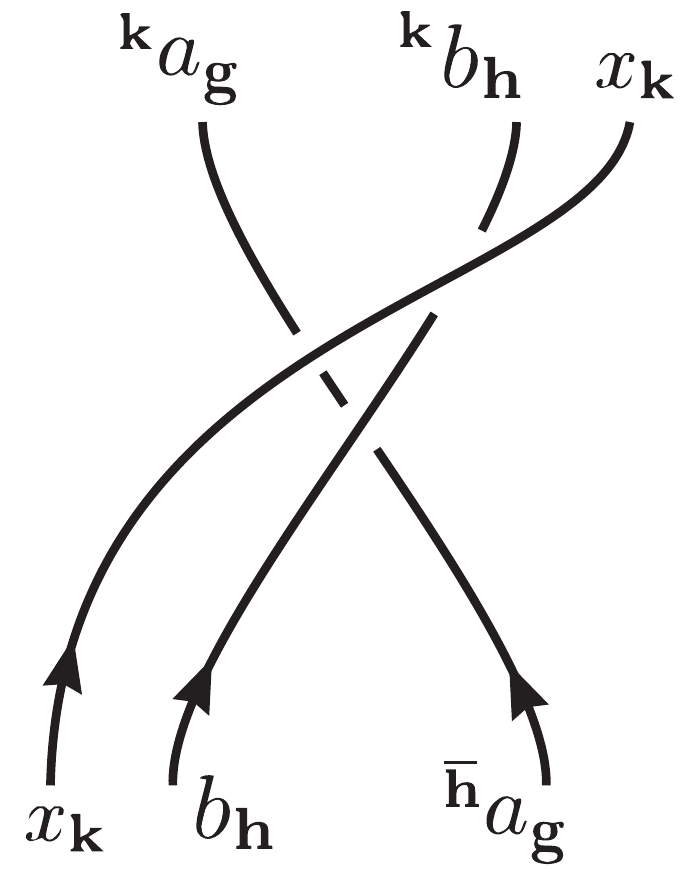}{2.25}
=\frac{\eta_{\,^{\bf k}a}({\bf kh\bar{k}},{\bf k}) }{\eta_{\,^{\bf k}a}({\bf k},{\bf h})}
\gineq{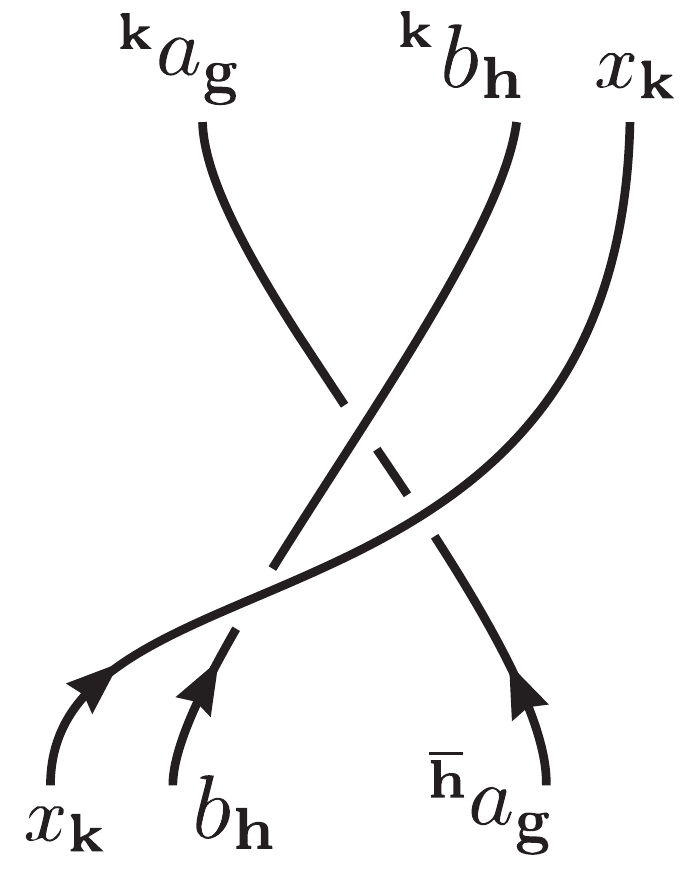}{2.25}
\label{eq:G-crossed_Yang_Baxter}
\end{equation}
and
\begin{equation}
\label{eq:R_invariant}
\frac{\eta_{\,^{\bf k}a_{\bf g}}({\bf kh\bar{k}},{\bf k}) }{ \eta_{\,^{\bf k}a_{\bf g}}({\bf k},{\bf h} )} \sum_{\mu',\nu'} \left[U_{\bf k}(\,^{\bf k}b_{\bf h} , \,^{\bf k \bar{h}}a_{\bf g} ; \,^{\bf k} c_{\bf gh} ) \right]_{\mu \mu'} \left[ R^{ \,^{\bf k}a_{\bf g} \,^{\bf k}b_{\bf h}}_{\,^{\bf k}c_{\bf gh}} \right]_{\mu' \nu'} \left[ U_{\bf k} (\,^{\bf k}a_{\bf g} , \,^{\bf k}b_{\bf h} ; \,^{\bf k}c_{\bf gh} )^{-1} \right]_{\nu' \nu} =   \left[ R^{a_{\bf g} b_{\bf h}}_{c_{\bf gh}} \right]_{\mu \nu}
.
\end{equation}
Finally, compatibility of sliding two vertices over and under each other yields
\begin{equation}
\label{eq:U_eta_consistency}
\sum_{\alpha, \beta} \left[ U_{\bf k}\left( a ,b ;c \right)^{-1} \right]_{\mu \alpha} \left[ U_{\bf l}\left( \,^{\bf \bar k}a ,\,^{\bf \bar k}b ;\,^{\bf \bar k}c \right)^{-1} \right]_{\alpha \beta} \left[ U_{\bf kl}\left( a ,b ;c \right) \right]_{\beta \nu}
= \frac{ \eta_{a}\left({\bf k}, {\bf l} \right) \eta_{b}\left({\bf k}, {\bf l} \right)}{\eta_{c}\left({\bf k}, {\bf l} \right)} \delta_{\mu \nu}
.
\end{equation}

BTCs have important gauge invariant quantities known as the topological $S$-matrix and the topological twists.
For $G$-crossed BTCs, the similarly defined quantities are no longer gauge invariant (in the case of the defects), but they remain important.
The topological twists are phases defined by
\begin{equation}
\theta_{a_{\bf g}}
= \frac{1}{d_{a_{\bf g}}} \, 
\gineq{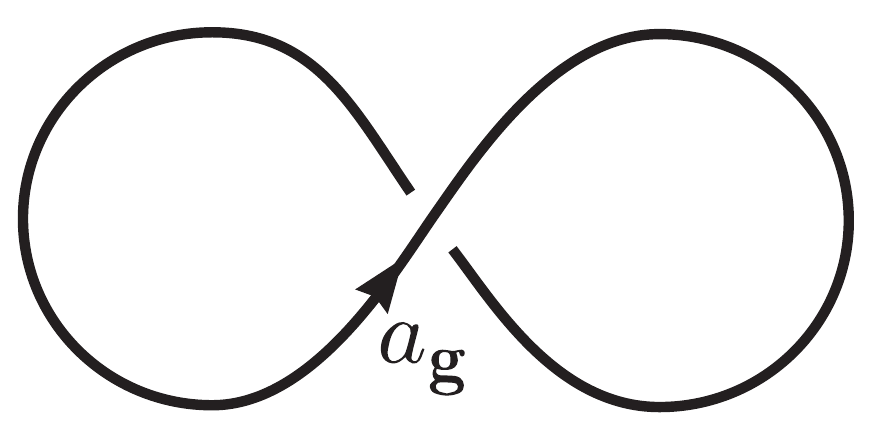}{2.5}
=\sum\limits_{c_{\bf gg},\mu } \frac{d_{c_{\bf gg}}}{d_{a_{\bf g}}}\left[ R_{c_{\bf gg}}^{a_{\bf g} a_{\bf g}}\right] _{\mu \mu }
.
\end{equation}
The topological $S$-matrix is defined as
\begin{eqnarray}
\label{eqn:topoSmatrix}
S_{a_{\bf g} b_{\bf h} }&=&\frac{1}{\mathcal{D}_\mb{0}} \, 
\gineq{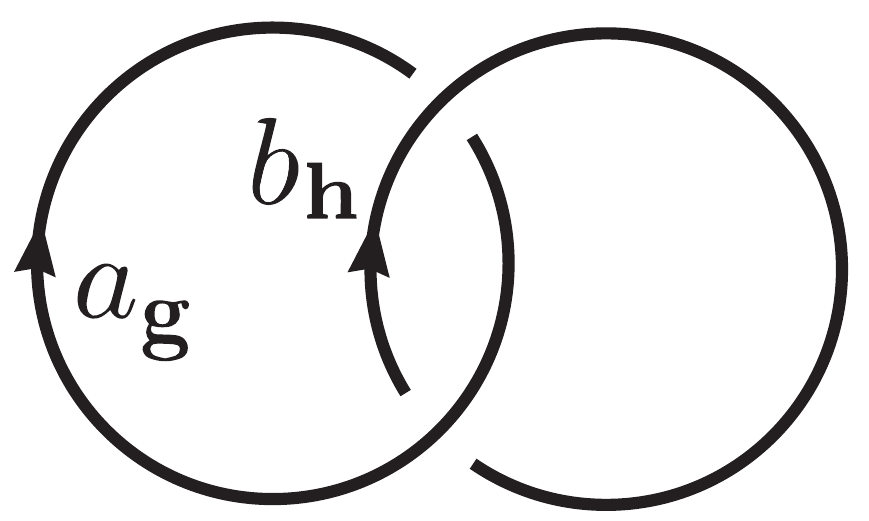}{2}
\notag \\
&=&\frac{1}{\mathcal{D}_\mb{0}}\sum_{c_{\bf \bar{g}h},\mu,\nu}d_{c_{\bf \bar{g}h}} \left[ R^{b_{\bf h} \overline{a_{\bf g}} }_{c_{\bf \bar{g}h}} \right]_{\mu \nu} \left[R_{c_{\bf \bar{g}h}}^{ \overline{a_{\bf g}} b_{\bf h}} \right]_{\nu\mu}
\notag\\
&=&\frac{1}{\mathcal{D}_\mb{0}}\sum_{c_{\bf \bar{g}h},\mu} d_{c_{\bf \bar{g}h}} 
\frac{\theta_{c_{\bf \bar{g}h}}}{\theta_{\overline{a_{\bf g}} }\theta_{b_{\bf h}}} \frac{\left[ U_{\bf \bar{g}h}(\overline{a_{\bf g}},b_{\bf h};c_{\bf \bar{g}h})\right]_{\mu \mu} }{\eta_{\overline{a_{\bf g}}}({\bf \bar{g},h}) \eta_{b_{\bf h}}({\bf h , \bar{g} })}
.
\end{eqnarray}
One should note that the quanitity above is only well-defined if ${\bf gh = hg }$, otherwise the loops will not be able to close back upon themselves, as the topological charge values would change upon braiding.
In the third equality, we have used the $G$-crossed ribbon identity 
\begin{equation}
\sum_\lambda \left[R^{b_{\bf h} \,^{\bf \bar{h}}a_{\bf g}}_{c_{\bf gh}}\right]_{\mu \lambda} \left[R^{a_{\bf g} b_{\bf h}}_{c_{\bf gh}} \right]_{\lambda \nu} =
\frac{\theta_{c_{\bf gh}}}{\theta_{a_{\bf g}}\theta_{b_{\bf h}}} \frac{\left[ U_{\bf gh}(a_{\bf g},b_{\bf h};c_{\bf gh})\right]_{\mu\nu} }{\eta_{a_{\bf g}}({\bf g,h}) \eta_{b_{\bf h}}({\bf h},\,^{\bf \bar{h}}{\bf g})}
.
\label{eq:G-crossed_ribbon}
\end{equation}
Additionally, we define the ``punctured torus'' $S^{(z)}$-matrix (corresponding to a torus with boundary carrying topological charge $z$) as 
\begin{align}
S_{(a_{\bf g},\mu)(b_{\bf h},\nu)}^{({\bf g,h};z_{\bf \bar{h}gh\bar{g}})} &=
\frac{1}{\mathcal{D}_{\bf 0} \sqrt{d_{z_{\bf \bar{h}gh\bar{g}}}} } \,
\gineq{Equations/Defs/inner_prod_low.pdf}{3.5} 
\notag \\
&= \frac{1}{\mathcal{D}_\mb{0}} \sum_{\substack{c_{\bf \bar{g}h} \\ \alpha,\beta,\gamma, \lambda}}
d_{a} d_{b}
\left[(F^{\bar{z} \,^{\bf \bar{h}g}b \,^{\bf \bar{h}g}\bar{b}}_{\bar{z}})^{-1}\right]_{0 (b,\alpha,\nu)}
\left[F^{a \,^{\bf \bar{h}}\bar{a} \,^{\bf \bar{h}g}b}_{b}\right]_{(\bar{z},\mu,\alpha) (c,\beta,\gamma)}
\notag \\
& \qquad \qquad \qquad \times
\frac{\theta_{c}}{\theta_{\bar{a} }\theta_{b}} \frac{\left[ U_{\bf \bar{g}h}(\bar{a},b;c)\right]_{\mu \mu} }{\eta_{\bar{a}}({\bf \bar{g},h}) \eta_{b}({\bf h , \bar{g} })}
\left[(F^{a \bar{a} b}_{b})^{-1} \right]_{(c,\lambda,\gamma) 0}
.
\label{eq:S^z}
\end{align}
In this case, ${\bf g}$ and ${\bf h}$ are not required to commute.
It is straightforward to see that $S^{(0)} = S$.

The definition of the $S$-matrix gives the property
\begin{equation}
\label{eq:loop_removal}
\gineq{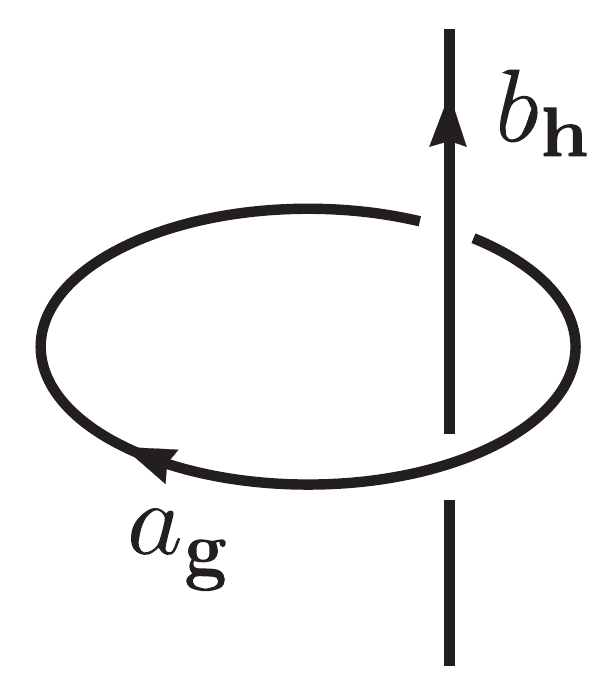}{1.8}
=\frac{S_{a_{\bf g} b_{\bf h}}}{S_{0 b_{\bf h}} }
\, \gineq{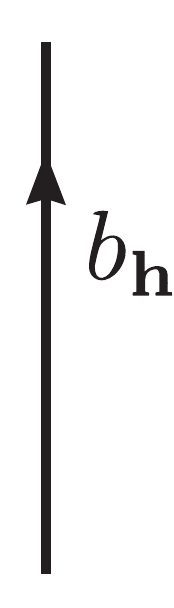}{0.6}
.
\end{equation}
Applying this twice gives the relation
\begin{align}
\frac{S_{x_{\bf k} a_{\bf g}}}{S_{x_{\bf k} 0}} \frac{S_{x_{\bf k} b_{\bf h}}}{S_{x_{\bf k} 0}}
= \sum_{c \in \mathcal{B}_{\bf gh}^{\bf k} , \mu}
\frac{[U_{\bf \bar{k}}(a_{\bf g}, b_{\bf h} ; c_{\bf gh})]_{\mu \mu }}{\eta_{\overline{x_{\bf k}} }({\bf g}, {\bf h})}
\frac{S_{x_{\bf k} c_{\bf gh}}}{S_{x_{\bf k} 0}}
,
\label{eq:G_pre_Verlinde}
\end{align}
when $^{\bf k}a_{\bf g} = a_{\bf g}$, $^{\bf k}b_{\bf h} = b_{\bf h}$, and $^{\bf g}x_{\bf k} = ^{\bf h}x_{\bf k} = x_{\bf k}$.

We will end this review section with a collection of additional derived properties that will be useful for our calculations.

Eq.~(\ref{eq:G-crossed_F_consistency}) with $e=f=0$ yields the relation
\begin{equation}
\frac{\varkappa_{\,^{\bf k}a} }{ \varkappa_{a} } =\frac{ \left[F_{\,^{\bf k}a}^{ \,^{\bf k}a \,^{\bf k}\bar{a} \,^{\bf k}a}\right]_{00} } {\left[F_a^{a\bar{a}a}\right]_{00}}
=\frac{U_{\bf k}(\,^{\bf k}\bar{a},\,^{\bf k}a ;0)}{ U_{\bf k}(\,^{\bf k}a, \,^{\bf k}\bar{a};0) }
.
\label{eq:G-crossed_FS}
\end{equation}
When $a = \bar{a}$, it follows from Eq.~(\ref{eq:G-crossed_FS}) that  $\varkappa_{a}=\varkappa_{\,^{\bf k}a}$.
When $^{\bf k}a = a$ is ${\bf k}$-invariant, it follows from Eq.~(\ref{eq:G-crossed_FS}) that
\begin{align}
U_{\bf k}(a,\bar{a};0) = U_{\bf k}(\bar{a},a;0).
\label{eq:U_k_relation}
\end{align}

Using Eq.~(\ref{eq:G-crossed_Yang_Baxter}) with the definition of the twist, we find the general relation between $\theta _{a}$ and $\theta _{\,^{\bf k}a}$ is
\begin{equation}
\theta_{a_{\bf g}} = \frac{\eta_{\,^{\bf k}a_{\bf g}}({\bf kg \bar{k} , k}) }{\eta_{\,^{\bf k}a_{\bf g}}({\bf k, g})} \theta_{\,^{\bf k}a_{\bf g}} = \frac{\eta_{a_{\bf g}}({\bf \bar{k} , kg\bar{k} }) }{\eta_{a_{\bf g}}({\bf g, \bar{k}})} \theta_{\,^{\bf k}a_{\bf g}}
.
\label{eqn:action_on_twist}
\end{equation}
When $^{\bf k}a = a$, it follows that
\begin{equation}
\eta_{a_{\bf g}}({\bf g, k}) = \eta_{a_{\bf g}}({\bf k, g}).
\label{eqn:etasym}
\end{equation}
We also note that Eq.~(\ref{eq:eta_consistency}) gives $\eta_{^{\bf k}x} (\mb{k},\mb{ \bar{k} }) = \eta_{x} (\mb{ \bar{k} } , \mb{k})$ for any $x$ and ${\bf k}$, so we also have
\begin{equation}
\eta_{a_{\bf g}}({\bf k,\bar{k}}) = \eta_{a_{\bf g}}({\bf \bar{k},k})
\label{eqn:etasym_2}
\end{equation}
when $^{\bf k}a = a$.

The topological $S$-matrix for defects can also transform nontrivially under symmetry action, as it satisfies the relation
\begin{equation}
S_{\,^{\bf k}a_{\bf g} \,^{\bf k} b_{\bf h} }= \frac{\eta_{\,^{\bf k}\bar{a}}({\bf k,h}) \eta_{\,^{\bf k}b}({\bf k , \bar{g} })} {\eta_{\,^{\bf k}\bar{a}}({\bf kh\bar{k},k}) \eta_{\,^{\bf k}b}({\bf k\bar{g}\bar{k} , k }) }S_{a_{\bf g} b_{\bf h} }
.
\end{equation}

Unlike a BTC, it is not necessarily the case that $\theta_{a_{\bf g}}$ and $\theta_{\overline{a_{\bf g}}}$ are equal in a $G$-crossed BTC. 
In particular, we have
\begin{equation}
\theta_{a_{\bf g}}= U_\mathbf{g}(\overline{a_{\bf g}},a_{\bf g};0) \eta_{\overline{a_{\bf g}}}(\mathbf{\bar{g}},{\mb{g}}) \theta_{\overline{a_{\bf g}}} .
\label{eq:lemmatwist}
\end{equation}
and
\begin{eqnarray}
\theta_{a_{\bf g}} &=& U_\mathbf{g}(a_{\bf g},\overline{a_{\bf g}};0) \varkappa_{a_{\bf g}} \left( R^{ \overline{a_{\bf g}} a_{\bf g} }_{0}\right)^{-1} 
= \eta_{a_{\bf g}}({\bf g} , {\bf \bar{g}})^{-1} \varkappa_{a_{\bf g}}^{-1} \left( R^{a_{\bf g} \overline{a_{\bf g}}  }_{0}\right)^{-1}
\label{eq:theta_R}
\end{eqnarray}

Eq.~\eqref{eq:U_eta_consistency} when ${\bf l}={\bf \bar{k}}$ gives
\begin{eqnarray}
    [U_{\bf \bar{k}}(^{\bf \bar{k}}a, ^{\bf \bar{k}}b; ^{\bf \bar{k}}c)]_{\mu \nu} = \frac{\eta_c({\bf k, \bar{k}})}{\eta_a({\bf k, \bar{k}}) \eta_b({\bf k, \bar{k}})} [U_{\bf k}(a, b; c)^{-1}]_{\mu \nu} \label{eq:U&eta_consistency_1}
\end{eqnarray}
and when $c=0$ gives
\begin{eqnarray}
    \eta_{a}({\bf k , l})\eta_{\bar{a}}({\bf k,l}) = \frac{U_{\bf kl}(a, \bar{a};0)}{U_{\bf k}(a, \bar{a};0) U_{\bf l}(^{\bf \bar{k}}a, ^{\bf \bar{k}}\bar{a};0)} .
    \label{eq:U&eta_consistency_2}
\end{eqnarray}

Eqs.~\eqref{eqn:topoSmatrix}, \eqref{qd_rel}, and \eqref{eq:lemmatwist} yield the property
\begin{equation}
\sum_{a_{\bf g}} \frac{d_a \theta_a}{ U_\mathbf{g}(\bar{a},a;0)}
\gineq{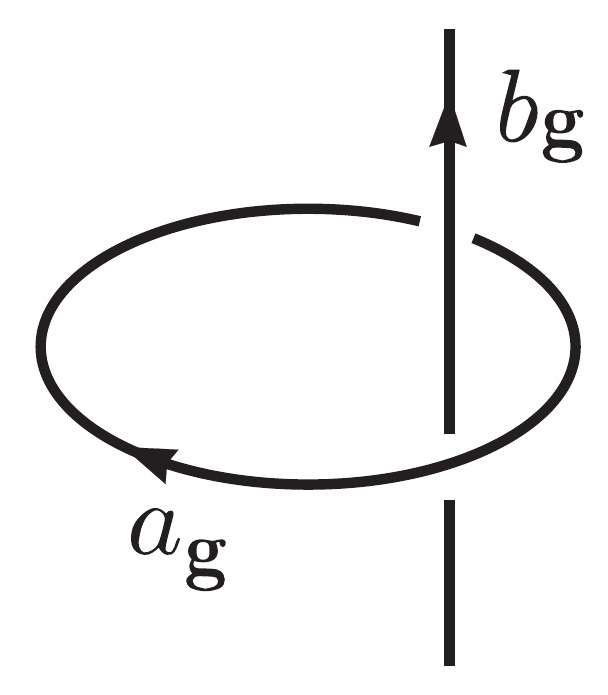}{1.5}
= \frac{ \mathcal{D}_\mb{0} \Theta_{\bf 0} }{ \eta_b({\bf g , \bar{g} }) \theta_{b} }
\gineq{Equations/Defs/chargeline_bg.pdf}{0.5}
\label{eq:loop_twist}
.
\end{equation}

\section{Sequences of Relations for Section~\ref{sec:MCG_relations}}

\begin{figure}[H]
    \centering
    \includegraphics[width=15cm]{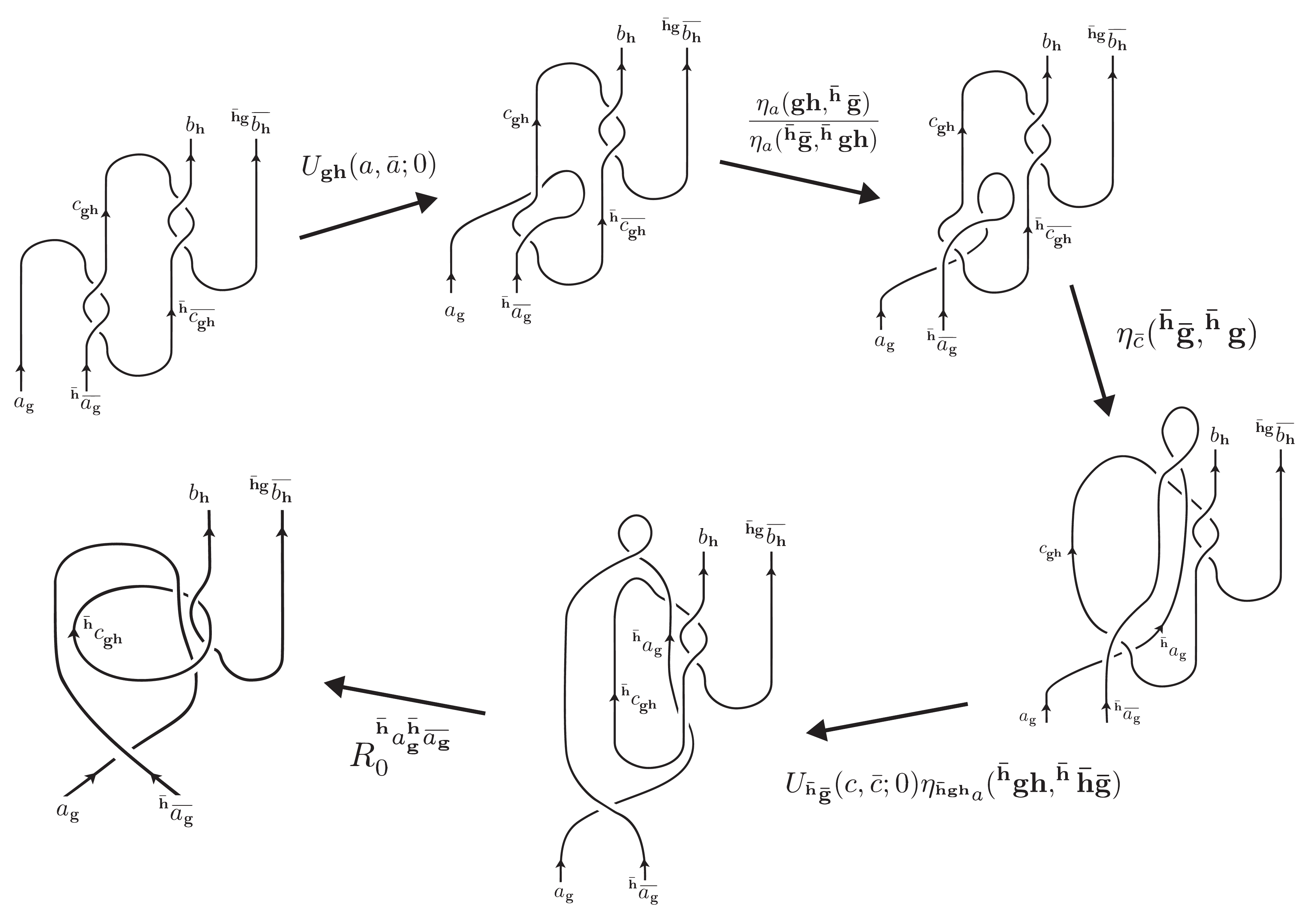}
    \caption{Diagrammatic steps used in deriving Eq.~\eqref{eq:ST3_1}.}
    \label{fig:ST3_1}
\end{figure}

\begin{figure}[H]
    \centering
    \includegraphics[width=13cm]{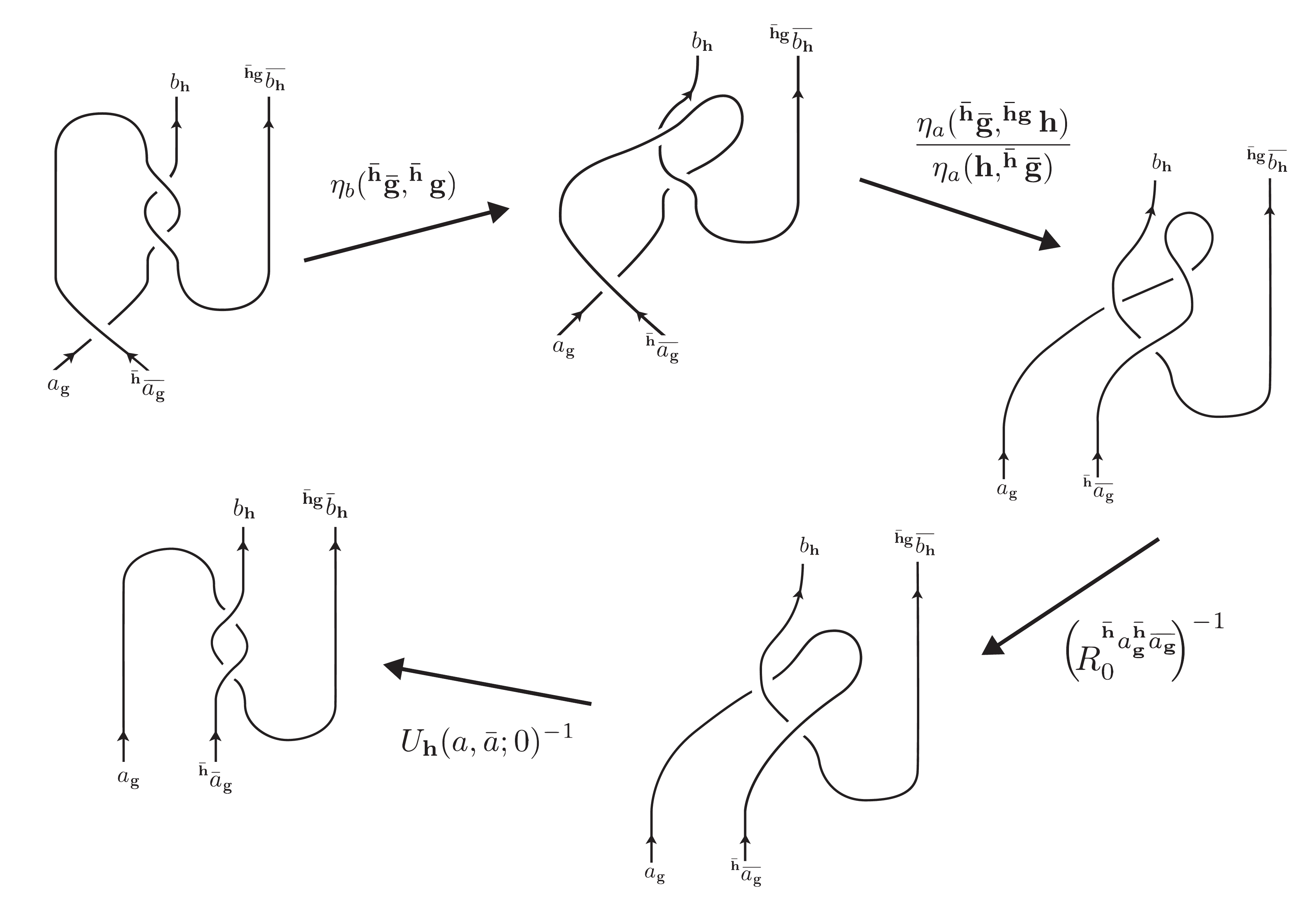}
    \caption{Diagrammatic steps used in deriving Eq.~\eqref{eq:ST3_final}.}
    \label{fig:ST3_2}
\end{figure}

\begin{figure}[H]
\centering
    \includegraphics[width=14cm]{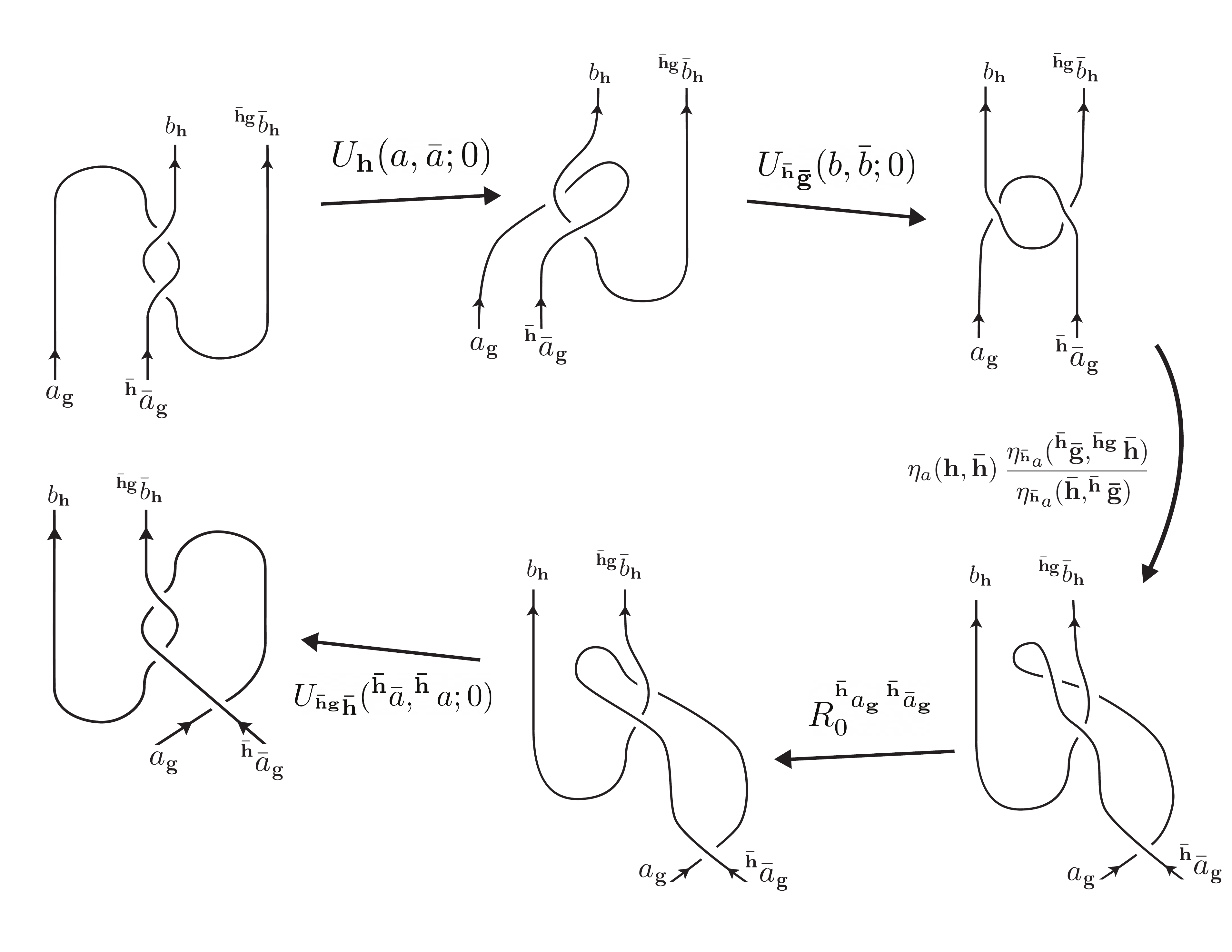}
    \caption{Diagrammatic steps used in deriving Eq.~\eqref{eq:SCS_1}.}
    \label{fig:SCS}
\end{figure}

\begin{figure}[H]
    \centering
    \includegraphics[width=13cm]{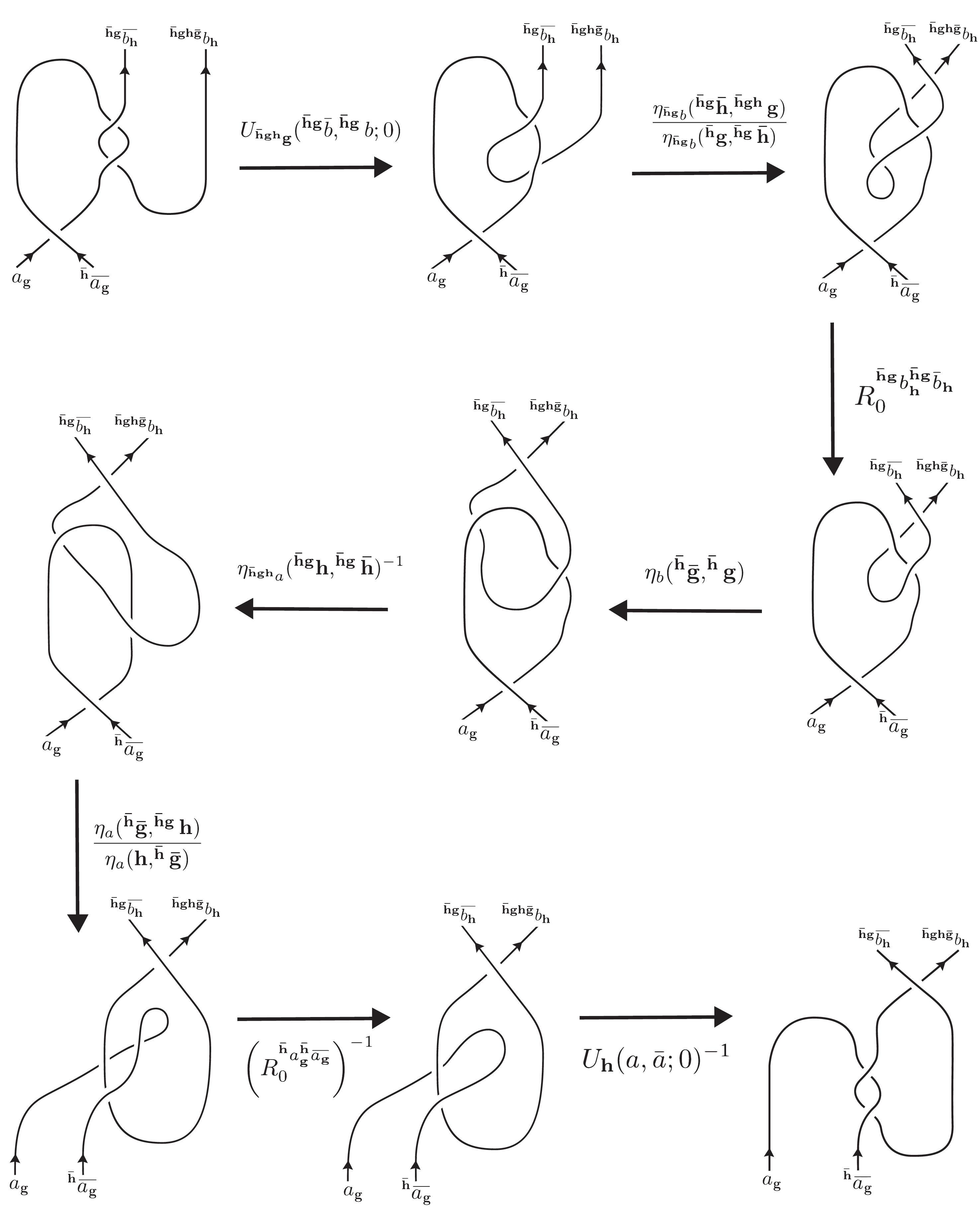}
    \caption{Diagrammatic steps used in deriving Eq.~\eqref{eq:SCCS}.}
    \label{fig:SCCS}
\end{figure}


\bibliographystyle{apsrev}
\bibliography{main}

\begin{thebibliography}{11}
\expandafter\ifx\csname natexlab\endcsname\relax\def\natexlab#1{#1}\fi
\expandafter\ifx\csname bibnamefont\endcsname\relax
  \def\bibnamefont#1{#1}\fi
\expandafter\ifx\csname bibfnamefont\endcsname\relax
  \def\bibfnamefont#1{#1}\fi
\expandafter\ifx\csname citenamefont\endcsname\relax
  \def\citenamefont#1{#1}\fi
\expandafter\ifx\csname url\endcsname\relax
  \def\url#1{\texttt{#1}}\fi
\expandafter\ifx\csname urlprefix\endcsname\relax\def\urlprefix{URL }\fi
\providecommand{\bibinfo}[2]{#2}
\providecommand{\eprint}[2][]{\url{#2}}

\bibitem[{\citenamefont{Wen}(2004)}]{Wen04}
\bibinfo{author}{\bibfnamefont{X.-G.} \bibnamefont{Wen}},
  \emph{\bibinfo{title}{Quantum Field Theory of Many-body Systems: From the
  Origin of Sound to an Origin of Light and Electrons}}, Oxford Graduate Texts
  (\bibinfo{publisher}{Oxford University Press}, \bibinfo{address}{Oxford},
  \bibinfo{year}{2004}).

\bibitem[{\citenamefont{Nayak et~al.}(2008)\citenamefont{Nayak, Simon, Stern,
  Freedman, and Sarma}}]{Nayak08}
\bibinfo{author}{\bibfnamefont{C.}~\bibnamefont{Nayak}},
  \bibinfo{author}{\bibfnamefont{S.~H.} \bibnamefont{Simon}},
  \bibinfo{author}{\bibfnamefont{A.}~\bibnamefont{Stern}},
  \bibinfo{author}{\bibfnamefont{M.}~\bibnamefont{Freedman}}, \bibnamefont{and}
  \bibinfo{author}{\bibfnamefont{S.~D.} \bibnamefont{Sarma}},
  \bibinfo{journal}{Rev. Mod. Phys.} \textbf{\bibinfo{volume}{80}},
  \bibinfo{pages}{1083} (\bibinfo{year}{2008}), \eprint{arXiv:0707.1889}.

\bibitem[{\citenamefont{Moore and Seiberg}(1989)}]{Moore89b}
\bibinfo{author}{\bibfnamefont{G.}~\bibnamefont{Moore}} \bibnamefont{and}
  \bibinfo{author}{\bibfnamefont{N.}~\bibnamefont{Seiberg}},
  \bibinfo{journal}{Commun. Math. Phys.} \textbf{\bibinfo{volume}{123}},
  \bibinfo{pages}{177} (\bibinfo{year}{1989}).

\bibitem[{\citenamefont{Turaev}(1994)}]{Turaev94}
\bibinfo{author}{\bibfnamefont{V.~G.} \bibnamefont{Turaev}},
  \emph{\bibinfo{title}{Quantum Invariants of Knots and 3-Manifolds}}
  (\bibinfo{publisher}{Walter de Gruyter}, \bibinfo{address}{Berlin, New York},
  \bibinfo{year}{1994}).

\bibitem[{\citenamefont{Bakalov and Kirillov}(2001)}]{Bakalov01}
\bibinfo{author}{\bibfnamefont{B.}~\bibnamefont{Bakalov}} \bibnamefont{and}
  \bibinfo{author}{\bibfnamefont{A.}~\bibnamefont{Kirillov}},
  \emph{\bibinfo{title}{Lectures on Tensor Categories and Modular Functors}},
  vol.~\bibinfo{volume}{21} of \emph{\bibinfo{series}{University Lecture
  Series}} (\bibinfo{publisher}{American Mathematical Society},
  \bibinfo{year}{2001}).

\bibitem[{\citenamefont{Turaev}()}]{Turaev2000}
\bibinfo{author}{\bibfnamefont{V.}~\bibnamefont{Turaev}},
  \eprint{math/0005291}.

\bibitem[{\citenamefont{Turaev}(2010)}]{turaev2010}
\bibinfo{author}{\bibfnamefont{V.}~\bibnamefont{Turaev}},
  \emph{\bibinfo{title}{Homotopy Quantum Field Theory}}
  (\bibinfo{publisher}{European Mathematical Society}, \bibinfo{year}{2010}).

\bibitem[{\citenamefont{Kirillov}()}]{Kirillov2004}
\bibinfo{author}{\bibfnamefont{A.~J.} \bibnamefont{Kirillov}},
  \eprint{math/0401119}.

\bibitem[{\citenamefont{Barkeshli et~al.}(2019)\citenamefont{Barkeshli,
  Bonderson, Cheng, and Wang}}]{Barkeshli2019}
\bibinfo{author}{\bibfnamefont{M.}~\bibnamefont{Barkeshli}},
  \bibinfo{author}{\bibfnamefont{P.}~\bibnamefont{Bonderson}},
  \bibinfo{author}{\bibfnamefont{M.}~\bibnamefont{Cheng}}, \bibnamefont{and}
  \bibinfo{author}{\bibfnamefont{Z.}~\bibnamefont{Wang}},
  \bibinfo{journal}{Physical Review B} \textbf{\bibinfo{volume}{100}},
  \bibinfo{pages}{115147} (\bibinfo{year}{2019}), \eprint{arXiv:1410.4540}.

\bibitem[{\citenamefont{{Bonderson} et~al.}(2017)\citenamefont{{Bonderson},
  {Knapp}, and {Patel}}}]{Bonderson2017}
\bibinfo{author}{\bibfnamefont{P.}~\bibnamefont{{Bonderson}}},
  \bibinfo{author}{\bibfnamefont{C.}~\bibnamefont{{Knapp}}}, \bibnamefont{and}
  \bibinfo{author}{\bibfnamefont{K.}~\bibnamefont{{Patel}}},
  \bibinfo{journal}{Annals of Physics} \textbf{\bibinfo{volume}{385}},
  \bibinfo{pages}{399} (\bibinfo{year}{2017}), \eprint{arXiv:1706.09420}.

\bibitem[{\citenamefont{Freedman et~al.}(2006)\citenamefont{Freedman, Nayak,
  and Walker}}]{Freedman06a}
\bibinfo{author}{\bibfnamefont{M.}~\bibnamefont{Freedman}},
  \bibinfo{author}{\bibfnamefont{C.}~\bibnamefont{Nayak}}, \bibnamefont{and}
  \bibinfo{author}{\bibfnamefont{K.}~\bibnamefont{Walker}},
  \bibinfo{journal}{Phys. Rev. B} \textbf{\bibinfo{volume}{73}},
  \bibinfo{pages}{245307} (\bibinfo{year}{2006}), \eprint{cond-mat/0512066}.

\end{thebibliography}

\end{document}